\documentclass{cimento}

\newif\ifarxiv\arxivtrue

\usepackage{tikz}
\usepackage{graphicx}
\usepackage{amsmath}
\usepackage{amssymb}
\usepackage{booktabs}
\usepackage{tabularx}
\usepackage{physics} 
\usepackage{comment}
\usepackage[numbers,sort&compress]{natbib}
\usepackage{xcolor}  
\usepackage{url}              \providecommand{\href}[2]{#2}

\definecolor{darkred}{RGB}{175,0,0}
\definecolor{darkblue}{RGB}{14,0,185}
\definecolor{salmon}{RGB}{255,160,105}

\newcommand{\beq}{\begin{equation}}
\newcommand{\eeq}{\end{equation}}

\newcommand{\ho}{\hat{\Omega}}
\newcommand{\hp}{\hat{p}}

\newcommand{\sect}[1]{sect.~\ref{#1}}
\newcommand{\fig}[1]{fig.~\ref{#1}}
\newcommand{\eq}[1]{eq.~\eqref{#1}}
\newcommand{\tab}[1]{tab.~\ref{#1}}

\usepackage{amsthm}
\theoremstyle{plain}

\theoremstyle{definition}

\newcommand{\lmaxrec}{\ell_{\max}^{\mathrm{rec}}}
\newcommand{\lmaxGWB}{\ell_{\max}^{\mathrm{GWB}}}

\title{Pulsar timing arrays: the emerging gravitational-wave landscape}

\shortauthor{C.~M.~F.~Mingarelli \textit{et al.}}

\author{C.~M.~F.~Mingarelli\from{ins:yale_phys} \from{ins:CCA}\thanks{E-mail: chiara.mingarelli@yale.edu},
        J.~A.~Casey-Clyde\from{ins:uconn}\from{ins:yale_phys},
        Y.~T.~Chang\from{ins:yale_astro},
        E.~Eisenberg\from{ins:yale_astro},
        F.~Hutchison\from{ins:yale_phys} 
        N.~Khusid\from{ins:stony},
        B.~Larsen\from{ins:yale_phys},
        A.~Moran\from{ins:columbia},
        F.~Semenzato\from{ins:padova},
        L.~Willson\from{ins:uconn}
        \atque
        Q.~Zheng\from{ins:yale_phys}}

\instlist{
  \inst{ins:yale_phys} Department of Physics, Yale University - New Haven, CT, USA
\inst{ins:CCA} Center for Computational Astrophysics, Flatiron Institute, 162 5th Avenue, New York, NY, 10010, USA
  \inst{ins:uconn} Department of Physics, University of Connecticut - Storrs, CT, USA
  \inst{ins:yale_astro} Department of Astronomy, Yale University - New Haven, CT, USA
  \inst{ins:stony} Department of Physics and Astronomy, Stony Brook University - Stony Brook, NY, USA
  \inst{ins:columbia} Department of Astronomy, Columbia University - New York, NY, USA
  \inst{ins:padova} Dipartimento di Fisica e Astronomia ``G. Galilei'', Universit\`a degli Studi di Padova - Padova, Italy
}

\PACSes{
  \PACSit{04.30.-w}{Gravitational waves}
  \PACSit{04.80.Nn}{Gravitational wave detectors and experiments}
  \PACSit{97.60.Gb}{Pulsars}
}

\ifarxiv
  \makeatletter\let\cimento@ssect\@ssect\makeatother 
  \usepackage[hidelinks,linktocpage=true]{hyperref}
  \usepackage{cleveref}
  \makeatletter\AtBeginDocument{\let\@ssect\cimento@ssect}\makeatother
\fi

\begin{document}

\maketitle

\begin{abstract}
Pulsar Timing Array (PTA) experiments {have found} evidence for a {nanohertz} gravitational wave {(GW)} background (GWB). {The challenge is now to turn evidence for a GWB into an understanding of the sources and physics that produce it. This review connects the physics of the Hellings--Downs correlation to the interpretation of the GWB spectrum and anisotropy, asking whether supermassive black hole binary populations consistent with electromagnetic observations can explain the signal. Searches for individual binaries take these questions further: multi-messenger observations and pulsar terms can reveal the evolution of these binary systems. Realizing this potential requires reliable noise models and candidate validation. We explain how pulsar noise, unresolved sources, and incomplete sky coverage limit these inferences. The same PTA observations also probe cosmic strings and ultralight dark matter, while fitted timing parameters extend GW searches to picohertz frequencies and enable measurements of Galactic acceleration. Rapid data combination and the larger, more precisely timed arrays expected from the Square Kilometre Array Observatory and Deep Synoptic Array will increase the sensitivity of these searches and measurements, sharpening our view of the emerging GW landscape.}
\end{abstract}

\tableofcontents

\newpage

\begin{center}
{\Large\bfseries Executive Summary}\\[6pt]

\end{center}

\medskip
\hrule
\smallskip

Six pulsar timing arrays (PTAs)---the North American Nanohertz Observatory for Gravitational Waves, the European PTA, the Indian PTA, the Parkes PTA, the Chinese PTA, and the MeerKAT PTA ---have reported evidence for a {nanohertz-frequency} gravitational wave (GW) background (GWB), opening the lowest-frequency window of GW astronomy. This review surveys this emerging {nanohertz} GW landscape.

\textbf{The GWB} (\S\ref{sec:stochastic_background}--\S\ref{sec:modeling_gwb}).
We derive the PTA response from first principles, including the Hellings--Downs correlation that distinguishes a GWB from correlated noise. The characteristic strain $h_c(f)$ directly probes the supermassive black hole binary merger rate and chirp mass distribution, with chirp mass as the dominant factor governing the amplitude. {In one illustrative binary population, the number of binaries per frequency bin decreases from $\sim10^6$ at 2~nanohertz to $\sim10^3$ at 20~nanohertz; at the highest frequencies, there may not be a single binary per bin.} {Updated population models can reproduce the measured amplitude, but agreement with electromagnetic constraints depends on the assumed black hole demographics and growth history.}

\textbf{GWB anisotropy} (\S\ref{sec:anisotropy}).
{The anisotropy of a GWB produced by supermassive black hole binaries reflects both galaxy clustering and source discreteness. Measurements and upper limits therefore test the GWB's origin and constrain binary population models. Cross-correlations with galaxy catalogs can reveal the clustering signal if GWB maps have sufficient angular resolution and loud sources can be removed. Relative shot-noise anisotropy generally increases with frequency as fewer binaries contribute, but its detectability also depends on the GWB signal strength and pulsar noise.} We review search methods from spherical harmonics to pixel-based approaches. 

\textbf{Individual binaries} (\S\ref{sec:CW}).
{All-sky and electromagnetically targeted continuous wave searches seek individual binaries; joint-likelihood analyses combine timing data and electromagnetic light curves to constrain their properties.} The systematic targeted search of 114 {candidate binaries in active galactic nuclei} {illustrates a suite of validation tests, including} signal coherence, robustness against a correlated GWB model, {and} dropout analyses. Strong gravitational lensing and bright-siren cosmography are explored as future science goals.

\textbf{{Angular resolution}} {(\S\ref{sec:resolution}).}
{We distinguish the separation of nearby sources, the width of a reconstructed point-source response, and the uncertainty in a continuous wave source's position. These also differ from the number of sources or angular modes an array can constrain. Their limits depend on pulsar coverage, signal-to-noise ratio, and whether pulsar distances are known precisely enough to model the pulsar terms coherently.}

\textbf{New physics} (\S\ref{sec:new_physics}).
{The GWB spectrum may encode signals from cosmic strings, phase transitions and primordial GWs. Self-interacting dark matter can help drive supermassive black hole binaries through the final parsec and produce a low-frequency turnover in the GWB spectrum. Dedicated timing and polarization searches can also probe ultralight dark matter. Searches for non-Einsteinian GW polarizations offer a direct test of general relativity in the nanohertz band.}

\textbf{Custom noise models} (\S\ref{subsec:noise}).
Accurate per-pulsar noise characterization underpins all PTA results. Custom noise models now treat achromatic red noise, chromatic interstellar medium variations, and instrumental systematics {with components tailored to each pulsar}. Getting the noise right determines the inferred GWB amplitude, the significance of continuous wave candidates, and the reliability of anisotropy searches.

\textbf{{International PTA and data combination}} (\S\ref{sec:ipta}).
The {International Pulsar Timing Array} combines data from all regional collaborations into a single global array. We review its data releases, mock data challenges, and complementary science, and discuss the long timescales inherent in full data combination. {Rapid methods such as Lite and FrankenStat provide earlier combined constraints and motivate full data combination, which also improves pulsar timing and noise models.}

\textbf{{Complementary science}} {(\S\ref{sec:complementary}).}
{Pulsar timing also probes slowly varying GWs through their contributions to fitted timing parameters, extending searches into the picohertz band. Binary orbital period changes provide measurements of Galactic acceleration and the local total mass density. These applications connect PTA observations to wider supermassive black hole binaries, cosmological signals and Galactic structure.}

{The Square Kilometre Array Observatory and Deep Synoptic Array will expand these opportunities through improved timing precision and a larger pulsar sample.} Evidence for the GWB is just the beginning---PTAs are poised to become a cornerstone of multi-messenger GW astronomy.

\newpage

\section{Introduction}\label{sec1}

\textit{Historical Context.} -- Pulsar Timing Arrays (PTAs) have a rich history at the interface of physics and astronomy. {Pirani~\cite{Pirani1956,Pirani1957} {showed how gravitational waves (GWs) affect} the relative motion of freely falling particles, while Kaufmann~\cite{Kaufmann1970} derived their effect on photon redshifts.} The idea of using precise radio links to detect GWs then took shape in the mid-1970s, when Estabrook and Wahlquist~\cite{EstabrookWahlquist1975} proposed that Doppler tracking of interplanetary spacecraft could reveal the signature of passing GWs. Motivated by high-precision tracking data from the Pioneer~10 mission, they calculated the effect of plane GWs on the two-way Doppler shift of a signal sent by a distant spacecraft, and suggested that cross-correlating the Doppler residuals from several spacecraft could isolate a GW signal from plasma and clock noise. In practice, however, the frequency stability of the best available clocks---hydrogen masers---was insufficient to detect the expected GW amplitudes. Moreover, the technique probed a frequency band  of $\sim 10^{-4}$--$10^{-1}$\,Hz where loud astrophysical sources were not anticipated at the time. Meanwhile, the discovery of the binary pulsar PSR~B1913+16 by Hulse and Taylor~\cite{HulseTaylor1975}, and the subsequent measurement of its orbital decay in precise agreement with {general relativity (GR)~}\cite{TaylorWeisberg1982}, provided the first indirect evidence for GW emission and galvanized the broader effort to detect GWs directly.

The breakthrough came with the realization that nature provides far better clocks than any spacecraft. Sazhin~\cite{sazhin} and Detweiler~\cite{det79} independently proposed that the regular radio pulses from pulsars could serve the same role, but with dramatically improved precision and at much lower frequencies ($\sim$\,nHz) set by the multi-year observing baselines. Building on this idea, Hellings and Downs~\cite{HD83}---working at {the National Aeronautics and Space Administration's} {Jet Propulsion Laboratory}, where spacecraft Doppler tracking was pioneered---submitted a landmark paper on October~1, 1982, using pulsar timing residuals from four ordinary (slow) pulsars to place an upper limit on the isotropic gravitational-wave background (GWB). More importantly, they showed that a true GWB would imprint a characteristic quadrupolar-like pattern of correlations between pulsar pairs as a function of their angular separation on the sky. This is now known as {the Hellings--Downs (HD) curve}, and its detection remains the ``smoking gun'' that distinguishes a GWB signal from correlated noise. The paper was accepted just nineteen days later, on October~20, 1982.

Unbeknownst to Hellings and Downs, a discovery was already underway that would transform the significance of their work. On September~25, 1982---six days before the Hellings-Downs paper was even submitted---Backer et al.~\cite{Backer1982} had detected a fleeting signal at the Arecibo Observatory: two harmonics of a 1.558\,ms periodicity from the steep-spectrum radio source 4C21.53. The signal appeared for only three minutes and could not be reproduced the following day. It was not until November~7, weeks after the Hellings-Downs paper was already accepted, that the source was confirmed as PSR~1937+21, the first millisecond pulsar. Millisecond pulsars---recycled pulsars spun up to periods of $\sim$\,1--10\,ms through accretion from a binary companion, with timing stabilities rivaling atomic clocks---would prove to be the ideal clocks that the Hellings and Downs cross-correlations method needed. Together, these two developments gave rise to the concrete possibility of a PTA: a Galaxy-scale GW detector built from an array of precisely timed millisecond pulsars.

\begin{figure*}[t!]
\centering
\resizebox{\textwidth}{!}{%
\begin{tikzpicture}[x=1.65cm, y=1cm]
\draw[thick, ->, >=stealth] (-0.3,0) -- (12.2,0);
\foreach \x/\yr in {0/1975, 0.88/1978, 1.43/1979, 2.31/1982, 2.86/1983, 3.74/1990, 4.62/1994, 5.50/2001, 6.38/2004, 7.26/2010, 8.14/2016, 9.02/2018, 9.90/2019, 10.78/2020, 11.66/2023} {
  \draw[thick] (\x, -0.1) -- (\x, 0.1);
  \node[below, font=\footnotesize\bfseries] at (\x, -0.15) {\yr};
}
\draw[thin, gray] (0, 0.1) -- (0, 2.6);
\node[above, text width=2.0cm, align=center, font=\scriptsize] at (0, 2.6) {Estabrook \&\\Wahlquist:\\spacecraft\\Doppler tracking};
\draw[thin, gray] (0.88, 0.1) -- (0.88, 1.0);
\node[above, text width=2.0cm, align=center, font=\scriptsize] at (0.88, 1.0) {Sazhin:\\pulsar timing\\for GWs};
\draw[thin, gray] (1.43, 0.1) -- (1.43, 2.6);
\node[above, text width=2.0cm, align=center, font=\scriptsize] at (1.43, 2.6) {Detweiler:\\pulsar timing\\for GWs};
\draw[thin, gray] (2.31, 0.1) -- (2.31, 1.0);
\node[above, text width=2.0cm, align=center, font=\scriptsize] at (2.31, 1.0) {{First millisecond\\pulsar\\(PSR 1937+21)}};
\draw[thin, gray] (2.86, 0.1) -- (2.86, 2.6);
\node[above, text width=2.0cm, align=center, font=\scriptsize] at (2.86, 2.6) {{Hellings--Downs\\published}};
\draw[thin, gray] (3.74, 0.1) -- (3.74, 1.0);
\node[above, text width=2.0cm, align=center, font=\scriptsize] at (3.74, 1.0) {Foster \&\\Backer:\\first PTA};
\draw[thin, gray] (4.62, 0.1) -- (4.62, 2.6);
\node[above, text width=2.1cm, align=center, font=\scriptsize] at (4.62, 2.6) {Kaspi et al.:\\first stringent\\GWB limit};
\draw[thin, gray] (5.50, 0.1) -- (5.50, 1.0);
\node[above, text width=2.1cm, align=center, font=\scriptsize] at (5.50, 1.0) {Phinney 2001:\\$h_c \propto f^{-2/3}$};
\draw[thin, gray] (6.38, 0.1) -- (6.38, 2.6);
\node[above, text width=2.1cm, align=center, font=\scriptsize] at (6.38, 2.6) {Regional PTAs\\form (EPTA,\\PPTA, NANOGrav)};
\draw[thin, gray] (7.26, 0.1) -- (7.26, 1.0);
\node[above, text width=2.0cm, align=center, font=\scriptsize] at (7.26, 1.0) {IPTA\\formed};
\draw[thin, gray] (8.14, 0.1) -- (8.14, 2.6);
\node[above, text width=1.5cm, align=center, font=\scriptsize] at (8.14, 2.6) {{IPTA}\\{{first data\\release}}};
\draw[thin, gray] (9.02, 0.1) -- (9.02, 1.0);
\node[above, text width=1.5cm, align=center, font=\scriptsize] at (9.02, 1.0) {{Bayes}\\{Ephem}};
\draw[thin, gray] (9.90, 0.1) -- (9.90, 2.6);
\node[above, text width=1.5cm, align=center, font=\scriptsize] at (9.90, 2.6) {{IPTA}\\{{second data\\release}}};
\draw[thin, gray] (10.78, 0.1) -- (10.78, 1.0);
\node[above, text width=2.1cm, align=center, font=\scriptsize] at (10.78, 1.0) {Common red\\noise: NANOGrav,\\PPTA, EPTA, IPTA};
\draw[thin, red!70!black] (11.66, 0.1) -- (11.66, 2.6);
\filldraw[red!70!black] (11.66, 0) circle (3pt);
\node[above, text width=2.2cm, align=center, font=\scriptsize\bfseries] at (11.66, 2.6) {GWB evidence:\\NANOGrav,\\EPTA+InPTA,\\PPTA, CPTA};
\end{tikzpicture}%
}
\caption{{\textbf{Timeline of key milestones in PTAs.} From the first proposals for GW detection via spacecraft Doppler tracking (1975) and pulsar timing (1978--79), the near-simultaneous development of the {HD} correlation framework (1983) and the discovery of the first millisecond pulsar (1982), to the construction of the first PTA by Foster \& Backer (1990)~\cite{FosterBacker1990} , the first stringent GWB upper limit by Kaspi et al. (1994)~\cite{Kaspi1994}, Phinney's GWB strain spectrum (2001)~\cite{phinney_practical_2001}, the formation of PTA collaborations, {{the first and second IPTA data releases (DR1 and DR2; 2016, 2019)}~\cite{IPTA_DR1, IPTA_DR2},} the independent detections of a common red-noise process (2020--2022)~\cite{NG12p5_gwb,Goncharov2021_CRN,Chen2021,Antoniadis2022}, {the introduction of Bayesian Solar-System ephemeris modeling (BayesEphem; 2018)~\cite{Arzoumanian2018, Vallisneri2020},} and the landmark 2023 evidence for a GWB. {Events are spaced for legibility, not in proportion to elapsed time.}}}
\label{fig:timeline}
\end{figure*}
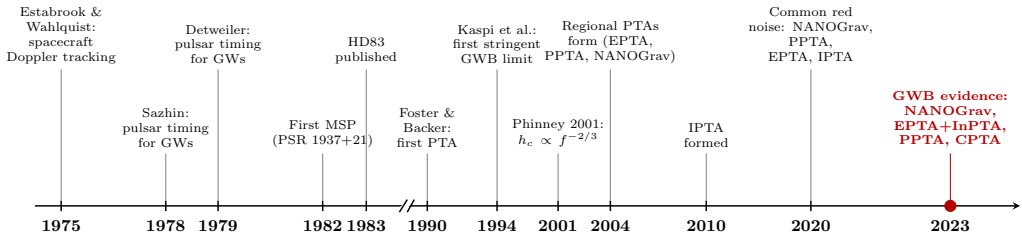

Fig. \ref{fig:timeline} summarizes the key milestones from these early theoretical foundations through to the current era. Foster \& Backer~\cite{FosterBacker1990} took the critical next step by constructing the first PTA: a dedicated program at {the National Radio Astronomy Observatory 43\,m telescope} timing PSRs~1620$-$26, 1821$-$24, and 1937+21 over a two-year baseline beginning in 1987. Their 1990 paper formalized the use of a spatial array of millisecond pulsars to simultaneously provide a long-timescale time standard, detect perturbations of the Earth's orbit, and search for a GWB. The observing equipment of the era limited their timing residuals to $\sim 1\,\mu$s, so while they demonstrated the method and placed an early GWB constraint, the first truly stringent limit came from Kaspi, Taylor \& Ryba~\cite{Kaspi1994}, who used over eight years of high-precision Arecibo data on PSRs~B1855+09 and B1937+21 (with residuals of $\sim 0.2$--$0.8\,\mu$s) to set an upper limit of $\Omega_g h^2 < 6 \times 10^{-8}$. McHugh et al.~\cite{McHugh1996} subsequently reanalyzed the same data within a Bayesian framework, demonstrating the practical power of millisecond pulsar timing for GW searches. A key theoretical advance came with Phinney~\cite{phinney_practical_2001}, who derived a simple relation between the GWB spectrum, the time-integrated energy spectrum of individual sources, and the comoving number density of their remnants. This result established the canonical $h_c \propto f^{-2/3}$ power-law prediction for a population of {circular, GW-driven supermassive black hole binaries (SMBHBs)}, providing the standard spectral benchmark against which all subsequent PTA measurements have been compared.

Through the late 1990s and 2000s, three regional PTA collaborations took shape. In Australia, a long-running pulsar timing program at the Parkes 64\,m telescope was formalized as the Parkes PTA (PPTA)~\cite{Manchester2013}, which pioneered many of the noise modeling and data analysis techniques now standard across the field, including the \textsc{Tempo2} timing package~\cite{Hobbs2006}. In Europe, observations from five major radio telescopes were combined under the European PTA (EPTA)~\cite{Kramer2013}. In North America, the North American Nanohertz Observatory for Gravitational Waves (NANOGrav)~\cite{Jenet2009} {unified efforts in the United States and Canada} around the Arecibo Observatory and the Green Bank Telescope. These three programs motivated the formation of the International PTA (IPTA) in 2010~\cite{Hobbs2010}, which combines data across all regional arrays (see \sect{sec:ipta} for a full discussion). The Indian PTA (InPTA) and the {African Pulsar Timing array\footnote{\url{https://africanpulsartiming.github.io}}} subsequently joined as full members. More recently, the Chinese PTA (CPTA) and the MeerKAT PTA (MPTA) have begun contributing independent datasets, though they are not yet formal IPTA members.

The search for nanoHertz-frequency GWs has now reached a historic milestone. After decades of precision monitoring, PTAs worldwide have uncovered evidence for a stochastic GWB. The first indication came from the NANOGrav 12.5-yr dataset~\cite{NG12p5_gwb}, which identified a common-spectrum red-noise process across its pulsars but could not yet confirm {HD} correlations. {Contemporaneous analyses by the PPTA~\cite{Goncharov2021_CRN}, EPTA~\cite{Chen2021}, and IPTA~\cite{Antoniadis2022} also found evidence for a common-spectrum red-noise process, with the IPTA analysis combining earlier regional datasets.} The breakthrough came in 2023, when four collaborations reported evidence for spatial correlations consistent with a GWB: NANOGrav~\cite{NG15_gwb}, the EPTA and InPTA~\cite{EPTA2023}, the PPTA~\cite{Reardon2023}, and the CPTA~\cite{Xu2023}. {Preparation of the IPTA's third data release (DR3), combining regional data with unified timing and noise models, is underway~\cite{3P+paper} (see \sect{sec:data_combination}).} Crucially, these recent datasets have begun to reveal the distinctive {HD} spatial correlations required to confirm the low-frequency PTA signal's GW origin.

\begin{figure}[b!]
\vspace{-1\baselineskip}
    \centering
    \includegraphics[width=\textwidth]{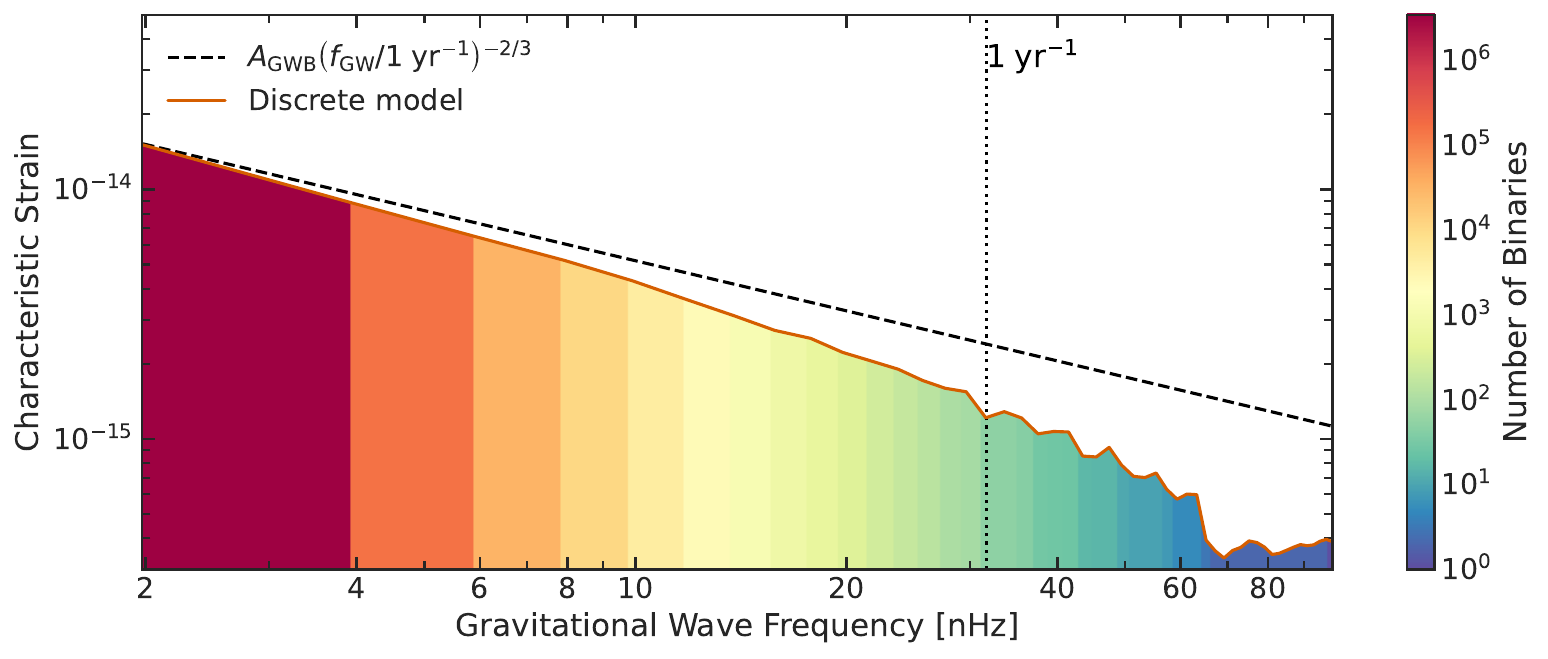}
    \caption{{\bf The GWB in the PTA frequency band.} The characteristic strain spectrum of the GWB from a cosmic population of SMBHBs. At low frequencies the background is well approximated by a power law $h_c \propto f^{-2/3}$, representing the Phinney~\cite{phinney_practical_2001} ensemble average (i.e.\ the mean over many realizations). Since we observe a single realization of the Universe, the measured spectrum will be jagged, with excursions above and below this mean; the median spectrum (from discrete realizations of the GWB) may better represent what any one realization looks like, see e.g. ~\cite{NG15_discreteness}. {For the population illustrated in Fig.~\ref{fig:fig1},} the number of binaries per frequency bin, $\Delta N(f)$, drops steeply with frequency (\eq{eq:delta_N}): from $\sim 10^6$ sources at 2\,nHz to $\sim 10^3$ at 20\,nHz, {because circular, GW-driven binaries evolve more quickly at higher frequencies, $\dot f\propto f^{11/3}$.} {This decline increases the relative importance of individual binaries and can produce a more anisotropic background. However, {detecting individual CW sources} (\sect{sec:CW}) or anisotropy (\sect{sec:anisotropy}) depends on PTA sensitivity.}}
    \label{fig:fig1}
    \vspace{-1\baselineskip}
\end{figure}

This evidence of a GWB opens a new window onto the dynamical Universe. The leading interpretation is that this background arises from the cosmic population of {SMBHBs} with masses of $\sim10^8$--$10^{10}\,M_\odot$. These binaries form as a natural consequence of hierarchical galaxy mergers: following a merger, dynamical friction drives the two central black holes toward the center of the newly formed galaxy on timescales of $\sim$100\,Myr - Gyrs. At parsec-scale separations, the binary must shed energy and angular momentum---via three-body stellar scattering or interactions with a circumbinary gas disk---to reach the sub-parsec regime where GW emission becomes efficient. {This transition was long considered a potential bottleneck---the ``final parsec problem''~\cite{Begelman1980}. If the GWB is produced by SMBHBs, it implies that at least part of the population reaches separations probed by PTAs; constraints on merger delays and the fraction of stalled binaries depend on the assumed population and evolutionary models~\cite{NG15_astro}. At these separations, these systems emit GWs with periods of years to decades.} {For the population illustrated in Fig.~\ref{fig:fig1}, the number of binaries per frequency bin decreases from $\sim10^6$ at 2\,nHz to $\sim10^3$ at 20\,nHz, because circular, GW-driven binaries evolve more quickly at higher frequencies, $\dot f\propto f^{11/3}$.} While SMBHBs are the standard astrophysical interpretation, the signal could also herald ``New Physics,'' potentially originating from cosmic strings, phase transitions in the early Universe, or primordial fluctuations~\cite{Afzal2023,Sesana2025}.

PTAs can detect these faint spacetime ripples by using millisecond pulsars in our own Galaxy as a GW detector. A passing GW with strain $h \sim 10^{-15}$ induces timing residuals of order 100\,ns over a decade-long baseline. Millisecond pulsars can reach this precision thanks to their rapid, stable rotation and low intrinsic spin-down noise, making them ideal clocks for GW detection. The sensitivity of a PTA is bounded by its total observation time ($f_{\mathrm{min}} \sim 1/T_{\mathrm{obs}}$) and observing cadence ($f_{\mathrm{max}} \sim 1/\Delta t$), uniquely positioning PTAs to probe physics on timescales not accessible by any other means.

\subsection*{State of the Art}

In anticipation of a detection, the IPTA Detection Committee established a detection checklist~\cite{Allen2023_checklist}. Shortly after, independent analyses by NANOGrav~\cite{NG15_gwb}, EPTA+InPTA~\cite{EPTA2023}, PPTA~\cite{Reardon2023}, CPTA~\cite{Xu2023}, all reported the presence of a common-spectrum process with the expected {HD} angular correlations. {The MPTA reported evidence for {HD} correlations using 4.5 years of data from 83 pulsars, although the significance depended strongly on the adopted noise model and treatment of noise uncertainties~\cite{Miles2025_noise} (see \sect{subsec:noise}).} Should these emerging PTAs join the IPTA, the inclusion of their additional southern-hemisphere pulsars with small residuals would be particularly valuable for improving the sky coverage, {detection prospects for continuous waves (CWs)}, and angular resolution of a combined analysis. {Joint data will improve sky coverage and sensitivity to GWB anisotropy, helping test astrophysical and primordial interpretations of the signal (\sect{sec:anisotropy}).}

A key advance since the first GWB evidence claims has been the development of increasingly sophisticated custom noise models~\cite{NG15_detchar,Goncharov2021,EPTA_noise}. These models treat each pulsar's red noise spectrum, {chromatic interstellar medium (ISM) variations}, and potential instrumental systematics self-consistently within hierarchical Bayesian frameworks. {Accurate noise characterization is essential: for a fixed $h_c\propto f^{-2/3}$ spectrum, reanalysis of the NANOGrav 15-year data with customized chromatic noise models shifts the median strain amplitude at $1\,\mathrm{yr}^{-1}$ from $2.4\times10^{-15}$ to $2.1\times10^{-15}$~\cite{NG15_gwb,NG15_CNM_GWB}. This modest reduction eases tension with some astrophysical population models~\cite{NG15_CNM_GWB}.} This apparent tension has motivated new theoretical work on the demographics and merger rates of SMBHBs, including the effects of environmental coupling, eccentricity evolution, and black hole mass scaling relations, e.g. Liepold and Ma (2024)~\cite{LiepoldMa2024}. {It has also motivated cosmological interpretations involving certain cosmic-string models or {first-order phase transitions (FOPTs) in} the early Universe~\cite{Afzal2023}.}

\begingroup

Errors in the {Solar System ephemeris}, particularly in Jupiter's orbit, perturb Earth's inferred position relative to the {Solar System barycenter (SSB) and} induce a dipolar timing signature. In a finite PTA with non-uniform sky coverage and unequal pulsar sensitivities, this signature is imperfectly separable from the correlations sought in a GWB search~\cite{Tiburzi2016,Arzoumanian2018,Vallisneri2020}. In the NANOGrav 11-year dataset, evidence for a common red process depended on the adopted ephemeris, and marginalizing over ephemeris uncertainties with BayesEphem removed that preference~\cite{Arzoumanian2018,Vallisneri2020}. This motivated explicit treatment of ephemeris uncertainty in GWB inference. Conversely, the same timing sensitivity allows PTAs to constrain planetary masses and ephemeris parameters~\cite{Caballero2018_solar}.
\par
\endgroup

\begingroup

Recent astronomical discoveries have added further importance to the GWB evidence. High-redshift quasars observed at $z \gtrsim 6$ host black holes with masses approaching or exceeding $10^9\,M_\odot$~\cite{Banados2018,Fan2023}, challenging models of early black-hole growth. At still higher redshift, {Chandra observations of UHZ1, a galaxy at $z \approx 10$ identified by the James Webb Space Telescope, imply} a black-hole mass of $\sim10^7$--$10^8\,M_\odot$, assuming Eddington-limited accretion~\cite{Bogdan2024,Natarajan2024}. {James Webb Space Telescope studies} have found accreting black holes in early galaxies~\cite{Maiolino2024,Maiolino2024_JADES,Greene2024} and inferred unusually high black-hole-to-host mass ratios in some systems~\cite{Pacucci2023,Furtak2024,Juodzbalis2024}, although selection effects and measurement uncertainties complicate comparisons with local scaling relations~\cite{Li2025}. The masses and physical nature of little red dots remain debated, with revised luminosity estimates and alternative emission models suggesting lower black-hole masses than initially inferred~\cite{Greene2026_LRDluminosities,Sun2026_LRDmasses}. Nearby {ultraviolet-bright} stellar companions have also motivated direct-collapse interpretations, although the causal connection remains uncertain~\cite{Baggen2026_A383,Baggen2026_ConnectingDots}. These observations provide new constraints on early black-hole formation; their implications for the nanohertz GWB depend on the subsequent growth and merger histories of the population. Population models illustrate this connection: Somalwar \& Ravi~\cite{SomalwarRavi2025} show that a substantial $z\gtrsim1$ SMBHB population can increase the predicted GWB amplitude.
\par
\endgroup

Together with stochastic background analyses, PTAs have entered a new phase of targeted searches for CWs from individual SMBHBs. These efforts represent a major conceptual leap from the early era, when PTAs could only place upper limits on a candidate binary in 3C 66B~\cite{3c66b}, to today coordinated campaigns testing specific binary hypotheses in well-identified active galactic nuclei (AGN). {The recent NANOGrav targeted search~\cite{Agarwal2026} examined 114 AGN that may host SMBHBs and used detailed follow-up of two targets, both consistent with noise, to illustrate a detection protocol.} To date, no CW detection has been made by any GW detector across the entire frequency spectrum. {Rapid data-combination methods can enable joint PTA analyses before a fully combined IPTA data release becomes available~\cite{LarsenEtAl2025,WrightEtAl2025FrankenStat} (\sect{sec:data_combination}).}

\textit{Scope of this review.} -- Several complementary reviews of nanoHertz GW science have appeared recently. Sesana \& Figueroa~\cite{Sesana2025} provide a thorough treatment of the GWB source physics, covering both the astrophysical signal from SMBHBs, including eccentricity, environmental coupling, and sparse sampling, as well as early-Universe backgrounds from inflation, phase transitions, and topological defects. Taylor~\cite{Taylor2025_review} gives a broad overview of PTA data analysis and astrophysical interpretation in the post-evidence era. Shaifullah~\cite{Shaifullah2025_review} focuses on the observational and instrumental landscape, summarizing the techniques behind sub-microsecond timing precision and the challenges posed by noise modeling, interstellar propagation, and Solar system ephemeris uncertainties. Liu \& Chen~\cite{LiuChen2025_review} survey the past decade of PTA progress, covering advances in instrumentation, data-analysis techniques, and the path from progressively tighter upper limits to the 2023 {HD} detection. The present review is intended to complement these works. Our emphasis is on the mathematical formalism underlying PTA searches---the spatial correlation signatures of GWs, and their harmonic decomposition---and on the methodologies for detecting and characterizing both the stochastic background and individual sources. We give particular attention to GWB anisotropy, multi-messenger CW searches, and the role of the IPTA. We also discuss new physics, including alternative GW polarizations {and ultralight dark matter (ULDM),} and the observational prospects opened by next-generation instruments.

\begin{table}[b!]
\centering
\caption{List of acronyms used in this review.}
\label{tab:acronyms}
\small
\begin{tabular}{@{} >{\raggedright\arraybackslash}p{.10\textwidth} @{\hspace{.02\textwidth}} >{\raggedright\arraybackslash}p{.34\textwidth} @{\hspace{.04\textwidth}} >{\raggedright\arraybackslash}p{.14\textwidth} @{\hspace{.02\textwidth}} >{\raggedright\arraybackslash}p{.34\textwidth} @{}}
\hline\hline
\textbf{Acronym} & \textbf{Definition} & \textbf{Acronym} & \textbf{Definition} \\
\hline
AGN & Active Galactic Nucleus & InPTA & Indian Pulsar Timing Array \\
ARN & Achromatic Red Noise & IPTA & International Pulsar Timing Array \\
BiPOSH & Bipolar Spherical Harmonic & ISM & Interstellar Medium \\
{BWM} & {Burst with Memory} & LSS & Large-Scale Structure \\
CPTA & Chinese Pulsar Timing Array & {MDC} & {Mock Data Challenge} \\
{CRTS} & {Catalina Real-time Transient Survey} & MPTA & MeerKAT Pulsar Timing Array \\
CURN & Common Uncorrelated Red Noise & NANOGrav & {North American Nanohertz Observatory for Gravitational Waves} \\
CW & Continuous Wave & ORF & Overlap Reduction Function \\
DM & Dispersion Measure & {pN} & {Post-Newtonian} \\
{DMX} & {Piecewise-constant dispersion measure parameterization} & PPTA & Parkes Pulsar Timing Array \\
{DR} & {Data Release} & PTA & Pulsar Timing Array \\
{DSA} & {Deep Synoptic Array} & QLF & Quasar Luminosity Function \\
{EM} & {Electromagnetic} & SKAO & Square Kilometre Array Observatory \\
EPTA & European Pulsar Timing Array & SMBH & Supermassive Black Hole \\
FoM & Figure of Merit & SMBHB & Supermassive Black Hole Binary \\
FOPT & First-Order Phase Transition & S/N & Signal-to-Noise Ratio \\
GR & General Relativity & SSB & Solar System Barycenter \\
GSMF & Galaxy Stellar Mass Function & TOA & Time of Arrival \\
GW & Gravitational Wave & ULDM & Ultralight Dark Matter \\
GWB & Gravitational Wave Background & VDF & Velocity Dispersion Function \\
HD & {Hellings--Downs} & VLBI & Very Long Baseline Interferometry \\
\hline\hline
\end{tabular}
\end{table}

As PTA baselines lengthen, noise modeling becomes increasingly precise, and targeted searches mature into a detection framework, the field is transitioning from evidence for a GWB to astrophysical inference and toward the first direct discovery of a SMBHB in its host galaxy.
This review summarizes the status of the field in this new era. A list of common acronyms is given in tab. \ref{tab:acronyms}. In \sect{sec:stochastic_background}, we present an order-of-magnitude derivation of the GWB stochasticity before developing the formalism underlying GWB searches, including the derivation of the {HD} curve and the cross-correlation methods used to extract the signal. Sect. \ref{sec:gwb_history} reviews predictions and current observational constraints on the GWB amplitude and spectrum while \sect{sec:modeling_gwb} covers astrophysical modeling of the SMBHB population. Sect. \ref{sec:anisotropy} addresses how we search for GWB anisotropy and how it scales. in \sect{sec:CW}, we turn our attention to targeted CW searches for individual binaries, including multi-messenger detection frameworks, and in  \sect{sec:resolution} we discuss the angular resolution of PTAs. Sect. \ref{sec:new_physics} examines prospects for probing new physics. Turning our focus to the detector aspects of PTAs, in \sect{subsec:noise} we examine pulsar noise models, while \sect{sec:data_combination} describes the IPTA and new rapid data combination methods. Sect. \ref{sec:complementary} highlights some (certainly not all) complementary science, and we conclude with an outlook for the field in \sect{sec:conclusion}.

{Throughout this review, we use geometrized units, with Newton's constant and the speed of light set to unity ($G=c=1$). Numerical values are quoted in the units indicated.}

\section{The Gravitational Wave Background}
\label{sec:stochastic_background}

The search for a stochastic GWB relies on identifying a common signal correlated across the PTA. We begin in \sect{sec:stochastic_why} with an original order-of-magnitude derivation of the GWB stochasticity, showing from first principles why the nanoHertz background is indeed a confusion-limited signal. Several scaling laws here may be useful for the reader in general. We then derive the response of a single pulsar to a GW, construct the cross-correlation statistic that isolates the GWB from intrinsic noise, and derive the expected spectral and spatial signatures of the signal.

\subsection{Why the nanoHertz GWB is stochastic}
\label{sec:stochastic_why}

A useful way to distinguish a genuinely stochastic (confusion) background from a set of individually resolvable binaries is to compute the duty cycle (or occupancy) per frequency bin, $\Delta N(f)$. Here $\Delta N(f)$ is the expected number of binaries whose gravitational radiation lies in a single Fourier bin centered at frequency $f$, at any given time. This basic resolvability criterion, and its relation to Gaussian versus non-Gaussian (``popcorn'') regimes, is widely discussed in the literature, see e.g. \citep{Sesana2008PTAoccupancy,KocsisSesana2011,Rosado2011Overlap,Regimbau2012Popcorn}. If $\Delta N \sim1$, bins are typically empty or contain a single system. Individual sources are, in principle, resolvable. If $\Delta N \gg 1$, many systems contribute to each bin. {When many sources contribute with uncorrelated phases and no small subset dominates the signal power, their superposition has approximately Gaussian statistics (by the Central Limit Theorem; see e.g. Ref.~\citep{cornishromano2015,Rosado2011Overlap,Regimbau2012Popcorn}).}

\begingroup

At the level of an order-of-magnitude estimate, the expected number of binaries in a frequency bin is the product of the cosmic merger rate and the time a binary spends in that bin:
\begin{equation}
    \Delta N(f) \sim \underbrace{\dot{N}}_{\text{Cosmic merger rate}} \times \underbrace{\tau_{\mathrm{res}}(f)}_{\text{Residence time}} \,.
\end{equation}
To build intuition, suppose each galaxy undergoes approximately one major merger over the age of the Universe, and that the resulting black-hole binaries reach the PTA band. Take a rough galaxy census of $N_{\mathrm{gal}}\sim10^{11}$~\citep{Conselice2016GalaxyCensus} and a cosmic age of $t_U\sim10^{10}\,\mathrm{yr}$~\citep{Planck2018Cosmo}. These are deliberately rounded scales. Writing $N_{\mathrm{merger}}$ for the average number of major mergers per galaxy over $t_U$, the average merger rate is then, to order of magnitude,
\[
    \dot N\sim\frac{N_{\mathrm{gal}}N_{\mathrm{merger}}}{t_U}\sim\frac{10^{11}\times1}{10^{10}\,\mathrm{yr}}\sim10\,\mathrm{yr}^{-1}.
\]
Thus, although a major merger is rare for any individual galaxy, mergers across the entire population occur frequently.

A binary appears effectively monochromatic if its frequency drift over the observing span is small compared to the Fourier resolution. For an observation time $T_{\mathrm{obs}}\sim10\,\mathrm{yr}$, the bin width is
\begin{equation}
    \Delta f = \frac{1}{T_{\mathrm{obs}}} \approx 3 \times 10^{-9} \, \mathrm{Hz} \,.
\end{equation}
The residence time in one bin is therefore
\begin{equation}
    \tau_{\mathrm{res}} \approx \frac{\Delta f}{\dot{f}} = \frac{1}{T_{\mathrm{obs}}\dot{f}} \,.
\end{equation}
For this idealized estimate, let these mergers produce circular, GW-driven binaries with redshifted chirp mass $\mathcal{M}_z=(1+z)\mathcal{M}\sim10^9M_\odot$. {In GR,} the leading-order evolution of the observed GW frequency is
\begin{equation}
    \dot{f} = \frac{96}{5}\pi^{8/3}\mathcal{M}_z^{5/3}f^{11/3} \,.
\end{equation}
At $f\sim10\,\mathrm{nHz}$, this gives $\dot{f}\approx8.5\times10^{-5}\,\mathrm{nHz\,yr}^{-1}$. Substituting into the residence-time estimate,
\begin{equation}
    \tau_{\mathrm{res}} \approx \frac{3\times10^{-9}\,\mathrm{Hz}}{2.7\times10^{-21}\,\mathrm{Hz/s}} \approx 1.1\times10^{12}\,\mathrm{s} \approx35{,}000\,\mathrm{yr} \,.
\end{equation}
Although the observation spans only ten years, a binary can remain in a single frequency bin for tens of thousands of years. The expected occupancy in this simple estimate is therefore
\begin{equation}
    \Delta N\sim N_{\mathrm{gal}}N_{\mathrm{merger}}\frac{\tau_{\mathrm{res}}}{t_U}\sim10^{11}\times1\times\frac{35{,}000}{10^{10}}\sim3.5\times10^5 \,.
\end{equation}
Combining $\Delta N(f)\sim N_{\mathrm{gal}}N_{\mathrm{merger}}/(t_UT_{\mathrm{obs}}\dot f)$ with the quadrupole frequency evolution makes the dependence on each quantity explicit:
\begin{equation}
\resizebox{0.94\linewidth}{!}{$\displaystyle \Delta N(f)\sim4\times10^5\left(\frac{N_{\mathrm{gal}}}{10^{11}}\right)\left(\frac{N_{\mathrm{merger}}}{1}\right)\left(\frac{t_U}{10^{10}\,\mathrm{yr}}\right)^{-1}\left(\frac{\mathcal M_z}{10^9M_\odot}\right)^{-5/3}\left(\frac{f}{10\,\mathrm{nHz}}\right)^{-11/3}\left(\frac{T_{\mathrm{obs}}}{10\,\mathrm{yr}}\right)^{-1}$.}
\label{eq:delta_N}
\end{equation}
More galaxies or more mergers per galaxy increase the occupancy. More massive binaries and binaries at higher frequencies evolve faster, reducing their residence time. Longer observations resolve narrower frequency bins, so fewer binaries occupy each bin.
\par
\endgroup

{In this illustrative estimate, $\Delta N\approx3.5\times10^5\gg1$, so many binaries contribute to each frequency bin. This large occupancy motivates a stochastic description of the unresolved population, while sufficiently bright individual binaries may still be detectable above the background (\sect{sec:CW}). An individual CW source imprints a source-dependent cross-correlation fingerprint across pulsar pairs that differs from the {HD} pattern of an isotropic GWB, providing a way to distinguish the two~\cite{Mingarelli2026_fingerprints} (\sect{subsec:deterministic_fingerprints}).}

At higher frequencies, the steep dependence $\Delta N \propto f^{-11/3}$ drives the occupancy downward, eventually reaching $\Delta N \lesssim 1$. This marks the transition from a confusion-dominated regime to a regime where individual CW sources can become resolvable, see e.g. Ref.~\citep{Sesana2008PTAoccupancy,KocsisSesana2011}. The exact frequency at which this happens is the subject of current research.

\subsection{PTA Response to GWs}
\label{subsec:pta_response}

We define a computational frame where the Earth is at the origin and pulsars are located at distances $L$ in directions $\hat{p}$. The response of a PTA to a passing GW is quantified by the fractional frequency shift, $z(t, \hat{\Omega})$, induced in the timing residuals. This shift arises from the metric perturbation $h_{ij}$ at the Earth and the pulsar, resulting in a ``two-term'' response function~\cite{EstabrookWahlquist1975,det79,anholm2009}:
\begin{equation}\label{eq:pta_redshift}
  z(t, \hat{\Omega}) = \frac{1}{2} \frac{\hat{p}^i \hat{p}^j}{1 + \hat{\Omega} \cdot \hat{p}} \left[ h_{ij}(t, \hat{\Omega}) - h_{ij}(t_p, \hat{\Omega}) \right] \,.
\end{equation}
The physical mechanism of PTA detection relies on the fact that a GW perturbs the spacetime metric along the photon's entire path from the pulsar to the Earth. The total timing residual is formally the integral of the metric perturbation $h_{ij}$ along the line of sight. For a plane wave, this integral simplifies analytically to the difference between the metric perturbation at the two boundaries: the Earth and the Pulsar \cite{det79}. We can thus decompose the residual into two distinct components:
\begin{equation}
\label{eq:earth_pulsar}
    R(t) = R_E(t) - R_P(t_p) \,,
\end{equation}
where $R_E(t)$ is the Earth term and $R_P(t_p)$ is the Pulsar term. In practice, the Earth is not a fixed observatory: it orbits the Sun, introducing a time-varying geometric delay that would couple to the GW signal if left unmodeled. To remove this complication, {all measured times of arrival (TOAs)} are first transformed to the {SSB} using a Solar system ephemeris~\cite{Hobbs2006}. The SSB is where we could in principle balance the solar system on the tip of our finger. It serves as the common inertial reference point for a PTA, so that ``the Earth term'' in \eq{eq:earth_pulsar} really refers to the GW perturbation evaluated at the SSB.

In principle, once all TOAs are referred to the SSB, the array acts as a single detector at a common reference point: a passing GW perturbs the metric at the SSB, and this perturbation affects every pulsar's TOA at the same instant. The timing residuals of all pulsars jump together, modulated only by each pulsar's geometric antenna factor $\hat{p}^i \hat{p}^j / (1 + \hat{\Omega} \cdot \hat{p})$, which depends on the pulsar's sky position relative to the GW propagation direction. It is this shared perturbation that gives rise to the Hellings and Downs spatial correlations, and in the limit of infinite pulsar distances~\cite{MM18}, the GWB signal is essentially the auto-correlation of the Earth term.

In practice, achieving this common reference frame is a substantial undertaking, particularly for the IPTA, which must combine data from telescopes with different backends, calibration procedures, and TOA modeling choices (see \sect{sec:data_combination}).

The Pulsar term, $R_P(t_p)$, represents the imprint of the GW field at the location of the pulsar itself. However, due to the finite speed of light, we observe this imprint with a significant time delay. The time at which the GW passed the pulsar, $t_p$, is related to the observation time $t$ by:
\begin{equation}
    {t_p = t - \tau_a = t - L_a(1 + \hat{\Omega}\cdot\hat{p}),}
\end{equation}
where $L_a$ is the pulsar distance, $\hat{p}$ is the unit vector pointing to the pulsar, and $\hat{\Omega}$ is the propagation direction of the GW.

Since pulsars are located at typical distances of $L_a \sim 1$\,kpc ($\sim 3000$ light-years), the delay $\tau_a$ is on the order of thousands of years. Furthermore, because every pulsar is at a different distance, the phase of the GW field at each pulsar is effectively random relative to the others. Consequently, the pulsar terms are uncorrelated between different pulsars. In stochastic background searches, the pulsar term therefore acts as an additional source of variance that does not contribute to the cross-correlation signal.

Despite this, the pulsar term carries unique astrophysical information. Since the pulsar term samples the GW field thousands of years in the past when it transited the pulsar, comparing the Earth and pulsar terms for a resolved CW source amounts to measuring the binary's orbital evolution over thousands of years---effectively turning the PTA into a time machine (\sect{subsec:time_machine}). Furthermore, if pulsar distances can be measured to within a GW wavelength (e.g.\ {via very long baseline interferometry (VLBI) parallax}), the pulsar term can be phased coherently, roughly doubling the effective signal power~\cite{corbincornish2010} and transforming the array into a true interferometer with a Galactic-scale baseline~\cite{Mingarelli12}, dramatically improving sky localization (\sect{sec:resolution}).

\subsection{The General Plane Wave Framework}
\label{subsec:plane_wave}

We now formalize the above by writing the metric perturbation as a general plane wave expansion. Any GW signal---whether from a single source or a superposition of many---can be decomposed as:
\begin{equation}
    h_{ij}(t, \mathbf{x}) = \sum_A \int_{-\infty}^\infty df \int_{S^2} d\hat{\Omega}\, h_A(f,\hat{\Omega}) e^{i2\pi f(t-\hat{\Omega}\cdot\mathbf{x})} e^A_{ij}(\hat{\Omega}) ,
\end{equation}
where $f$ is the GW frequency, $A = \{+, \times\}$ labels the polarization degrees of freedom, $S^2$ indicates that the integral is performed on the two-sphere, $\hat{\Omega}$ is a unit vector identifying the GW propagation direction, $e^A_{ij}$ is the polarization tensor, and $h_A$ is the amplitude. In GR the sum runs over the two tensor polarizations $A = \{+, \times\}$; alternative theories of gravity can introduce up to four additional modes (scalar and vector), each with a distinct polarization tensor and antenna pattern. These non-Einsteinian polarizations modify the Overlap Reduction Function (ORF) and are discussed in \sect{sec:non_gr_polarizations}.

The two-point function of a stationary, Gaussian, and unpolarized background reads:
\begingroup

\begin{equation}\label{eq:hcorr}
    \left\langle h^*_A(f,\hat{\Omega})h_{A'}(f',\hat{\Omega}')\right\rangle = \delta^D(\hat{\Omega}-\hat{\Omega}')\delta^K_{AA'}\delta^D(f-f')\frac{h_c^2(|f|)}{16\pi|f|}P(\hat{\Omega}).
\end{equation}
Here $\delta^K$ and $\delta^D$ are Kronecker and Dirac deltas, respectively, and the ensemble average $\langle\dots\rangle$ is over realizations of the source population. The characteristic strain $h_c(f)$ describes the spectral content, while $P(\hat{\Omega})$ describes the angular distribution of GW power, normalized so that $\int_{S^2}P(\hat{\Omega})\,d\hat{\Omega}=4\pi$. We assume that the frequency and angular dependence separate~\cite{m13}. An isotropic background has $P(\hat{\Omega})=1$; deviations from isotropy are the subject of \sect{sec:anisotropy}. For circular, GW-driven binaries, $h_c\propto f^{-2/3}$~\cite{phinney_practical_2001}, but environmental hardening mechanisms---stellar scattering and circumbinary gas torques---modify the spectral shape at low frequencies by altering the binary residence time in each frequency bin. These effects and the resulting $da/dt$ contributions are detailed in \sect{sec:modeling_gwb}.
\par
\endgroup

The timing residual induced in pulsar $a$ by this background is:
\begin{equation}\label{eq:Rt}
R_a^{\mathrm{GW}}(t) = \int df \frac{1}{{2\pi i f}} \int_{S^2} d\hat{\Omega} \sum_A h_A(f,\hat{\Omega}) F^A_{a}(\hat{\Omega}) \left(1-e^{-2\pi i f L_a(1+\hat{\Omega}\cdot\hat{p}_a)}\right) e^{2\pi i f t}\,,
\end{equation}
where $L_a$ is the pulsar distance, $\hat{p}_a$ is the direction to the pulsar, and
\begin{equation}\label{eq:beam_pattern}
    F^A_{a}(\hat{\Omega}) \equiv \frac{\hat{p}_a^i \hat{p}_a^j}{2(1+\hat{\Omega}\cdot \hat{p}_a)} e^A_{ij}(\hat{\Omega})
\end{equation}
is the antenna beam pattern function~\cite{m13}, which encodes the geometric sensitivity of the pulsar to GWs from direction $\hat{\Omega}$. The term in parentheses in \eq{eq:Rt} is the difference between the Earth term (unity) and the Pulsar term (exponential) discussed above.

\begingroup

We define the one-sided cross-power spectrum of the timing residuals by~\cite{ani2020}
\[
    \left\langle R_a(f)R_b^*(f')\right\rangle=\frac{1}{2}\delta^D(f-f')S_{ab}(f),\qquad f>0.
\]
Using the pulsar response above, we define {the normalized ORF}
\begin{equation}\label{eq:orf_general}
    \Gamma_{ab}(f)=\frac{3}{8\pi}\int_{S^2}d\hat{\Omega}\,P(\hat{\Omega})\kappa_{ab}(f,\hat{\Omega})\sum_A F_a^A(\hat{\Omega})F_b^A(\hat{\Omega}),
\end{equation}
\par
\endgroup
where $\kappa_{ab}$ encodes the Earth--Pulsar term interference:
\begin{equation}\label{eq:pulsterm}
    \kappa_{ab}(f,\hat{\Omega}) \equiv \left(1-e^{-2\pi i f L_a(1+\hat{\Omega}\cdot\hat{p}_a)}\right)\left(1-e^{2\pi i f L_b(1+\hat{\Omega}\cdot\hat{p}_b)}\right)\,.
\end{equation}
\begingroup

The cross-power spectrum is then
\begin{equation}\label{eq:crosspower}
    S_{ab}(f)=\frac{h_c^2(f)}{12\pi^2f^3}\,\Gamma_{ab}(f).
\end{equation}
In the short-wavelength (incoherent) limit $fL_a\gg1$, the pulsar terms average out in the sky integral apart from the autocorrelation contribution, giving the factor $(1+\delta_{ab})$~\cite{MS14,MM18}. For an isotropic background, $P(\hat{\Omega})=1$, and $\Gamma_{ab}$ reduces to the usual {HD} function of pulsar angular separation.
\par
\endgroup
This connects the observable cross-correlated residuals directly to the GWB strain spectrum and the geometric response of the array. Note that $S_{ab}$ has dimensions of $\mathrm{s}^3$ (or equivalently $\mathrm{yr}^3$), which is perhaps more intuitive when read as $\mathrm{s}^2/\mathrm{Hz}$: it is a power spectral density of a quantity (timing residuals) with dimensions of time. It is here that the two common spectral parameterizations present themselves: for a GW-driven circular binary population $h_c(f) \propto f^{-2/3}$ (spectral index $\alpha = -2/3$), the $1/f^3$ prefactor converts this to a power spectral density $S_{ab} \propto f^{-13/3}$ (spectral index $\gamma = 13/3$). {In free-spectrum plots, {the root-mean-square timing-residual amplitude} per frequency bin is $\sqrt{S_{aa}(f)/T_{\mathrm{obs}}}$, where $S_{aa}$ is the GW contribution to the auto-power spectrum of pulsar $a$.}

Eq.~\ref{eq:crosspower} makes explicit that the cross-power factorizes into two independently measurable components: the ORF $\Gamma_{ab}$, which encodes the spatial correlations between pulsar pairs as a function of their angular separation, and the characteristic strain spectrum $h_c(f)$, which encodes the spectral content of the background. Both must be detected to claim a GWB: a common red-noise spectrum across pulsars without the {HD} spatial correlations could be mimicked by spatially correlated noise, while the angular pattern alone without a consistent spectrum would lack physical interpretation. We derive the ORF in the following section (\sect{sec:HD}), and then turn to the characteristic strain spectrum in \sect{sec:hc}.

\begin{figure}
    \centering
    \includegraphics[width=0.8\linewidth]{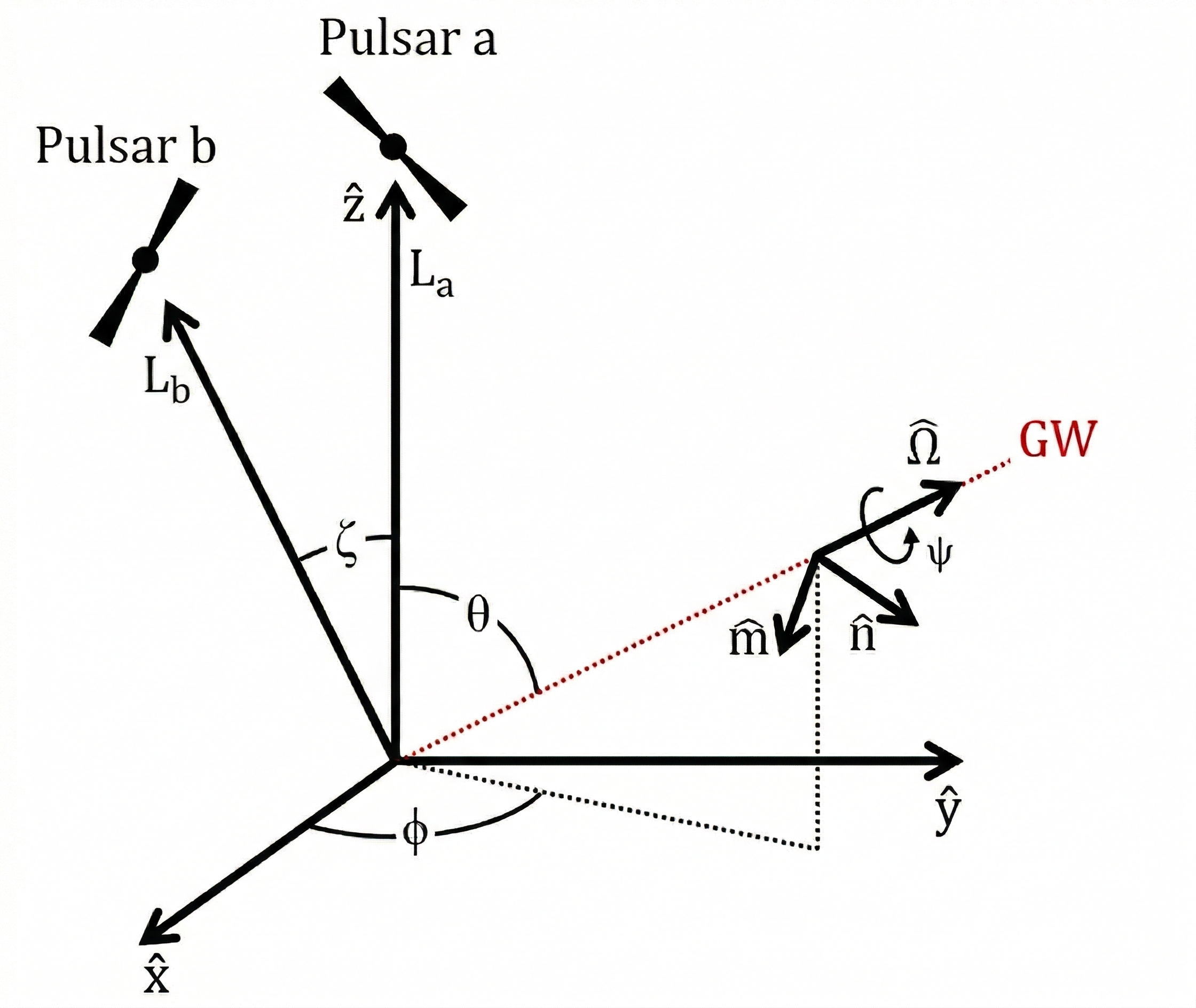}
    \caption{Computational frame used to define the ORF(s). Pulsar $a$ lies at distance $L_a$ on the $+\hat{\mathbf{z}}$ axis and
pulsar $b$ lies at distance $L_b$ in the $x$--$z$ plane at an angular separation $\zeta$. The GW
propagation direction $\hat{\boldsymbol{\Omega}}$ is specified by polar angle $\theta$ and
azimuthal angle $\phi$ with respect to $+\hat{\mathbf{z}}$. The principal polarization axes $\hat{m}$ and $\hat{n}$ are perpendicular to $\hat{\boldsymbol{\Omega}}$, with $\hat{m} \times \hat{n} = \hat{\boldsymbol{\Omega}}$, and define the $+$ and $\times$ polarization tensors that enter the antenna beam pattern $F_a^{A}(\hat{\boldsymbol{\Omega}})$ (\eq{eq:beam_pattern}). The polarization angle $\psi$ rotates $(\hat{m}, \hat{n})$ about $\hat{\boldsymbol{\Omega}}$ (\sect{sec:signal_model}). This geometry
determines the antenna pattern functions $F_a^{A}$ and $F_b^{A}$ and hence any
spatial correlation.}
\label{fig:ptaGeometry}
\end{figure}

\subsection{Derivation of the Hellings and Downs Curve}
\label{sec:HD}
The ORF, $\Gamma_{ab}$, introduced in \eq{eq:crosspower}, quantifies the geometric response of the array. To derive it, we define a frame where pulsar $a$ lies on the $z$-axis and pulsar $b$ lies in the $x-z$ plane at an angular separation $\zeta$ (see \fig{fig:ptaGeometry}). The GW propagation direction $\hat{\Omega}$ is defined by polar angles $(\theta, \phi)$.

The coordinate system is defined as:
\begin{align*}
  \hat{m} &= (\sin\phi, -\cos\phi, 0), \\
  \hat{n} &= (\cos\theta \cos\phi, \cos\theta \sin\phi, -\sin\theta), \\
  \hat{\Omega} &=  (\sin\theta \cos\phi, \sin\theta \sin\phi, \cos\theta), \\
  \hat{p}_a &= (0,0,1), \\
  \hat{p}_b &= (\sin\zeta, 0, \cos\zeta).
\end{align*}

The response of a pulsar to a GW is governed by the antenna beam patterns $F^+(\hat{\Omega})$ and $F^\times(\hat{\Omega})$. For pulsar $a$ on the $z$-axis:
\begin{align}
  F^{+}_a &= -\frac{1}{2}(1 - \cos \theta) \\
  F^{\times}_a &= 0
\end{align}
For pulsar $b$, located at angle $\zeta$, the antenna patterns are more complex due to the rotation:
\begin{align}
  F^{\times}_b &= \frac{(\sin \phi \sin \zeta)(\cos \theta \sin \zeta \cos \phi - \sin \theta \cos \zeta)}{1 + \cos \theta \cos \zeta + \sin \theta \sin \zeta \cos \phi},  \\
  F^{+}_b &= \frac{1}{2} \frac{(\sin \phi \sin \zeta)^2 - (\sin \zeta \cos \theta \cos \phi - \sin \theta \cos \zeta)^2}{1 + \cos \theta \cos \zeta + \sin \theta \sin \zeta \cos \phi}.
\end{align}

It is important to note that these specific patterns arise from the transverse-traceless tensor nature of GWs in GR. Alternative theories of gravity can predict up to four additional polarization modes (scalar and vector), which would possess distinct antenna patterns $F^S(\hat{\Omega})$ and $F^V(\hat{\Omega})$. These alternative modes result in different angular correlation structures (e.g., monopolar or dipolar correlations) when integrated over the sky. We discuss the search for these beyond-GR signatures, including scalar ``breathing'' modes, in \sect{sec:non_gr_polarizations}.

\begingroup

For an isotropic background in the incoherent limit, the normalized ORF is
\begin{equation}
  \Gamma_{ab}=\frac{3(1+\delta_{ab})}{8\pi}\int_{S^2}d\hat{\Omega}\left(F_a^+F_b^++F_a^\times F_b^\times\right).
\end{equation}
\par
\endgroup
The integrand---the product of beam patterns evaluated at a fixed sky direction $\hat{\Omega}$---is the single-source ORF $\Upsilon_{ab}(\hat{\Omega})$, the deterministic ``fingerprint'' of an individual GW source on the pulsar pair. The {HD} curve is thus the sky average of these fingerprints. We derive a closed-form expression for $\Upsilon_{ab}$ in \sect{subsec:deterministic_fingerprints}.

Substituting the geometric expressions and integrating over $\theta \in [0, \pi]$ and $\phi \in [0, 2\pi]$ yields the standard analytic form for the Hellings and Downs correlation \cite{HD83}:
\begin{equation}
  \Gamma_{ab} = (1+\delta_{ab})\frac{1}{2} - \frac{1}{4} \left( \frac{1-\cos \zeta}{2} \right) + \frac{3}{2} \left( \frac{1-\cos \zeta}{2} \right) \ln \left( \frac{1-\cos \zeta}{2} \right) \,.
\end{equation}
This form is normalized to 0.5 at $\zeta\sim0$ for distinct pulsars so that the autocorrelation yeilds 1.0, consistent with the standard literature.

\begin{figure}[t]
    \centering
    \includegraphics[width=0.45\textwidth]{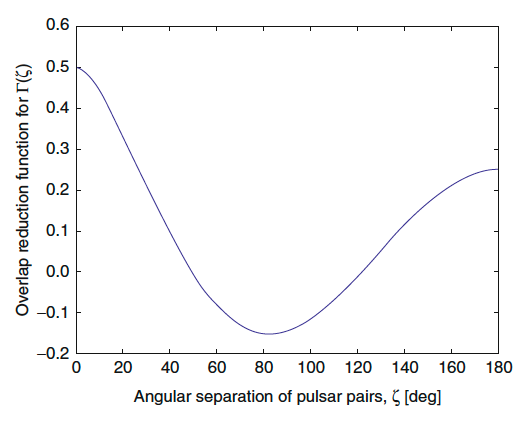} 
    \includegraphics[width=0.52\linewidth]{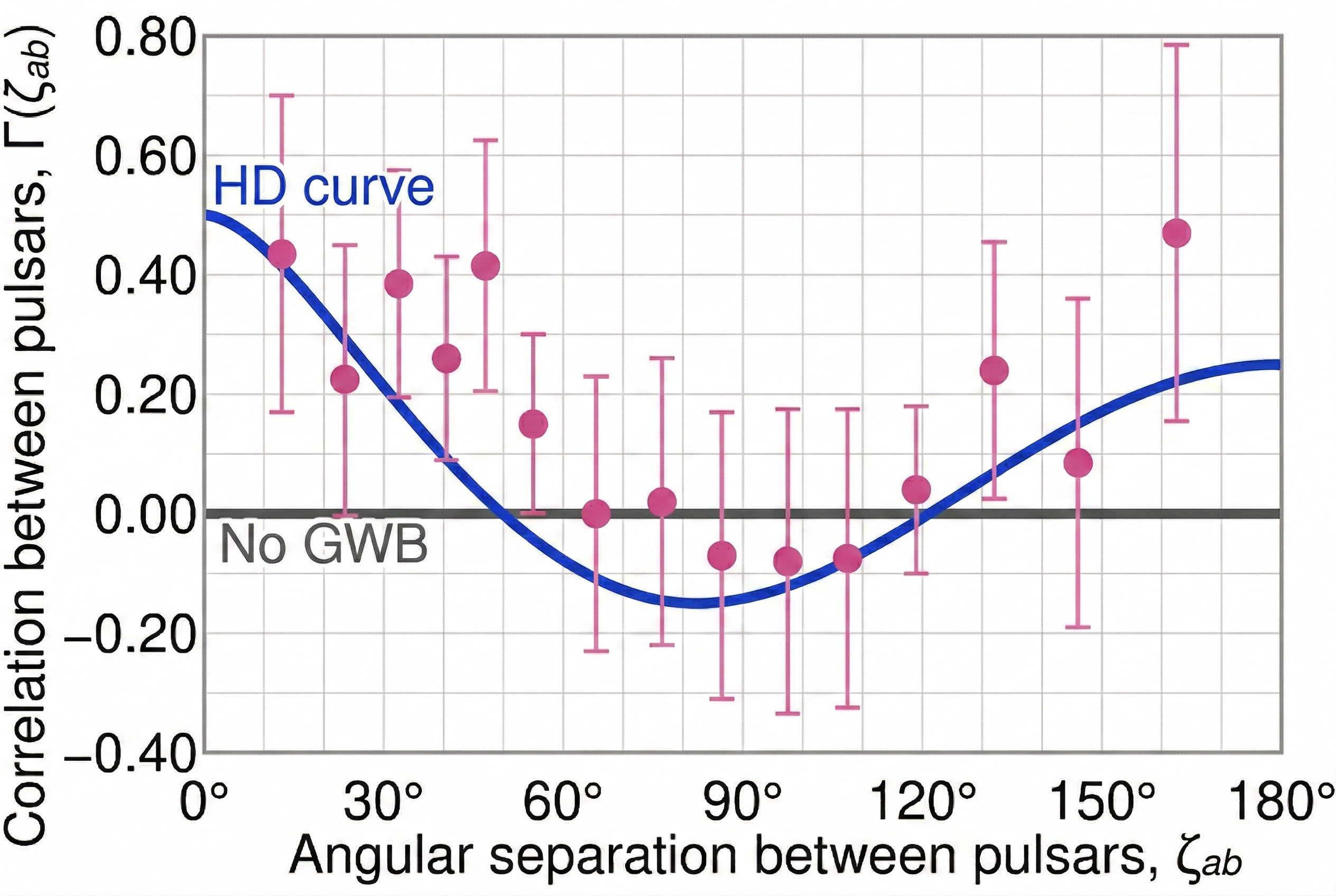}
    \caption{Left: The ORF for an isotropic stochastic GWB, called the Hellings and Downs curve. Right: The optimal statistic~\cite{anholm2009,Chamberlin2015} result for the GWB search from Agazie et al.~\cite{NG15_gwb}. The quantity actually plotted on the $y$-axis is $\hat{A}^2 \Gamma_{ab}$, where $\hat{A}^2 \sim 10^{-30}$ is the maximum-likelihood estimator of the squared GWB amplitude; the axis label shows only $\Gamma(\zeta_{ab})$ for presentation clarity. The binning of pulsar pairs by angular separation is for visual clarity only and has no bearing on the result.}
    \label{fig:HDcurve}
\end{figure}

This curve (\fig{fig:HDcurve}) is the ``smoking gun'' signature of a GWB. It features a maximum correlation of 0.5 at $\zeta\sim0^\circ$ (for distinct pulsars) and a minimum near $\zeta \approx 90^\circ$. For pulsars with small angular separations, the timing residuals are positively correlated (the pulsars appear to ``bob'' in sync). As the separation increases, the correlation drops, becoming negative at separations of $\sim 90^\circ$, before becoming positive again at wider angles.

Standard analyses often model the ORF assuming the pulsar term contribution averages to zero~\cite{MM18}. However, a rigorous treatment, as detailed in \cite{MS14}, includes the interference effects from the pulsar term, denoted as $\kappa_{ab}$. It has been shown that at low frequencies, where the gravitational wavelength is comparable to the separation between pulsars, the pulsar term does not simply vanish. Instead, it introduces Bessel function oscillations around the standard {HD} solution at small angles. While current PTAs often neglect this effect for typical angular separations, it becomes non-negligible for pulsars with very small separations, such as those found in globular clusters or in double pulsar systems where both pulsars are being timed.

\begingroup

An alternative and concise derivation of the Hellings and Downs curve, avoiding any choice of coordinate frame, was given by Ali-Ha\"imoud, Smith \& Mingarelli~\cite{ani2020}. Defining $x\equiv\hp_a\cdot\ho$, $y\equiv\hp_b\cdot\ho$, and $\mu\equiv\cos\zeta=\hp_a\cdot\hp_b$, their sky integral can be written using the same ORF normalization as above:
\begin{equation}
\label{eq:HD_split}
\Gamma_{ab}=\frac{3}{8}\left[\mathcal{J}(\mu)+1+\frac{\mu}{3}\right]+\frac{\delta_{ab}}{2},
\end{equation}
\par
\endgroup
where the first term collects the nontrivial part of the integral:
\begin{equation}
\label{eq:J_integral}
\mathcal{J}(\mu) \equiv 2 \int \frac{d^2\ho}{4\pi}\, \frac{x^2 + y^2 - 2\mu x y - (1-\mu^2)}{(1+x)(1+y)}\,.
\end{equation}
The key is to change variables from sky coordinates to $(x,y)$, which are constrained to an ellipse $x^2 + y^2 - 2\mu x y < 1-\mu^2$, with area element $d^2\ho = 2\,dx\,dy/\sqrt{1-\mu^2 - x^2 - y^2 + 2\mu x y}$. This reduces $\mathcal{J}$ to a factorized double integral: for fixed $x$, the inner integral over $y$ evaluates to $|x+\mu| - (1+\mu x)$, and the remaining outer integral over $x$ yields $\mathcal{J}(\mu) = 2(1-\mu)\ln[(1-\mu)/2]$. Combining gives the Hellings and Downs function in the compact form
\begingroup

\begin{equation}
\label{eq:HD_AliHaimoud}
\Gamma_{ab}=\frac{3+\mu}{8}+\frac{3}{4}(1-\mu)\ln\!\left(\frac{1-\mu}{2}\right)+\frac{\delta_{ab}}{2}.
\end{equation}
\par
\endgroup
This frame-independent approach also generalizes naturally to anisotropic backgrounds, as discussed in \sect{sec:anisotropy}.

While the standard analysis assumes the specific shape of the {HD} curve, a more agnostic approach is to decompose the spatial correlations into a basis of Legendre polynomials, $P_\ell(\cos \zeta_{ab})$.  Roebber \& Holder~\cite{Roebber2017} showed that the angular power spectrum of the PTA redshift map is the harmonic transform of the {HD} curve, demonstrating that the quadrupole carries the dominant sensitivity. Building on these foundations, Nay et al.~\cite{Nay2024} developed a complete harmonic analysis framework for PTAs that enables model-independent reconstruction of the ORF directly from the timing residuals. By measuring the Legendre coefficients $C_\ell$, one can rigorously test for the presence of the quadrupolar ($\ell=2$) signature predicted by GR without assuming it \textit{a priori}. This formalism was already mature by the time the NANOGrav 15-yr GWB evidence was published, and was cited in the GWB evidence paper itself~\cite{NG15_gwb}.

The NANOGrav collaboration applied this harmonic analysis to the 15-yr dataset~\cite{NANOGravHarmonic2024}, providing an important and independent measurement of the GWB. They found that the data disfavor a purely monopolar or dipolar structure and are consistent with the {HD} prediction, although the recovery of the exact quadrupolar shape is currently limited by the number of unique pulsar pairs at small angular separations. This independent measurement technique offers direct insight into the structure of the {HD} curve itself: by reconstructing the ORF multipole by multipole, harmonic analysis can reveal deviations from GR, constrain alternative polarization content, and probe frequency-dependent modifications to the spatial correlations that would otherwise be invisible. It can also distinguish the GWB from correlated noise sources like clock errors ($\ell=0$) or ephemeris drift ($\ell=1$), see e.g. Tiburzi et al.~\cite{Tiburzi2016}, providing an independent verification of the GWB evidence.

\subsection{The Characteristic Strain Spectrum}
\label{sec:hc}
{For a circular SMBHB, {the root-mean-square strain} averaged over binary orientations is~\cite{Thorne1987,BurkeSpolaor2019}}
\begingroup

\begin{equation}
\label{eq:SMBHB_h}
    h = \sqrt{\frac{32}{5}}\,\frac{\mathcal{M}^{5/3}}{D}\,(\pi f_r)^{2/3},
\end{equation}
where $\mathcal{M}$ is the source-frame chirp mass, $D$ is the comoving distance, and $f_r=(1+z)f$ is the source-frame GW frequency, with $f$ the observed GW frequency.
\par
\endgroup

The total energy density of the background $\rho_{\mathrm{GW}}$ is related to the characteristic strain by \cite{phinney_practical_2001, Sesana2008PTAoccupancy}:
\begin{equation}
    \label{eq:strain_spectrum_energy_density}
    \frac{d \rho_{\mathrm{GW}}}{d \ln f} = \frac{\pi}{4} f^{2} h_{c}^{2}(f) \, .
\end{equation}
By integrating the comoving number density of mergers $\phi_{\mathrm{BHB}}(z)$, we arrive at the standard power-law representation for the characteristic strain:
\begin{equation}
    \label{eq:strain_spectrum_circular_bhbs}
    h_{c}^{2}(f) = \frac{4}{3 \pi^{1 / 3}} \frac{1}{f^{4 / 3}} \iint \frac{\mathcal{M}^{5 / 3}}{(1 + z)^{1 / 3}} \dot{\phi}_{\mathrm{BHB}}(\mathcal{M}, z) \frac{dt_{\mathrm{r}}}{dz} d \log_{10} \mathcal{M} \, dz \, .
\end{equation}
Measuring $h_c$ is therefore a direct probe of the physics in the double integral above, \eq{eq:strain_spectrum_circular_bhbs}: the SMBHB merger rate density $\dot{\phi}_{\mathrm{BHB}}$ and, critically, the chirp mass distribution~\cite{Mingarelli2019,MingarelliCaseyClyde2022,Mingarelli2025_KITP}. {The steep $\mathcal{M}^{5/3}$ weighting makes the GWB particularly sensitive to the high-mass end of the SMBHB population~\cite{CaseyClyde2022}.}
This integration yields the characteristic power-law slope $h_c(f) = A_{\mathrm{yr}} (f/f_{\mathrm{yr}})^{-2/3}$  for astrophysical GWB searches.

\subsection{A discrete population of SMBHBs}
\label{sec:discrete_population}
{Phinney's smooth spectrum follows from the ensemble-averaged GW power, calculated using expected source counts that can be fractional~\cite{phinney_practical_2001}. Each simulated population contains integer numbers of binaries; sampling many such realizations gives the distribution of spectra and its median. At high frequencies, rare massive binaries can contribute strongly to the ensemble average while being absent from most realizations, causing the median spectrum to fall below the smooth prediction, as illustrated in Fig.~\ref{fig:fig1}~\cite{NG15_discreteness}.}

Sesana et al.~\cite{Sesana2008PTAoccupancy} first identified this spectral break, referring to it as a ``knee'' in the characteristic strain spectrum where the transition from the continuous to discrete regime occurs, predicting $f_{\mathrm{knee}} = 37^{+15}_{-13}$~nHz.

{Roebber et al.~\cite{Roebber2016} used cosmological simulations and empirical relations linking dark matter halos, galaxies, and black holes to distinguish astrophysical model uncertainty from the intrinsic scatter between realizations of the SMBHB population. In their models, variations in these relations mainly shifted the overall GWB amplitude, whereas source discreteness produced scatter between realizations that increased toward higher frequencies. Above roughly 20~nHz, this Poisson scatter exceeded the modeled astrophysical uncertainty. The median spectrum fell below the smooth $h_c\propto f^{-2/3}$ prediction, while rare bright binaries produced large upward excursions. This scatter is an intrinsic consequence of observing only one realization of the population.}

Using a semi-analytic SMBHB population model calibrated to the NANOGrav 15-yr GWB amplitude, Agazie et al.~\cite{NG15_discreteness} carried out the first search for discreteness in real PTA data. {A broken power law provided a better fit to the simulated GWB spectra than a pure $f^{-2/3}$ power law, yielding $f_{\mathrm{knee}} = 25^{+42}_{-19}$~nHz, consistent with the earlier prediction by Sesana et al.} {Agazie et al. also identified a modest excess at 16~nHz and explored a possible single-binary interpretation~\cite{NG15_discreteness}. Subsequent analysis with customized chromatic noise models found that this excess was no longer significant, linking its reduction to improved modeling of chromatic transients in PSR J1713+0747~\cite{NG15_CNM_GWB,LarsenCNM2025}. This highlights the importance of pulsar-specific noise models when interpreting spectral irregularities (see \sect{subsec:noise}). A loud SMBHB could nevertheless first reveal itself as an excess in the GWB spectrum, before being individually resolved in a CW search~\cite{NG15_discreteness}.}

{The statistical properties of spectral irregularities offer a way to test whether the GWB is produced by a discrete SMBHB population.} Importantly, signs of discreteness in the GWB spectrum may be observable before either GWB anisotropy (\sect{sec:anisotropy}) or individually resolved CW sources, making spectral irregularities a near-term probe of the astrophysical nature of the signal. {In this regime, the background can exhibit net polarization because the polarizations of a few dominant binaries need not average out (see \sect{sec:gwb_polarization}).}

\section{GWB Predictions and Limits}
\label{sec:gwb_history}

The path to the current GWB evidence has been shaped by a complex interplay between astrophysical theory and observational constraints. For nearly a decade, the field grappled with an apparent tension: while early models predicted a loud background, pulsar timing limits seemed to rule them out, driving theorists toward increasingly pessimistic scenarios. {Subsequent work showed that pulsar-noise modeling and prior choices can make GWB upper limits overly stringent, changing how these limits should be compared with astrophysical predictions~\cite{Hazboun2020}.}

\subsection*{The Foundational Era (1995--2013)}
The framework for the stochastic background was established in the 1990s and early 2000s. Foundational analytic work by Rajagopal \& Romani~\cite{Rajagopal1995} and Jaffe \& Backer~\cite{Jaffe2003} demonstrated that the incoherent superposition of signals from cosmic binaries would result in a red noise spectrum with a characteristic strain $h_c(f) \propto f^{-2/3}$~\cite{phinney_practical_2001}. These early models, along with Wyithe \& Loeb~\cite{Wyithe2003}, predicted amplitudes in the range $A_{\mathrm{yr}} \sim 10^{-15}$, setting the sensitivity targets for the first generation of PTA experiments.

The statistical toolkit for PTA searches was also rapidly maturing. {van Haasteren et al.~(2009)~\cite{vanHaasteren2009} developed a Bayesian framework for GWB searches and demonstrated it on simulated pulsar timing data.} On the frequentist side, Anholm et al. (2009)~\cite{anholm2009} derived the optimal cross-correlation statistic in the frequency domain. Beyond detection, the groundwork for characterizing the GWB was also laid during this period. Mingarelli et al. (2013)~\cite{m13} developed the spherical harmonic formalism for GWB anisotropy, quantifying the expected angular power spectrum from a discrete population of SMBHBs, while Taylor \& Gair~(2013)\cite{TaylorGair2013} introduced practical Bayesian search methods for these spherical harmonics. {Furthermore, Lee, Jenet \& Price (2008)~\cite{Lee2008} showed that PTAs are sensitive to non-Einsteinian GW polarizations---scalar and vector modes predicted by alternative theories of gravity. Lee et al. (2010)~\cite{LeeJenetPrice2010} subsequently showed that PTAs can constrain the graviton mass through its effect on the angular correlation between pulsar timing residuals.}

{Sesana et al. (2008)~\cite{Sesana2008PTAoccupancy} explored SMBHB populations using halo merger trees, while Sesana (2013)~\cite{sesana_systematic_2013} combined {observed galaxy stellar mass functions (GSMFs) and pair fractions} with merger timescales and empirical black-hole--host relations. These studies predicted amplitudes of order $A_{\mathrm{yr}}\sim10^{-15}$, with substantial astrophysical uncertainty.}

\subsection*{The ``Stalling'' Crisis (2014--2018)}
As PTAs collected more data without a detection, theoretical models shifted downward to accommodate the lack of a signal. It became clear that the transition from kpc-scale galaxy mergers to pc-scale binaries is not instantaneous. If the mechanisms for hardening the binary---dynamical friction, stellar scattering, and gas drag---are inefficient, the binary may stall at parsec separations (the ``Final Parsec Problem'').

Even if binaries do merge, environmental coupling can attenuate the signal. Ravi et al.~\cite{Ravi2015}, Sampson et al.~\cite{Sampson2015}, and Dvorkin \& Barausse~\cite{Dvorkin2017} demonstrated that strong environmental coupling accelerates the inspiral at low frequencies, depleting the number of binaries radiating in the lower end of the PTA band. {Using Illustris-based populations and models of binary evolution, Kelley et al.~\cite{Kelley2017_DynamicalEnvironments} predicted $A_{\mathrm{yr}}\simeq0.4\times10^{-15}$ for a conservative fiducial model.}

An important theoretical development during this period was the realization that {triple supermassive black hole (SMBH) interactions} can resolve the final parsec problem. In hierarchical structure formation, sequential galaxy mergers naturally deliver a third black hole to the nucleus before the existing binary has coalesced. Bonetti et al.~\cite{Bonetti2018} modeled the {post-Newtonian (pN) evolution} of massive black hole triplets in galactic nuclei and showed that three-body interactions efficiently shrink the inner binary's orbit, driving it into the GW-dominated regime. They established a robust lower limit of $A_{\mathrm{yr}} \gtrsim 10^{-16}$ for the stochastic GWB, demonstrating that even in the most pessimistic scenarios for stellar and gas-driven hardening, triplet interactions alone guarantee a detectable signal floor. {Independently, Ryu et al.~\cite{Ryu2018} used simulations of multiple SMBHs in merging galaxies and showed that interactions among the black holes could drive coalescence even when stellar hardening stalled.} Their predicted GWB amplitude was consistent with the Bonetti et al. floor, reinforcing the conclusion that the final parsec problem does not prevent SMBHBs from reaching the GW-driven regime.

\subsection*{The Noise Modeling Breakthrough (2020--2021) }
\label{sec:noise_breakthrough}
\begin{figure}
    \centering
    \includegraphics[width=\linewidth]{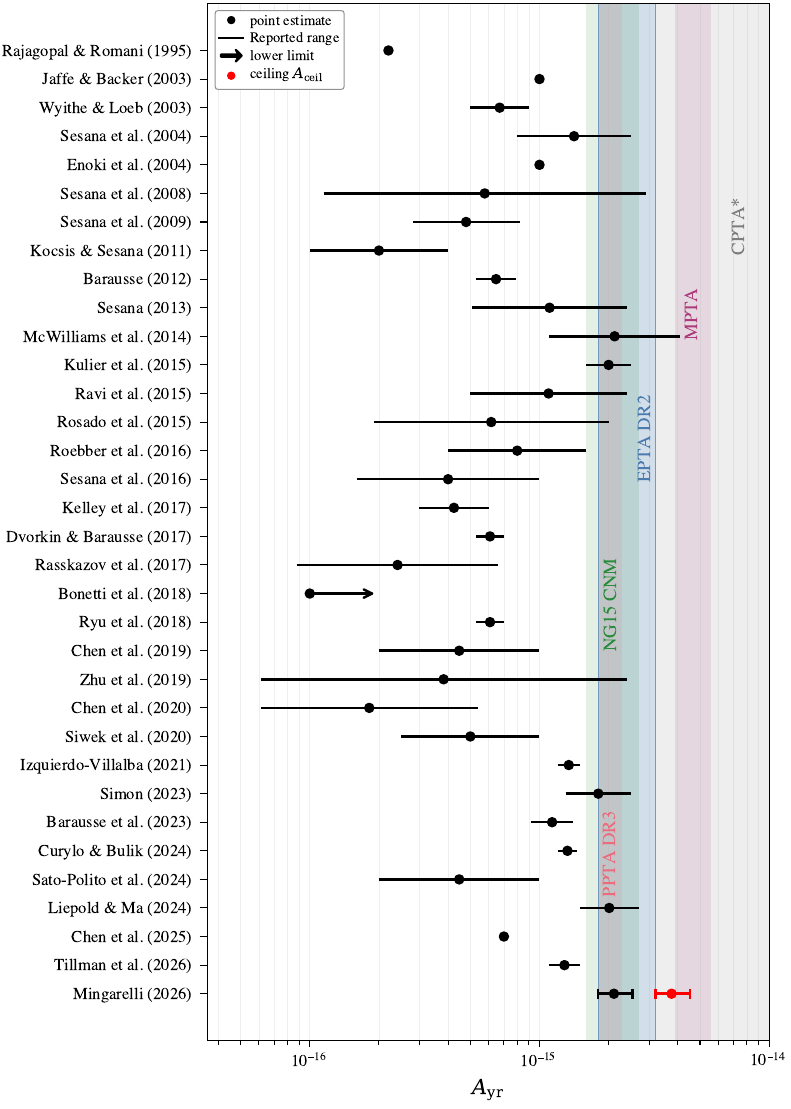}
    \caption{  Predicted GWB characteristic strain amplitudes at $f = 1\,\mathrm{yr}^{-1}$ from 33 astrophysical models spanning three decades of
  work~\cite{Rajagopal1995, Jaffe2003, Wyithe2003, Sesana2004, Enoki2004, Sesana2008PTAoccupancy, SVV2009, KocsisSesana2011, Barausse2012,
  Sesana13, McWilliams2014, Kulier2015, Ravi2015, RosadoSesana2015, Roebber2016, sesana2016, Kelley2017, Dvorkin2017, Rasskazov2017,
  Bonetti2018, Ryu2018, chen_constraining_2019, Zhu2019, Chen2020, Siwek2020, Izquierdo2021, Simon2023, Barausse2023, Curylo2024,
  SatoPolito2024_bigBH, LiepoldMa2024, Chen2025_ASTRID, Tillman2026}. {The bottom row shows the SMBHB energetic benchmark ($A_{\rm bench}$, black) and ceiling ($A_{\rm ceil}$, red) with 90\% intervals~\cite{Mingarelli2026}. {The NANOGrav band shows the 90\% credible interval from the 15-yr analysis with custom noise models} {with $\gamma$ fixed to $13/3$}~\cite{NG15_CNM_GWB}.} *CPTA upper limits span three order of magnitude in amplitude and has been truncated for clarity.}
    \label{fig:astro-limits-GWB}
\end{figure}

While theorists were refining environmental models, a parallel breakthrough occurred in data analysis. For years, PTA upper limits (e.g., from PPTA and NANOGrav) appeared to rule out amplitudes above $A_{\mathrm{yr}} \sim 10^{-15}$, reinforcing the pessimistic theoretical outlook (see {\fig{fig:astro-limits-GWB} for model predictions compared with current PTA measurements}). {However, Hazboun et al. (2020)~\cite{Hazboun2020} showed that noise models and priors can make GWB upper limits overly stringent, while Goncharov et al. (2022)~\cite{Goncharov2022} developed a test to distinguish a common spectrum from similar but distinct pulsar-noise spectra.}

Hazboun et al. (2020) identified the problem from the prior side. They demonstrated that Bayesian priors on intrinsic pulsar red noise are critical to the recovered GWB upper limit. In previous analyses, {standard uniform priors on the red noise amplitude} allowed the per-pulsar noise models to absorb power that was, in fact, common across the array. The noise models were effectively ``eating'' the GWB signal, biasing the recovered amplitude downward and producing overly stringent upper limits. {By systematically varying the noise models and prior choices, Hazboun et al. showed that the NANOGrav 11-year upper limit could shift upward by about a factor of two depending on the noise model.}

Goncharov et al. (2022) approached the problem from the model-selection side using PPTA data. {Focusing on priors,} they showed that intrinsic spin noise in individual pulsars can be covariant with {{a common uncorrelated red-noise (CURN) process}}. {When the noise model does not account for this covariance,} power can leak between the two components in either direction: the common process can be absorbed into individual noise terms (suppressing the GWB), or conversely, {pulsar noise with similar spectra can mimic evidence for a CURN process}. {With the common-process spectral index fixed at $\gamma=13/3$, their test found the PPTA data consistent with a CURN process.}

{These studies showed that noise models and priors can affect both GWB upper limits and evidence for a CURN process (see also Goncharov et al.~\cite{Goncharov2021_CRN} for an analysis of HD correlations).} {Improvements in noise modeling and prior choices helped recover CURN signals that earlier analyses had absorbed into noise terms~\cite{Hazboun2020,NG12p5_gwb}. In retrospect, the GWB had been present in the data all along.}

\subsection*{The Modern Era (2021--Present)}
With the observational ceiling lifted, the theoretical landscape has realigned around high-amplitude models. The current GWB evidence is best explained by models that emphasize the contribution of the most massive black holes ($M_{\mathrm{BH}} > 10^9 M_{\odot}$). {In the models compared by Simon~\cite{Simon2023}, predictions based on the Velocity Dispersion Function (VDF) have higher amplitudes than GSMF-based predictions because the VDF-based populations contain more massive binaries at higher redshifts (see \sect{sec:modeling_gwb} for a detailed comparison).}

Taking a complementary approach, Casey-Clyde et al.~\cite{casey-clyde_quasarbased_2022} developed an inverse model anchored to the quasar luminosity function (QLF): rather than predicting $h_c$, they take the observed GWB amplitude as input and infer the underlying SMBHB population---including the local number density, mass distribution, and {an upper limit on the fraction of local SMBHBs associated with quasars}. This is a concrete realization of the program outlined in \eq{eq:strain_spectrum_circular_bhbs}: measuring $h_c$ constrains the physics in the double integral, and the QLF model exploits this by using the GWB measurement to extract the merger rate density $\dot{\phi}_{\mathrm{BHB}}$ via the observed quasar population rather than galaxy-merger assumptions (see \fig{fig:quasarVmm} for a schematic comparison).

Meanwhile, large-volume simulations like ASTRID~\cite{Ni2024_ASTRID} now capture the rare, ultra-massive $z \sim 0.3$ binaries missed by smaller boxes. {They report $h_c \approx 2.8\times10^{-15}$ at $4\,\mathrm{nHz}$, corresponding to $A_{\mathrm{yr}}\approx7\times10^{-16}$ when extrapolated to $1\,\mathrm{yr}^{-1}$ using $h_c\propto f^{-2/3}$.} However, their fiducial GWB predictions remain below the measured amplitude (e.g.\ $A_{\mathrm{yr}} \approx 7 \times 10^{-16}$ for ASTRID), which may reflect AGN feedback prescriptions, or other modeling choices~\cite{Tillman2026}. More broadly, Lapi et al.~\cite{Lapi2026} use a semi-empirical framework to highlight a persistent tension between local SMBH demographics and the GWB amplitude, suggesting that reconciling the two may require revisions to the assumed mass function, accretion history, or both.


\subsection*{The Astrophysical GWB Ceiling}
\label{subsec:energetic_ceiling}

\begingroup

Mingarelli~\cite{Mingarelli2026} relates the GWB amplitude to the available source mass, merger history and emitted spectrum. For SMBHBs, $\rho_{\rm src}=f_{\rm merge}\rho_\bullet$, where $\rho_\bullet$ is the local SMBH mass density~\cite{YuTremaine2002,ShankarWeinberg2009,marconi_local_2004} and $\rho_{\rm src}$ is the cumulative total binary mass density entering merger events. The merger factor $f_{\rm merge}$ can exceed unity when the same mass participates in successive mergers. The integrated quasar light independently constrains the mass accumulated by accretion, given an assumed radiative efficiency~\cite{Soltan1982}.

Using the Liepold--Ma black-hole mass function~\cite{LiepoldMa2024}, based in part on the MASSIVE survey~\cite{Ma2014}, the benchmark scaling relation at $f_{\rm ref}=1\,{\rm yr}^{-1}$ is
\begingroup\small
\begin{equation}
A_{\rm bench}
=2.11\times10^{-15}
\left(\frac{\rho_\bullet}
{1.80\times10^{6}\,M_\odot\,{\rm Mpc}^{-3}}\right)^{1/2}
\!\left(\frac{f_{\rm merge}}{1}\right)^{1/2}
\!\left(\frac{\epsilon_{\rm gw}}{0.02}\right)^{1/2}
\!\left(\frac{\mathcal H}{\mathcal H_{\rm LM24}}\right)^{1/2}.
\label{eq:smbhb_benchmark_scaling}
\end{equation}
\endgroup
Here $\epsilon_{\rm gw}$ is the fraction of processed rest-mass energy radiated as GWs, and $\mathcal H(f)\equiv(\epsilon_{\rm gw}f_{\rm merge}\rho_\bullet)^{-1}\,{\rm d}\rho_{\rm gw}/{\rm d}\ln f$ is the present-day GW energy density per logarithmic frequency interval, normalized by the total emitted GW-energy budget. It incorporates the source spectrum and redshift distribution. The reference $\mathcal H_{\rm LM24}$ evaluates this quantity at $f_{\rm ref}$ for the adopted Liepold--Ma population. Assigning one progenitor merger to each present-day SMBH gives $f_{\rm merge}=1$ and $A_{\rm bench}=2.11^{+0.43}_{-0.32}\times10^{-15}$, neglecting GW mass loss and later accretion in the remnant-mass accounting.

The ceiling follows by holding the present-day mass function fixed, setting equal masses and $z\rightarrow0$, and including the infinite, mass-conserving merger hierarchy of Sato-Polito et al.~\cite{SatoPolito2024_bigBH}. These changes increase the fiducial benchmark's squared amplitude by $1.05\times1.12\times2.70\simeq3.18$, giving
\begin{equation}
A_{\rm ceil}
\simeq\sqrt{3.18}\,A_{\rm bench}
=3.76^{+0.76}_{-0.56}\times10^{-15}.
\label{eq:pta_ceiling}
\end{equation}
The quoted 90\% ranges propagate the adopted SMBH-density uncertainty. The hierarchy changes both the binary-mass distribution and $\mathcal H$; it is not simply an increase in $f_{\rm merge}$ at fixed population assumptions. Within the benchmark model, the revised NANOGrav amplitude is consistent with $f_{\rm merge}=1$, without requiring an additional uncounted SMBH population.
\par
\endgroup

\section{Modeling the GWB}
\label{sec:modeling_gwb}

The amplitude and spectral shape of the GWB contain a wealth of information about the history of galaxy formation and black hole growth. As derived in \sect{sec:stochastic_background}, the characteristic strain spectrum (\eq{eq:strain_spectrum_circular_bhbs}) is set by the SMBHB merger rate density and chirp mass distribution, and connects directly to the cross-power spectral density of the timing residuals (\eq{eq:crosspower}). To decode this information, we rely on population synthesis models that link the observable properties of galaxies (such as stellar mass and merger rate) to the invisible population of GW-emitting binaries.

Broadly, these models fall into three categories: (1) Major Merger Models, which track the hierarchical assembly of dark matter halos and galaxies; (2) Quasar-based Models, which use the population of AGN as tracers for merger activity; and (3) Astrophysics-agnostic Models, which parameterize the binary population directly without assuming specific host galaxy relations.

\begin{figure}[b]
    \centering
    \includegraphics[width=\columnwidth]{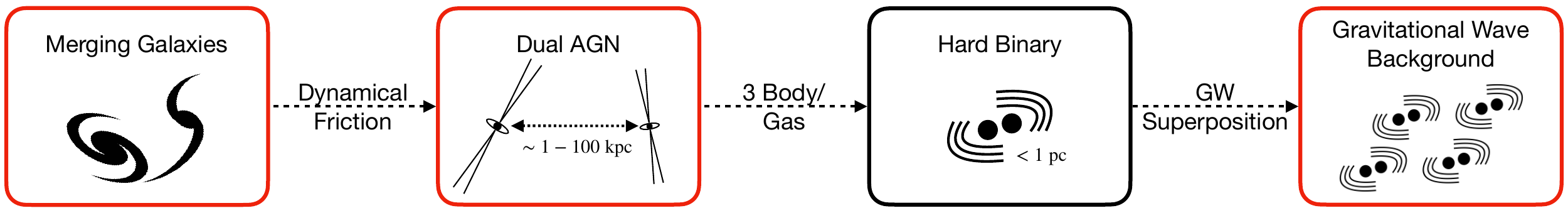} 
    \caption{\textbf{The SMBHB Lifecycle.}
    (Left) Following a galaxy merger, dynamical friction from the stellar and dark matter background drives the two SMBHs toward the center of the remnant, from $\sim$\,kpc separations down to $\sim$\,pc scales on timescales of $\sim 100$\,Myr--Gyr. The galaxy merger rate that initiates this process is described by \eq{eq:galaxy_pair_rate}.
    (Center) At kiloparsec-to-parsec scales, if both black holes are accreting, the system may be observable as a dual AGN~\cite{Goulding2019,Koss2023}. Further hardening requires three-body stellar scattering or circumbinary gas drag to overcome the ``final parsec problem.'' The physics governing these stages is captured by \eq{eq:binarylife}.
    (Right) Once the binary reaches milliparsec separations, GW emission becomes efficient, driving the system to coalescence. These binaries are potential CW sources (\sect{sec:CW}). The incoherent superposition of signals from this final phase constitutes the nanoHertz GWB (\sect{sec:stochastic_background}).}
    \label{fig:binary_lifecycle}
\end{figure}

Before describing the statistical models, it is useful to outline the physical journey of a SMBHB (see \fig{fig:binary_lifecycle}). The process begins when two galaxies, each hosting a central SMBH, become gravitationally bound. The SMBHs sink toward the center of the merger remnant via dynamical friction against the background of stars and dark matter on kiloparsec scales. Once the black holes form a bound pair, dynamical friction becomes inefficient. The binary must continue to shrink through interactions with its environment before GW emission can drive the system to coalescence. The total orbital decay rate is the sum of three contributions:
\begin{equation}
    \left(\frac{da}{dt}\right) = \left(\frac{da}{dt}\right)_{\mathrm{GW}} + \left(\frac{da}{dt}\right)_{\star} + \left(\frac{da}{dt}\right)_{\mathrm{gas}}
    \label{eq:binarylife}
\end{equation}

\begin{enumerate}
    \item \textbf{Stellar Hardening (pc scale):} At parsec scales, binaries harden by scattering individual stars from the loss cone. The hardening rate is typically parameterized as~\cite{Quinlan96, SesanaKhan2015}:
    \begin{equation}
      \left(\frac{da}{dt}\right)_{\star} = -\frac{ \rho H}{\sigma} a^2 \,,
    \end{equation}
    where $\rho$ is the stellar density, $\sigma$ is the velocity dispersion, and $H \approx 15-20$ is a dimensionless hardening parameter. This mechanism is efficient at low frequencies but fades as the binary shrinks ($da/dt \propto a^2$). If stellar hardening dominates, the residence time of binaries at low frequencies is reduced, causing a turnover or flattening of the GWB spectrum to $h_c(f) \propto f$ at $f \lesssim 10$~nHz. {The time spent in this phase contributes to the delay between the galaxy merger and SMBHB coalescence.}

    \item \textbf{Gas-Driven Migration (sub-pc scale):} In gas-rich environments, circumbinary disks can exert torques on the binary. While highly model-dependent, the migration rate is often scaled to the local accretion rate $\dot{M}$~\cite{Ivanov, KocsisSesana2011}:
    \begin{equation}
        \left(\frac{da}{dt}\right)_{\mathrm{gas}} \propto - \frac{\dot{M}}{\mu} a^{1/2} \,,
    \end{equation}
    where $\mu$ is the reduced mass. Gas-driven migration typically results in spectral indices flatter than the GW-only prediction {(e.g., $h_c(f) \propto f^{1/2}$ in the gas-dominated limit of the scaling above)}, depending on the specific disk viscosity and cavity interaction models~{\cite{KocsisSesana2011,Siwek2024_CBD}}.

    \item \textbf{Gravitational Wave Emission (milliparsec scale):} At small separations, GW emission becomes the dominant energy loss mechanism. For a circular binary of mass ratio $q=M_2/M_1$ and total mass $M_{\mathrm{tot}}$, the decay rate is given by Peters (1964)~\cite{P1964}:
    \begin{equation}
      \left(\frac{da}{dt}\right)_{\mathrm{GW}} = -\frac{64}{5} \frac{ M_1 M_2 M_{\mathrm{tot}}}{ a^3} \,.
    \end{equation}
    Since the frequency scales as $f \propto a^{-3/2}$, this leads to the rapid evolution $\dot{f} \propto f^{11/3}$ and the standard characteristic strain spectrum $h_c(f) \propto f^{-2/3}$.
\end{enumerate}

{PTAs probe the late inspiral, where environmental interactions may still compete with GW emission.} However, the environmental hardening mechanisms above shape the population of binaries that reach the PTA band, and their imprint on the GWB spectrum is observable. Recent PTA analyses search for these signatures by fitting for a {low-frequency turnover} or allowing the spectral index $\gamma$ to deviate from $13/3$~{\cite{NG15_gwb,ChenYifan2026}}. The population models below must account for the time elapsed during these earlier hardening phases to correctly predict the merger rate at redshift $z$.
\subsection{Major Merger Models}
The most direct approach models the SMBHB population semi-analytically by deriving the SMBHB merger rate from the galaxy merger rate, which is in turn derived from the observed galaxy pairing rate \citep{sesana_systematic_2013,sesana2016,chen_constraining_2019, middleton_massive_2021}.

The galaxy differential pairing rate per unit proper time, $t$, logarithmic galaxy stellar mass, $\log_{10} M_{*}$, and galaxy mass ratio, $q_{\mathrm{G}}$, is modeled as:
\begin{equation}
    \dot{\phi}_{\mathrm{G, pair}} = \frac{d^{3} \Phi_{\mathrm{G, pair}}}{dt\ d\log_{10} M_{*}\ dq_{\mathrm{G}}} = \phi_{*}(M_{*}, z) \frac{\mathcal{F}(M_{*}, z, q_{\mathrm{G}})}{\tau_{\mathrm{G}}(M_{*}, z, q_{\mathrm{G}})} \, ,
\label{eq:galaxy_pair_rate}
\end{equation}
where $\phi_{*}(M_{*}, z)$ is {the GSMF} describing the density of galaxies at a given mass and redshift, and $\mathcal{F}$ is the {differential} pair fraction {per unit galaxy mass ratio $q_{\mathrm{G}}$}. The critical astrophysical uncertainty lies in $\tau_{\mathrm{G}}$, the galaxy merger timescale. This is the proper time elapsed between the pairing of two galaxies at $z_{\mathrm{pair}}$ and their merger at $z_{\mathrm{merge}}$.

To translate galaxy mergers into black hole mergers, standard models rely on empirical scaling relations, typically chaining $M_* \rightarrow M_{\mathrm{bulge}} \rightarrow M_{\mathrm{BH}}$. In the standard GSMF framework, we scale from bulge mass to black hole mass using:
\begin{equation}
    \log_{10} M_{\mathrm{BH}} = \alpha + \beta \log_{10} \left(\frac{M_{\mathrm{bulge}}}{10^{11} \; \mathrm{M}_{\odot}}\right) \pm \Delta \, ,
\end{equation}
where $\Delta$ represents the intrinsic scatter. This scatter is physically important; it allows for the possibility that a moderate-mass galaxy hosts an unexpectedly massive black hole. Given the steep mass scaling of the GW signal ($h \propto \mathcal{M}^{5/3}$), even a small number of ``over-massive'' black holes scattered upward by $\Delta$ can dominate the background signal. We compute the final SMBHB merger rate, $\dot{\phi}_{\mathrm{BHB}}$, by convolving the galaxy merger rate with the probability {distribution of black hole masses} $P(\log_{10} M_{\mathrm{BH}} \vert \log_{10} M_{\mathrm{bulge}})$.

However, recent work has highlighted that the velocity dispersion ($\sigma$) of the host galaxy may be a more fundamental predictor of black hole mass than stellar mass. Simon~\cite{Simon2023} first explored the VDF as an alternative proxy for the SMBH mass function, comparing predictions from the GSMF, an inferred VDF, and a spectroscopic VDF measured directly from galaxy surveys. The VDF-based approaches predict a higher abundance of massive black holes and correspondingly larger GWB amplitudes than the standard GSMF chain, while also exhibiting less dependence on the {choice of} $M_{\mathrm{BH}}$--host relation.

{However, using the VDF does not always increase the predicted GWB amplitude.} Sato-Polito et al. (2024) {obtain similar amplitudes from their VDF- and GSMF-based populations, with both baseline predictions below PTA measurements} \cite{SatoPolito2024_bigBH}.

\begin{figure*}[t]
    \centering
    \includegraphics[width=\textwidth]{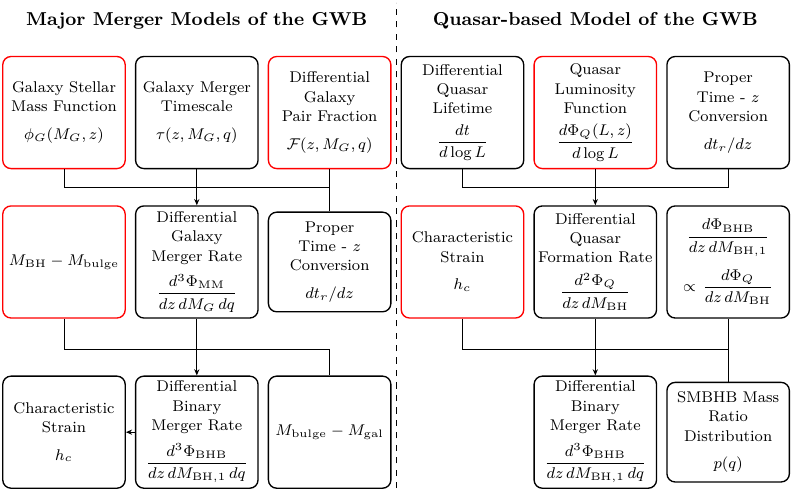}
    \caption{Comparison of model frameworks. {The illustrated} major merger models (left) are forward models: they start from {GSMFs and pair fractions} to predict the GWB amplitude. {The illustrated} quasar-based {model} (right) {is used for} population inference: {it takes} the observed GWB amplitude as input and {uses} the QLF to constrain the underlying SMBHB population. Red boxes indicate observational inputs; black boxes indicate theoretical components. The two approaches are complementary---forward models predict the signal, while QLF-based inference extracts the population physics from the measurement (\eq{eq:strain_spectrum_circular_bhbs}). Figure adapted from \cite{casey-clyde_quasarbased_2022}.}
    \label{fig:quasarVmm}
\end{figure*}

\subsection{Quasar-based SMBHB Models}
An alternative approach exploits the connection between SMBHBs and AGN. Galaxy major mergers are believed to trigger quasar activity by funneling gas to the central black holes \cite{sanders_ultraluminous_1988,volonteri_assembly_2003, hopkins_cosmological_2008}. 

\begin{figure}
    \centering
    \includegraphics[width=\columnwidth]{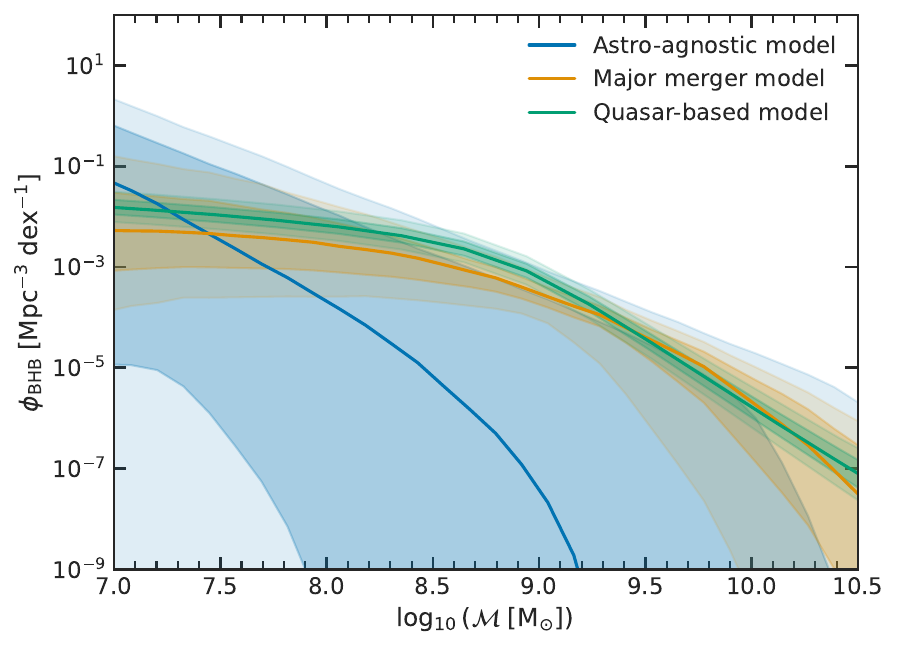}
    \caption{Comparison of three SMBHB merger rate models which all produce a GWB amplitude consistent with NANOGrav 15-yr results ($A_{\mathrm{GWB}} \approx 2.4 \times 10^{-15}$). Despite their different assumptions astro-agnostic (blue), major-merger (yellow), and quasar-based (green) all three can reproduce the observed signal, illustrating the current degeneracy in astrophysical interpretation.}
    \label{fig:bhb_models}
\end{figure}

Quasar-based models \cite{Goulding2019,casey-clyde_quasarbased_2022} use the observed quasar population as a proxy for the SMBHB population. This has the advantage of relying on the QLF, which is well-measured out to high redshifts \cite{hopkins_observational_2007,shen_bolometric_2020}.
In {a separable approximation}, the {quasar-based} binary merger rate density {can be written as}:
\begin{equation}
    \dot{\phi}_{\mathrm{BHB}}(\mathcal{M}, z) \frac{dt_{\mathrm{r}}}{dz} = \phi_{0} p_{\mathcal{M}}(\mathcal{M}) \hat{\phi}_{\mathrm{QSO}}(z) \,
\end{equation}
where $\phi_{0}$ is a normalization constant calibrated to the local universe ($z=0$), $p_{\mathcal{M}}$ is the chirp mass probability distribution, and $\hat{\phi}_{\mathrm{QSO}}(z)$ describes the redshift evolution of the quasar density. Alternatively, one can link the merger rate directly to the quasar triggering rate $\dot{\phi}_{\mathrm{QSO}}$:
\begin{equation}
    \hat{\phi}_{\mathrm{QSO}}(M_{\mathrm{BH}}, z) = \frac{\dot{\phi}_{\mathrm{QSO}}(M_{\mathrm{BH}}, z) \frac{dt_{\mathrm{r}}}{dz}}{\int \dot{\phi}_{\mathrm{QSO}}(M_{\mathrm{BH}}, z = 0) \left.\frac{dt_{\mathrm{r}}}{dz} \right\vert_{z = 0} d \log_{10} M_{\mathrm{BH}}}
\end{equation}
By anchoring the merger rate to the visible population of accreting black holes, these models provide an independent check on the major-merger scenarios derived from stellar mass functions (see \fig{fig:quasarVmm} for a schematic comparison).

\begin{figure}
    \centering
    \includegraphics[width=\columnwidth]{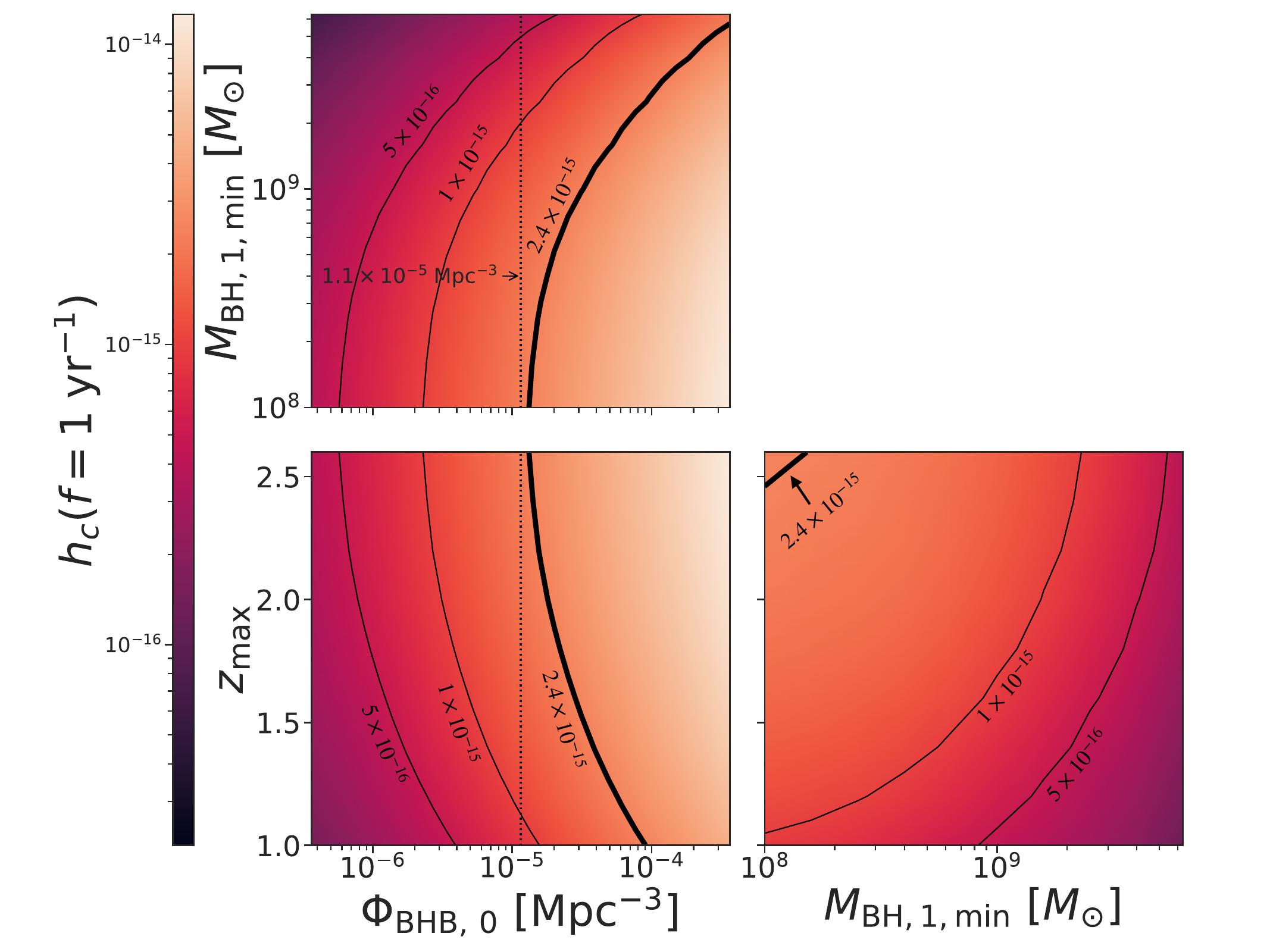}
    \caption{GWB characteristic strain amplitude $h_c(f = 1\;\mathrm{yr}^{-1})$ as a function of three parameters of the quasar-based SMBHB population model: the local binary number density $\Phi_{\mathrm{BHB},0}$, the minimum primary black hole mass $M_{\mathrm{BH},1,\mathrm{min}}$, and the maximum contributing redshift $z_{\mathrm{max}}$. Contours are labeled by the corresponding strain amplitude. The thick contour marks $h_c = 2.4 \times 10^{-15}$, corresponding to the NANOGrav 15-yr GWB measurement. The dotted vertical line indicates the local SMBHB number density of $1.1 \times 10^{-5}\;\mathrm{Mpc}^{-3}$. Figure adapted from \cite{casey-clyde_quasarbased_2022}, updated to the NANOGrav 15-yr amplitude.}
    \label{fig:corner_astro}
\end{figure}

The dependence of the predicted GWB on the underlying population parameters is illustrated in \fig{fig:corner_astro}. The amplitude is most sensitive to the local SMBHB number density $\Phi_{\mathrm{BHB},0}$ and the minimum primary black hole mass $M_{\mathrm{BH},1,\mathrm{min}}$, with a weaker dependence on $z_{\mathrm{max}}$. Reproducing the NANOGrav 15-yr amplitude requires either a high number density of binaries or a population dominated by massive systems. Dual AGN fractions can independently constrain $\Phi_{\mathrm{BHB},0}$~\cite{Goulding2019, Koss2023}, while CW detections of individual binaries probe the high-mass end of the population. Kis-T{\'o}th et al.~\cite{KisToth2025} further test this quasar--merger connection by computing the GWB from merger-triggered quasars using the observed QLF as a constraint on the SMBHB population, finding that the predicted signal is consistent with the NANOGrav amplitude for plausible merger parameters.

\subsection{Astrophysics-agnostic Models} 
Finally, we can adopt an ``astrophysics-agnostic'' approach. Rather than assuming specific scaling relations or galaxy pairing fractions, these models parameterize the merger rate $\dot{\phi}_{\mathrm{BHB}}$ directly using a flexible functional form, such as a modified Schechter function \citep{middleton_astrophysical_2016}:
\begin{equation}
    \dot{\phi}_{\mathrm{BHB}} = \dot{\phi}_{\mathrm{BHB}, 0} \left(\frac{\mathcal{M}}{10^{7} \; \mathrm{M}_{\odot}}\right)^{-\alpha_{\mathcal{M}}} e^{-\mathcal{M} / \mathcal{M}_{*}} (1 + z)^{\beta_{z}} e^{-z / z_{*}} \, .
\end{equation}
This method is useful for placing robust upper limits on the merger rate without being tied to the uncertainties of galaxy evolution physics. However, because the GWB spectrum integrates over the entire population, {analyses of the NANOGrav 12.5-yr data and {EPTA DR2 and IPTA DR2} with this model} primarily constrain the normalization $\dot{\phi}_{\mathrm{BHB}, 0}$, while leaving the shape parameters ($\alpha_{\mathcal{M}}, \beta_{z}$, etc.) largely unconstrained \citep{middleton_massive_2021}{, \citep{Smarra2025,Steinle2023_PTA_LISA}}. Despite their different assumptions, all three model classes can reproduce the observed GWB amplitude, illustrating the current degeneracy in astrophysical interpretation (\fig{fig:bhb_models}).


\section{Gravitational Wave Background Anisotropy}
\label{sec:anisotropy}

While the initial measurement of the GWB relies on the characteristic {HD} correlation curve, this curve represents the response to an \textit{isotropic} background. However, the physical GWB is formed by the superposition of discrete sources distributed throughout the Universe, and perfect isotropy is an approximation that must eventually break down. Mapping the angular distribution of GW power across the sky is therefore a major focus of PTA science.

The search for anisotropy is both important and interesting because it serves as a powerful discriminator of the GWB's origin. A background generated by cosmological processes, such as inflation or phase transitions, is expected to be isotropic to a very high degree. In contrast, an astrophysical background arising from the cosmic population of SMBHBs must inherit the spatial structure of the Universe. This angular structure depends strongly on frequency, creating distinct physical regimes for anisotropy searches.

The search for spatial anisotropy is also deeply intertwined with tests of GR. While the standard analysis assumes the quadrupolar {HD} correlation characteristic of tensor (spin-2) waves, alternative theories of gravity predict the existence of scalar (spin-0) and vector (spin-1) polarization modes. These non-Einsteinian modes generate distinct angular correlation patterns which modify the effective ORF.

From a data analysis perspective, this creates a problem: a deviation from the standard {HD} curve could arise either from a breakdown of isotropy (e.g., a clustered astrophysical background) or from a breakdown of GR (e.g., a scalar background component). Disentangling these effects requires a generalized map-making approach that simultaneously reconstructs the angular power spectrum for all possible polarization modes. We discuss the physical nature of these alternative polarizations and current observational limits in detail in \sect{sec:new_physics}.

In the following, we review the mathematical frameworks developed to detect and characterize angular structure in the GWB (\sect{sec:anisotropy_methods}), then derive the {frequency scaling of} intrinsic shot-noise anisotropy (\sect{sec:scaling_laws}). We examine the limiting case of a single dominant source and its distinct angular signature (\sect{sec:single_source_anisotropy}), and conclude with cross-correlations between PTA sky maps and galaxy catalogs as a probe of {the large-scale structure (LSS) traced by the GWB} (\sect{sec:cross_correlation}).

\medskip
\noindent\textit{A note on notation.}---Throughout this section, $\gamma_{pq}(\hat{\Omega})$ denotes the directional geometric kernel that encodes the antenna-pattern response of a pulsar pair $(p,q)$ to GW power arriving from direction $\hat{\Omega}$. This should not be confused with the scalar spectral index $\gamma$ (e.g., $\gamma=13/3$ for the power spectral density of circular SMBHB-driven residuals) used elsewhere in this review.

\subsection{Search Methods for Anisotropy}
\label{sec:anisotropy_methods}

Several complementary techniques have been developed to detect and characterize GWB anisotropy, including spherical harmonic decompositions, pixel-based mapping, square-root spherical harmonics, and Fisher matrix approaches. These frameworks have been extended to incorporate astrometric deflections~\citep{Hotinli2019} and circular polarization of the anisotropic background~\citep{SatoPolito2021}. We summarize the main approaches below.

\subsubsection{Spherical Harmonics}
Mingarelli et al.~(2013)~\citep{m13} introduced the spherical harmonic decomposition of GW power for PTAs, deriving generalized ORFs for arbitrary angular distributions. Taylor and Gair (2013)~\citep{TaylorGair2013} applied this framework to simulated data, and Taylor et al.~(2015)~\citep{taylor15} performed the first search using real data from EPTA DR1; see \sect{sec:anisotropy_results}.

As derived in \sect{subsec:plane_wave}, the cross-correlation power spectrum (\eq{eq:crosspower}) depends on the angular power distribution $P(\hat{\Omega})$. In the isotropic limit $P(\hat{\Omega}) = 1$, this reduces to the {HD} curve. For anisotropic searches, we define the geometric kernel $\gamma_{pq}(\hat{\Omega})$ (adopting the $(p,q)$ convention standard in the anisotropy literature~\cite{m13,TaylorGair2013}):
\begin{equation}
\label{eq:gamma_pq}
    \gamma_{pq}(\hat{\Omega}) = 2\frac{\left( \hat{p} \cdot \hat{q} - (\hat{p} \cdot \hat{\Omega}) (\hat{q} \cdot \hat{\Omega})\right)^2}{(1+ \hat{p} \cdot \hat{\Omega})(1 + \hat{q} \cdot \hat{\Omega})} - (1 - \hat{p}\cdot \hat{\Omega})(1 - \hat{q}\cdot \hat{\Omega})\,,
\end{equation}
\begingroup

For distinct pulsars in the incoherent limit, the one-sided cross-power spectrum is
\begin{equation}\label{eq:Rpowsp_simp}
    \mathcal{R}_{pq}^{\mathrm{GW}}(f)=\frac{h_c^2(f)}{128\pi^3f^3}\int_{S^2}d\hat{\Omega}\,P(\hat{\Omega})\gamma_{pq}(\hat{\Omega}).
\end{equation}
Here $\mathcal{R}_{pq}^{\mathrm{GW}}$ is the same one-sided GW timing-residual spectrum denoted $S_{pq}$ above; the geometric kernel satisfies $\gamma_{pq}=4\sum_A F_p^A F_q^A$.
\par
\endgroup
Map reconstruction methods essentially invert this integral equation. However, the number of statistically independent angular modes one can constrain is strictly bounded by the number of pulsar pairs, $N_{\mathrm{pair}} = N_{\mathrm{psr}}(N_{\mathrm{psr}}-1)/2$. Pixel-based approaches {(e.g., using Hierarchical Equal Area isoLatitude Pixelization)} often face ill-conditioned covariance matrices scaling as $\mathcal{O}(N^2)$, making basis decomposition methods preferred for current data. A further limitation is that spherical harmonics form an orthogonal basis on the full sky, implicitly assuming a uniform distribution of pulsars. In practice, regional PTAs have pulsars concentrated in their own hemisphere, so the spherical harmonics modes are no longer orthogonal over the sampled sky, leading to mode coupling and leakage of power between multipoles. This motivates the Fisher-matrix and pixel-based approaches discussed below, which naturally account for non-uniform sky coverage.

\subsubsection{Square-Root Spherical Harmonics}
The spherical harmonic basis provides a natural framework for decomposing $P(\hat{\Omega})$. However, a direct decomposition can yield unphysical negative power values. To ensure positive-definiteness ($P(\hat{\Omega}) \ge 0$) by construction, it is advantageous to decompose the \textit{square root} of the power \citep{Taylor:2020zpk,Pol:2022sjn}:
\begin{equation}
P(\hat{\Omega})^{1/2} = \sum_{L=0}^{\infty}\sum_{M=-L}^{L} a_{LM} Y_{LM}(\hat{\Omega}) \,.
\end{equation}
Squaring this expansion to obtain $P(\hat{\Omega})$ introduces non-linear coupling between the $a_{LM}$ coefficients, but the parameterization guarantees $P(\hat{\Omega}) \geq 0$ by construction, providing a robust physical prior for Bayesian searches. However, the underlying basis functions are still spherical harmonics, so the square-root decomposition inherits the same sensitivity to anisotropic pulsar distributions: mode coupling from incomplete sky coverage affects the $a_{LM}$ coefficients just as it does the standard $c_{\ell m}$. This basis was adopted for the NANOGrav 15-yr anisotropy search~\cite{Agazie2023_Anisotropy}; see \sect{sec:anisotropy_results}.

\subsubsection{Pixel-based Methods}
\label{sec:pixel_methods}

An alternative to the spherical harmonic decomposition is to map the GW power directly in a pixel basis~\cite{Cornish2014}. The angular distribution of power, $P(\hat{\Omega})$, is {represented on} $N_{\mathrm{pix}}$ disjoint sky regions {using the pixel basis of Pol et al.~\cite{Pol:2022sjn}}:
\begin{equation}
    P(\hat{\Omega}) = \sum_{{\hat{\Omega}'}} P_{{\hat{\Omega}'}} {\delta^2(\hat{\Omega},\hat{\Omega}')} \,,
\end{equation}
where $P_{{\hat{\Omega}'}}$ is the power in the {pixel centered on $\hat{\Omega}'$} and the sum runs over disjoint sky regions {defined by Hierarchical Equal Area isoLatitude Pixelization}. {Here $\delta^2(\hat{\Omega},\hat{\Omega}')$ selects that pixel in the discrete representation.} The cross-power spectral density (\eq{eq:Rpowsp_simp}) then becomes a sum over pixels, weighted by the {pixel response of the} geometric kernel $\gamma_{pq}(\hat{\Omega})$.

The primary advantage of pixel-based methods is their flexibility; they are agnostic to the global shape of the background and are particularly effective at identifying localized ``hotspots'' or anisotropic features that would require high-$\ell$ modes to resolve in a spherical harmonic basis. Furthermore, {non-negative power can be enforced in} the pixel basis {by imposing} $P_{{\hat{\Omega}'}} \geq 0$. While technically bounded by the number of unique baselines, the number of independent pixels one can robustly constrain is in practice limited by the intrinsically broad antenna pattern of the PTA: the geometric kernel $\gamma_{pq}(\hat{\Omega})$ varies slowly across the sky, making adjacent pixels highly degenerate and limiting the number of independent pixels the array can resolve. This is also referred to as the pixels being correlated. For current PTAs, this limits robust reconstructions to low spatial resolutions (see the EPTA {sensitivity forecasts in} \cite{ani2021}), although this will scale dramatically with the advent of next-generation facilities like {{the Deep Synoptic Array (DSA, formerly DSA-2000)}}~\cite{Hallinan2019_DSA2000} and {the Square Kilometre Array Observatory (SKAO)}~\cite{Shannon2025_SKAPTA}.

Full pixel-based reconstruction inverts the complete pixel-pixel response matrix via Bayesian inference or maximum-likelihood methods, jointly fitting all pixels and thereby deconvolving the array's beam pattern to produce a ``clean map.'' The PTA baseline configuration has blind spots, so the response matrix is typically ill-conditioned, requiring regularization (e.g., {truncation of the singular-value decomposition}) to suppress noise amplification in insensitive modes. Grunthal et al.~\cite{Grunthal2026_skymaps} show that accounting for spatially variable resolution in this regularization can improve the significance of localized features by up to a factor of two.

\subsubsection{The Radiometer}
\label{sec:radiometer}

The \textit{radiometer}, originally developed for ground-based GW detectors by Ballmer~\cite{Ballmer2006} and Mitra et al.~\cite{Mitra2008} and now widely adopted in PTA analyses (see \cite{Konstandin2026} for a systematic comparison of PTA search bases), takes the opposite limit to full pixel reconstruction: it inverts only the diagonal of the response matrix, estimating each pixel independently. For a target direction $\hat{\Omega}_0$, the radiometer estimator is
\begin{equation}
\label{eq:radiometer}
    \hat{P}(\hat{\Omega}_0) = \frac{\sum_{p<q} w_{pq} \, \gamma_{pq}(\hat{\Omega}_0) \, \hat{\rho}_{pq}}{\sum_{p<q} w_{pq} \, \gamma_{pq}^2(\hat{\Omega}_0)} \,,
\end{equation}
where $\hat{\rho}_{pq}$ is the measured cross-correlation for pair $(p,q)$ and $w_{pq}$ are inverse-variance weights. Scanning $\hat{\Omega}_0$ across the sky produces a ``dirty map'' of GW power, analogous to the dirty image in radio interferometry. By ignoring inter-pixel correlations, the radiometer is computationally inexpensive and numerically stable even for arrays with sparse coverage, making it well suited for rapid identification of localized hotspots. The trade-off is that the broad antenna patterns of the pulsar pairs smear localized sources across the sky---frequently producing antipodal sidelobes on the opposite side of the sky, a characteristic artefact of the PTA beam pattern---and the single-direction estimates are correlated through shared pulsars. This approach has been used in recent anisotropy searches by the MPTA~\cite{Grunthal2025} and PPTA~\cite{Chen2026}; see \sect{sec:anisotropy_results}.

\subsubsection{Fisher Matrix Formalism}
\label{sec:fisher_matrix}

Ali-Ha\"imoud, Smith \& Mingarelli~\cite{ani2020, ani2021} developed a Fisher matrix formalism for anisotropic GWB searches that addresses the limitations of the spherical harmonic and pixel bases described above. In the weak-signal limit, the Fisher information matrix for the band-integrated GWB intensity map is:
\begin{equation}
\label{eq:fisher_aniso}
\mathbf{F}_f(\hat{\Omega}, \hat{\Omega}') = \frac{1}{(4\pi f)^4} \sum_{p\neq q} \mathcal{T}_p(f) \mathcal{T}_q(f) \frac{2T_{pq}\Delta f}{\sigma_{p,f}^2\sigma_{q,f}^2} \gamma_{p q}(\hat{\Omega})\gamma_{p q}(\hat{\Omega}')\,,
\end{equation}
where $\mathcal{T}_p(f)$ is the transmission function encoding information loss from the timing model fit, $T_{pq}$ is the effective observation time for pair $(p,q)$, and $\sigma_{p,f}^{{2}}$ is the {band-integrated} noise {variance} of pulsar $p$. Unlike the spherical harmonic basis, which assumes uniform sky coverage, $\mathbf{F}$ is constructed directly from the discrete pulsar positions and noise properties, so it naturally accounts for the anisotropic distribution of pulsars in any real PTA.

The eigenvectors of $\mathbf{F}$ (eigenmaps) form an orthogonal, PTA-specific basis for the observable sky. At most $N_{\mathrm{pair}}$ eigenmaps have non-zero eigenvalues, setting a hard limit on the number of independent angular modes the array can constrain. Reconstructing the GWB in this basis {recovers its observable projection without imposing a spherical-harmonic cutoff on the true sky. Neglecting power beyond such a cutoff can bias the reconstructed large-scale sky} ({Semenzato et al.~\cite{semenzato2025}; see} \sect{sec:anisotropy_challenges}).

A key result of this formalism is that the principal maps are, in general, statistically correlated with the GWB monopole~\cite{ani2021}. For any finite, anisotropically distributed set of pulsars, searching for anisotropy simultaneously with the monopole degrades the sensitivity to both. This arises because the pairwise timing response functions $\gamma_{pq}(\hat{\Omega})$ are not orthogonal to an isotropic background, so the monopole amplitude is coupled to all angular modes. The severity of this coupling depends on the specific pulsar configuration: Ali-Ha\"imoud et al.~\cite{ani2021} showed that for the EPTA, the minimum detectable monopole amplitude increases by up to $\sim 30\%$ when simultaneously fitting for a hotspot, depending on its sky location.

A further insight concerns the {GWB contribution to} pulsar autocorrelations. The single-pulsar timing response $\gamma_{\hat{p}\hat{p}}(\hat{\Omega})$ is sensitive not only to the GWB monopole but also to a dipole projected along $\hat{p}$ and a quadrupole projected twice onto $\hat{p}$ (eq.~38 of \cite{ani2021}). Each pulsar therefore picks up a different combination of these low-order multipoles in its red noise spectrum. When using autocorrelations to set upper limits on the GWB amplitude, one must account for these anisotropic contributions; neglecting them biases the inferred monopole~\cite{ani2021}.

\subsubsection{{Timing-Residual Maps and Bipolar Spherical Harmonics}}
The methods above reconstruct the angular power distribution $P(\hat{\Omega})$ that enters the cross-power spectral density (\eq{eq:Rpowsp_simp}). A complementary approach asks a more fundamental question: is the GWB \textit{statistically} isotropic? Rather than fitting a map, one tests whether the covariance structure of the timing residuals is consistent with isotropy.

In the limit of a large, dense array, the timing residuals can be treated as a scalar field $z(\hat{\Omega})$ on the celestial sphere~\cite{Roebber2017}, which is the convolution of the true GW sky with the PTA response function. Decomposing this field into spherical harmonics,
\begin{equation}
  z(\hat{\Omega}) = \sum_{\ell=2}^\infty \sum_{m=-\ell}^{\ell} z_{\ell m} Y_{\ell m}(\hat{\Omega})\,,
\label{eq:expansion}
\end{equation}
an isotropic background produces a diagonal covariance: $\langle z_{\ell m} z_{\ell'm'}^* \rangle = C_{\ell}^z \delta_{\ell \ell'} \delta_{mm'}$. Anisotropy breaks this diagonality by coupling different $(\ell, m)$ modes, just as {gravitational lensing couples cosmic microwave background multipoles}. The Bipolar Spherical Harmonic (BiPOSH) formalism~\cite{Kumar2024} captures these off-diagonal correlations:
\begin{equation}
\langle z_{\ell m} z_{\ell'm'}^* \rangle = C_{\ell}^z \delta_{\ell \ell'} \delta_{mm'} + \sum_{L=1}^{\infty} \sum_{M=-L}^{L} (-1)^{m'} \langle \ell \, m \, \ell' \, -m' \mid L \, M \rangle A_{\ell \ell'}^{LM}\,,
\label{eq:bipoSHexp_review}
\end{equation}
where $\langle \ell \, m \, \ell' \, -m' \mid L \, M \rangle$ are the Clebsch-Gordan coefficients. The BiPOSH coefficients $A_{\ell \ell'}^{LM}$ vanish for an isotropic background and encode the angular scale and orientation of any anisotropy: $L=1$ captures dipolar structure (e.g., from a nearby dominant source or our motion relative to the GWB rest frame), while $L=2$ captures quadrupolar patterns arising from the large-scale distribution of galaxies. {The relative strength of shot-noise anisotropy generally increases toward higher GW frequencies as the GW power becomes concentrated in fewer binaries} (\sect{sec:scaling_laws}){, but its magnitude depends on the SMBHB population. Casey-Clyde et al. (2022)~\cite{CaseyClyde2022} showed that replacing the McConnell \& Ma (2013)~\cite{McConnellMa2013} scaling relation with Kormendy \& Ho (2013) yielded more massive binaries but fewer nearby systems in the PTA band because of their shorter residence times.} {Building on the calculation in \cite{m13},} Sato-Polito \& Kamionkowski~\citep{SatoPolito2024} {compared SMBHB populations with the same mean isotropic strain spectrum but different mass and redshift distributions, quantifying their different angular power spectra $C_\ell(f)$. They also modeled how stellar scattering and gas-driven hardening can suppress the low-frequency strain spectrum while increasing the relative anisotropy}.

The BiPOSH coefficients connect directly to the angular power distribution $P(\hat{\Omega})$ of \eq{eq:Rpowsp_simp}. Parameterizing the anisotropy as
\begin{equation}
  P(\hat{\Omega}) = 1 + \sum_{L>0} \sum_{M=-L}^{L} g_{LM} Y_{LM}(\hat{\Omega}) \,,
\end{equation}
one can construct a minimum-variance estimator for the physical anisotropy coefficients from the measured BiPOSH spectrum:
\begin{equation}
  \widehat{g_{LM}} = \frac{\sum_{\ell \ell'} (\widehat{g_{LM}})_{\ell \ell'} \, (\Delta g_{LM})_{\ell \ell'}^{-2}}{\sum_{\ell \ell'} (\Delta g_{LM})_{\ell \ell'}^{-2}}\,,
\end{equation}
where the sum runs over all accessible multipole pairs. This provides a model-independent statistic for testing whether the nanoHertz GW sky exhibits clustering or localized excess power.

\subsection{Scaling Laws and the Shot Noise Regime}
\label{sec:scaling_laws}

Having established the methods used to search for anisotropy, we now turn to the signal itself: what level of anisotropy does nature produce? The level of anisotropy can be quantified by the ratio $\sigma_{\mathrm{gw}}/\mu_{\mathrm{gw}}$, where $\mu_{\mathrm{gw}}$ is the expected GW energy density from all sources within a solid angle $d\Omega$, and $\sigma_{\mathrm{gw}}$ is the standard deviation of this energy density due to Poisson fluctuations in the number and distance of contributing sources. This was first calculated in \cite{m13} by considering the statistics of sources within a resolution element $d\Omega$. 

The original derivation (eq.~29 in \cite{m13}) approximated the solid angle of a multipole $\ell$ as $d\Omega \approx 4\pi/2\ell$. However, in standard spherical harmonic decompositions, the solid angle of a resolution element scales more steeply as $d\Omega \approx 4\pi/\ell^2$. Updating the derivation with this standard geometric mapping modifies the scaling from a square-root to a linear dependence on $\ell$. We thus present the updated scaling relation for the anisotropy level:

\begin{equation}
    \label{eq:mingarelli_29_updated}
    \frac{\sigma_{\mathrm{gw}}}{\mu_{\mathrm{gw}}} \approx 0.2 \left(\frac{f}{10^{-7} \mathrm{Hz}}\right)^{11/6} \left(\frac{5 \mathrm{yr}}{T_{\mathrm{obs}}}\right)^{-1/2} \left(\frac{\ell}{2}\right) \alpha^{1/2} \,,
\end{equation}
where $f$ is the GW frequency, $T_{\mathrm{obs}}$ is the total observation time, and $\alpha$ is a factor of order unity related to the redshift distribution of sources.
This result is fully consistent with recent analytical modeling by Sato-Polito \& Kamionkowski (2024) \cite{SatoPolito2024}, who derived the scaling of the relative angular power spectrum in the shot-noise limit. {In the low-frequency limit of their GW-driven model, they} found the power ratio scales as $C_{\ell}/C_0 \propto f^{11/3}$, which matches the square of our amplitude scaling in \eq{eq:mingarelli_29_updated} (since $(\sigma/\mu)^2 \propto f^{11/3}$).
\label{sec:freq_anisotropy}
{The relative importance of shot noise and spatial clustering depends on both GW frequency and the SMBHB population.}

{\emergencystretch=1em
{At low frequencies}, millions of binaries simultaneously contribute ($\Delta N \gg 1$; \eq{eq:delta_N}), suppressing shot-noise fluctuations. {Spatial clustering can also contribute to the anisotropy}: SMBHBs reside in massive galaxies that trace the cosmic web, and cross-correlating GWB maps with galaxy surveys can recover this LSS imprint (\sect{sec:cross_correlation}).\par}

As frequency increases, the bin occupancy $\Delta N$ (\eq{eq:delta_N}; \fig{fig:fig1}) drops steeply{. The same source discreteness responsible for the spectral departures discussed in \sect{sec:discrete_population} can also enhance angular anisotropy, as individual loud binaries contribute a larger fraction of the GW power. This connects stochastic background searches with searches for individual CW sources} (\sect{sec:CW}). 

The frequency at which this transition occurs depends on the SMBHB mass scale: for $\mathcal{M} = 10^{10}\,M_\odot$ binaries and $T_{\mathrm{obs}} = 20$~yr, \eq{eq:delta_N} gives $\Delta N \approx 6$ at $f = 60$~nHz, so only a handful of systems occupy each frequency bin. Recent evidence that SMBHBs may be more massive than previously assumed~\cite{LiepoldMa2024} would push this transition to even lower frequencies.

\subsubsection{Anisotropy from a Single Bright Source}
\label{sec:single_source_anisotropy}

In the extreme limit of the shot-noise regime, the ``background'' may be dominated by a single, deterministic CW source. Recent targeted searches in the NANOGrav 15-yr dataset have modeled this limit analytically~\cite{Agarwal2026}, and we sketch the derivation below.

We can model the characteristic strain squared field, $h_c^2(\hat{\Omega})$, as the sum of a uniform isotropic component and a single point source localized at $\hat{\Omega}_0$:
\begin{equation}
    h_c^2(\hat{\Omega}) = h_{c,\mathrm{iso}}^2 + h_{c,\mathrm{CW}}^2 \, \delta(\hat{\Omega} - \hat{\Omega}_0) \,.
\end{equation}
For all multipoles $\ell > 0$, the isotropic contribution vanishes, and the power spectrum is driven entirely by the point source. This produces a ``white'' angular power spectrum:
\begin{equation}
\label{eq:Cl_point_source}
    C_{\ell} = \frac{(h_{c,\mathrm{CW}}^2)^2}{4\pi} \quad \text{for } \ell > 0 \,.
\end{equation}
However, the ``isotropic'' background $h_{c,\mathrm{iso}}^2$ is itself a superposition of discrete sources, which introduces a stochastic shot noise floor. {Following Appendix A of \cite{Agarwal2026}, we neglect spatial clustering and average over realizations of the stochastic background. The resulting angular} power spectrum is:
\begin{equation}
    C_{\ell}^{\mathrm{tot}} = C_{\ell}^{\mathrm{stoch}} + \frac{(h_{c,\mathrm{CW}}^2)^2}{4\pi} + 2\delta_{\ell 0} \bar{I} h_{c,\mathrm{CW}}^2 \,.
\end{equation}
{Here $\bar{I}$ is the mean intensity of the stochastic background.}
{A point-source CW contributes equal angular power at every multipole (by the spherical harmonic addition theorem; see \cite{Agarwal2026}), producing} a flat $C_\ell$ for $\ell > 0$ (\eq{eq:Cl_point_source}); power concentrated at a single $\ell$ would not be consistent with a CW. The stochastic shot noise $C_{\ell}^{\mathrm{stoch}}$ from the background population is also approximately flat, so distinguishing the two contributions requires the CW signal to stand above this floor.

As the source population thins at high frequencies ($\Delta N \to 1$; \eq{eq:delta_N}), the background transitions to the single-source regime where this flat $C_\ell$ signature becomes the dominant feature. Agarwal et al.~\cite{Agarwal2026} exploit this in the first systematic targeted search for CW sources in SMBHB candidates, using the flat $C_{\ell>0}/C_0$ ratio as part of the CW detection protocol (\sect{sec:CW}) to assess whether individual binaries are detectable in the angular power spectrum.

\subsection{GWB Polarization from Anisotropies}
\label{sec:gwb_polarization}
The search for anisotropy is incomplete without considering the polarization of the signal. While {stochastic GWBs are often modeled as} unpolarized, {a few dominant} SMBHBs {could induce appreciable} local polarization. {Random binary orientations give zero mean polarization over realizations, but their contributions need not cancel in the one realization of the universe which we inhabit~\cite{Conneely2019}.}

{This polarization carries information beyond the intensity alone. Intensity and circular polarization contribute differently to pulsar timing correlations, allowing their separation when their angular patterns are known}~\cite{KatoSoda2016}. {To describe the spatial structure of the signal, angular spectra of the Stokes parameters can retain information about the source distribution that the usual amplitude spectra miss}~\cite{Conneely2019}. {In particular, uncorrelated source phases can produce flat amplitude spectra even when the sources have a structured distribution on the sky. Measuring intensity and polarization together can therefore help reveal whether many binaries contribute comparably or a few bright systems dominate. An unbiased estimator for anisotropic circular polarization, validated with simulated PTA data, provides a way to investigate this distinction}~\cite{SatoPolito2021}.

{Mapping intensity and polarization can reveal where the GW power is concentrated and how those concentrations are polarized. When a few bright binaries dominate, retaining this information could make individual sources stand out more clearly than in an analysis optimized for an isotropic GWB. An idealized single-source calculation using a quadratic estimator for polarized sky maps demonstrates this potential gain in detection significance relative to a {HD} analysis}~\cite{Jow2026}. {Such maps could therefore help identify candidate individual binaries within the GWB} (\sect{sec:CW}).

\subsection{Cross-Correlation Analysis}
\label{sec:cross_correlation}

As the GWB is generated by the incoherent superposition of signals from SMBHBs, it naturally inherits the clustering properties of their host galaxies. {The} GWB {can therefore act} as a biased tracer of the underlying dark matter distribution {\cite{Semenzato2024}}.

This connection enables a powerful probe of cosmic structure: cross-correlating GWB anisotropy maps {with electromagnetic (EM) galaxy surveys}. {In simulations,} this technique {can reveal the LSS imprint in the unresolved SMBHB} background {after loud sources are removed}~\cite{Semenzato2024}. While the auto-correlation of the GWB is often dominated by the ``shot noise'' of a few loud, nearby sources, the cross-correlation signal is driven by the bulk population of binaries that trace the LSS.

\begin{figure*}[t!] 
\centering
      \includegraphics[width=0.49\columnwidth]{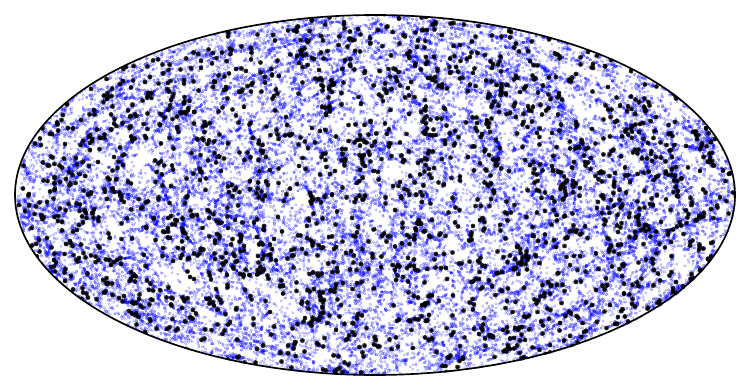}   
      \includegraphics[width=0.49\columnwidth]{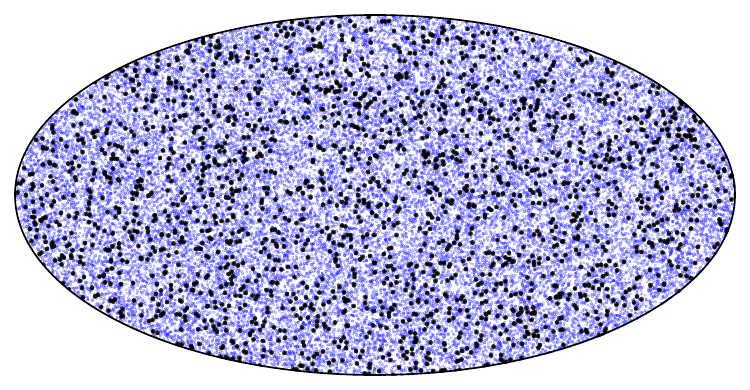}
\caption{\textbf{The GWB as a Tracer of LSS.} An illustration of how a GWB map informed by LSS (left) differs from one where source positions are isotropized (right). While only 1-4\% of galaxies host active SMBHBs, the resulting GWB anisotropy carries the imprint of the cosmic web. Detecting this structure requires cross-correlation with galaxy catalogs to overcome the shot noise of individual loud sources \cite{casey-clyde_quasarbased_2022, Semenzato2024}.}
  \label{fig:cross_maps}
\end{figure*}

Ref.~\cite{Semenzato2024} showed that by effectively filtering out these loud sources (e.g., by imposing a strain cut of $h_{\mathrm{res}} \sim 10^{-16}$), the underlying LSS imprint becomes retrievable.

Simulations indicate that with next-generation PTA sensitivity, cross-correlations could detect the LSS imprint at $>3\sigma$ significance for angular resolutions of $\ell_{\max} \ge 42$, providing further evidence for the astrophysical origin of the GWB (\fig{fig:cross_maps}).

\subsection{Challenges in Detecting Anisotropy}
\label{sec:anisotropy_challenges}

Despite the promise of anisotropy searches, significant observational and systematic hurdles remain. {The anisotropy signal-to-noise ratio (S/N) depends} on how uniformly the array samples the sky, since each pulsar-pair baseline couples to the angular modes of $P(\hat{\Omega})$ through a geometric response that varies across the sky~\cite{Lemke2025, Konstandin2026}. A further fundamental limitation is cosmic variance: even a statistically isotropic GWB will exhibit realization-dependent deviations from {HD} correlations{, which must be accounted for when assessing the significance of} anisotropy~\cite{Konstandin:2024fyo}. The most immediate practical limitation is the anisotropic distribution of pulsars. Current arrays like NANOGrav are heavily biased toward the Northern hemisphere and the Galactic plane, creating a window function that distorts the reconstructed sky maps. The IPTA will improve sky coverage, but until it does, analysis methods must account for this non-uniform sampling to avoid introducing spurious dipolar or quadrupolar power.

A fundamental limitation in reconstructing the GWB anisotropy arises from the finite number of pulsar pairs available to sample the continuous GW sky. While \eq{eq:Rpowsp_simp} describes the cross-power $\mathcal{R}_{pq}^{\mathrm{GW}}$ as an integral over the entire celestial sphere, current arrays can only effectively constrain a finite set of angular modes. This mismatch between the infinite degrees of freedom in the physical sky and the finite rank of the array's response {can lead} to {a recently identified systematic bias, termed ``small-scale leakage'' by Semenzato et al.~\cite{semenzato2025}}{, when small-scale power is left unmodeled}.

We can see how this problem emerges directly from the map-making formalism. By decomposing the angular power distribution $P(\hat{\Omega})$ into spherical harmonics $Y_{\ell m}(\hat{\Omega})$, the integral in \eq{eq:Rpowsp_simp} transforms into a linear system. For a specific pulsar pair indexed by $k$ (corresponding to pair $p,q$), the timing residual cross-power is:
\begin{equation}
    \mathcal{R}_k = \sum_{\ell=0}^{\infty} \sum_{m=-\ell}^{\ell} a_{\ell m} \, \Gamma_{k, \ell m} \,,
\end{equation}
where $a_{\ell m}$ are the harmonic coefficients of the GWB sky and $\Gamma_{k, \ell m}$ are the geometric projection factors derived from the ORF:
\begin{equation}
    \Gamma_{k, \ell m} \propto \int_{S^2} d\hat{\Omega} \, Y_{\ell m}(\hat{\Omega}) \, \gamma_{p_k q_k}(\hat{\Omega}) \,.
\end{equation}
This allows us to write the entire array's response in compact matrix notation. Let $\mathbf{R}$ be the vector of measured cross-correlations for all pairs, and $\mathbf{a}$ be the vector of sky coefficients. The measurement equation is simply $\mathbf{R} = \Gamma \mathbf{a}$.

Crucially, the vector $\mathbf{a}$ contains coefficients up to $\ell \to \infty$. However, we probe a finite amount of pulsar pairs, therefore we have the ability of constraining at most~$N_\mathrm{pairs}$ real spherical harmonic coefficients. We must therefore split the sky {at a chosen reconstruction cutoff $\lmaxrec$} into the {modeled} {large-scale modes (subscript $\mathrm{LS}$) and the unmodeled small-scale modes (subscript $\mathrm{SS}$)}:
\begin{equation}
    \mathbf{R} = \Gamma_{\mathrm{LS}} \mathbf{a}_{\mathrm{LS}} + \Gamma_{\mathrm{SS}} \mathbf{a}_{\mathrm{SS}} \,.
    \label{eq:leakage_split}
\end{equation}
Standard analysis pipelines operate under the assumption that the sky is effectively band-limited to $\ell_{\max}^{\mathrm{rec}}$, modeling the data as $\mathbf{R} \approx \Gamma_{\mathrm{LS}} \mathbf{a}_{\mathrm{LS}}$. Solving this inverse problem (e.g., via generalized least squares) yields an estimator $\mathbf{\hat{a}}_{\mathrm{LS}}$:
\begin{equation}
    \mathbf{\hat{a}}_{\mathrm{LS}} = (\Gamma_{\mathrm{LS}}^T \mathbf{C}^{-1} \Gamma_{\mathrm{LS}})^{-1} \Gamma_{\mathrm{LS}}^T \mathbf{C}^{-1} \mathbf{R} \,,
\end{equation}
where $\mathbf{C}$ is the covariance matrix of the {measured cross-correlations}. 
However, the physical data $\mathbf{R}$ contains the $\Gamma_{\mathrm{SS}} \mathbf{a}_{\mathrm{SS}}$ term defined in \eq{eq:leakage_split}. Substituting the full expression for $\mathbf{R}$ into the estimator reveals the bias:
\begin{equation}
    \mathbf{\hat{a}}_{\mathrm{LS}} = \mathbf{a}_{\mathrm{LS}} + \underbrace{(\Gamma_{\mathrm{LS}}^T \mathbf{C}^{-1} \Gamma_{\mathrm{LS}})^{-1} \Gamma_{\mathrm{LS}}^T \mathbf{C}^{-1} \Gamma_{\mathrm{SS}} \mathbf{a}_{\mathrm{SS}}}_{\text{Leakage Bias}} \,.
    \label{eq:leakage_estimator}
\end{equation}
The recovered coefficients are thus the sum of the true large-scale signal and a deterministic projection of the small-scale sky. The array's window function essentially aliases high-$\ell$ power into the low-$\ell$ map.

This bias propagates when quantifying the angular power spectrum $C_\ell$. Since the small-scale coefficients $\mathbf{a}_{\mathrm{SS}}$ are statistically independent of $\mathbf{a}_{\mathrm{LS}}$, the the measured power spectrum is inflated. The expectation value of the reconstructed power $\hat{C}_\ell$ becomes:
\begin{equation}
    \langle \hat{C}_\ell^{\mathrm{rec}} \rangle = C_\ell^{\mathrm{true}} + C_\ell^{\mathrm{leakage}} = C_\ell^{\mathrm{true}} + \sum_{\ell' > \ell_{\max}^{\mathrm{rec}}} M_{\ell \ell'} C_{\ell'}^{\mathrm{true}} \,.
    \label{eq:leakage_spectrum}
\end{equation}
Here, $M_{\ell \ell'}$ is a mode-mixing matrix determined entirely by the array geometry. It quantifies how much power from a physical mode $\ell'$ leaks into the reconstructed mode $\ell$.

\begin{figure}
    \centerline{
    \includegraphics[width=\textwidth]{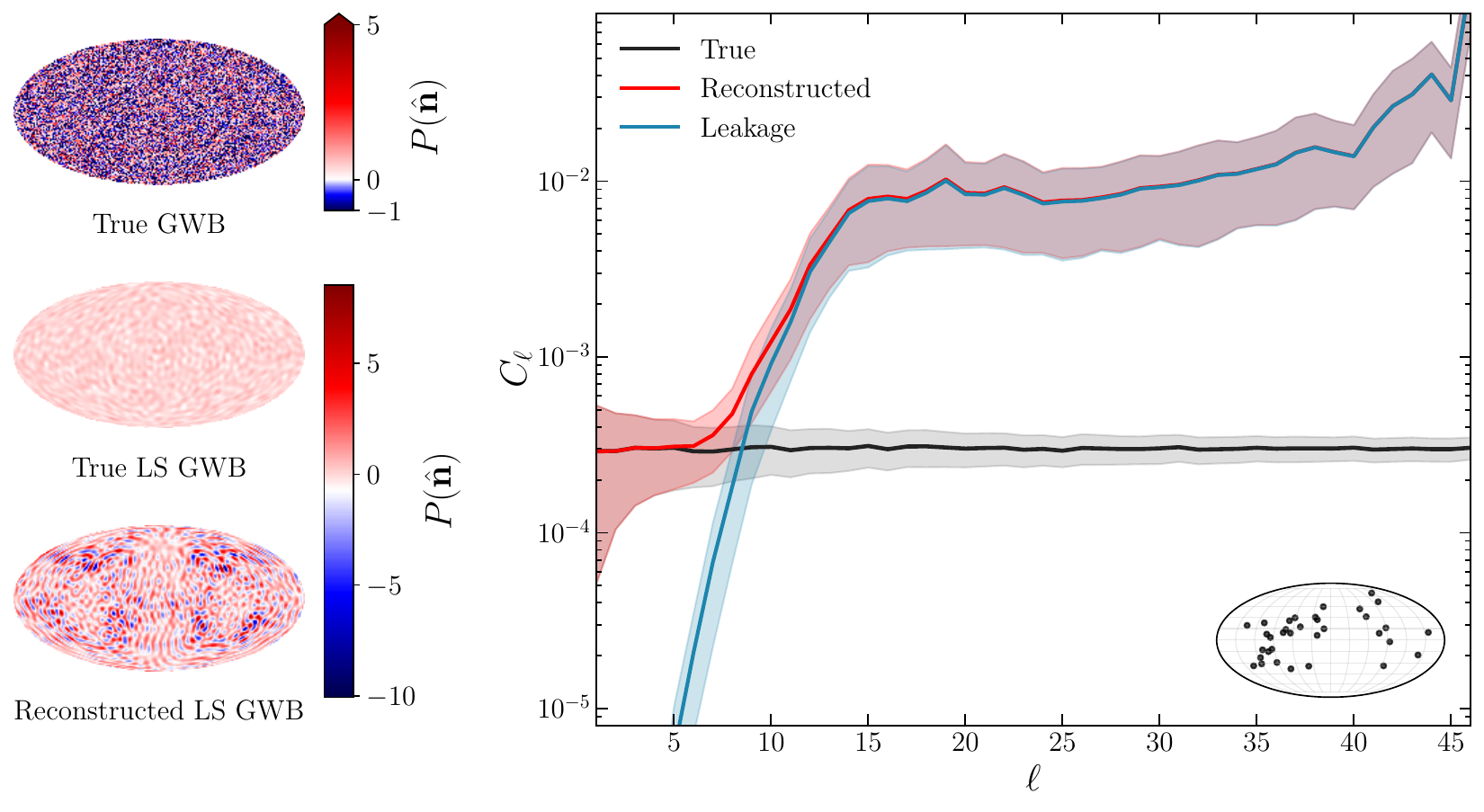}}
    \caption{\textit{Left panels}: individual realization of a true GWB with power up to~$\lmaxGWB=250$ (\textit{top panel}), its large-scale component with power up to~$\lmaxrec={46}$ (\textit{central panel}), and reconstructed large-scale background, including small-scale leakage, up to~$\lmaxrec={46}$ (\textit{bottom panel}). 
    \textit{Right panel}: envelopes of true (\textit{black}), reconstructed (\textit{red}), and leakage (\textit{blue}) correlated residual angular power spectra for~$10^3$ realizations of the GWB and the NANOGrav pulsar geometry.
    Both covariance and likelihood regularization are applied (see~\cite{semenzato2025} for further details).  Solid lines indicate the median value of angular power spectra.}
    \label{fig:SSL_pure_geometry}
\end{figure}

\fig{fig:SSL_pure_geometry} illustrates the impact of small-scale leakage for a realistic PTA configuration (NANOGrav 15-yr pulsar sky). The left panels show a single realization of a GWB with power up to~$\ell_{\max}^{\mathrm{GWB}} = 250$ (top), its true large-scale component up to $\ell_{\max}^{\mathrm{rec}} = {46}$ (middle), and the reconstructed large-scale map including leakage (bottom). The reconstructed map is visibly contaminated by spurious small-scale features not present in the true large-scale sky. The right panel quantifies this effect statistically over $10^3$ realizations. The reconstructed angular power spectrum (red) significantly exceeds the true large-scale spectrum (black) {at higher} multipoles, {where} small-scale leakage (blue) {dominates}.

For astrophysical backgrounds where power is distributed broadly across scales (e.g., a population of discrete binaries tracing LSS), the leakage term $\sum M_{\ell \ell'} C_{\ell'}$ is non-negligible. This term can overestimate the true large-scale anisotropy by an order of magnitude~\cite{semenzato2025}. 

The bias persists regardless of observational details. Whether the array samples the sky isotropically, uniformly, or inhomogeneously (as in NANOGrav), the mode-mixing structure $M_{\ell \ell'}$ remains qualitatively similar. Introducing regularization schemes to stabilize ill-conditioned matrices (such as ridge penalties or {truncation of the singular-value decomposition}) cannot eliminate the leakage. Regularization modifies the pseudoinverse $\Gamma_{\mathrm{LS}}^+$, trading the original leakage for a combination of regularization bias (which pulls the reconstructed large-scale coefficients away from their true values even in the absence of small-scale power) and residual small-scale contamination. Adding realistic timing noise to the covariance matrix does not fundamentally alter the leakage phenomenology, though it modifies the detailed form of $M_{\ell \ell'}$ by changing the effective weights in the least-squares fit.

The most serious consequence of small-scale leakage emerges when attempting to interpret anisotropy measurements in physical terms. The distorted maps and inflated $C_\ell$ values are fundamentally incompatible with the true LSS pattern of the SMBHB population. Cross-correlating PTA maps with galaxy catalogs to establish the astrophysical origin of the GWB (\sect{sec:cross_correlation}) relies on matching the true large-scale clustering signal. If the GW map contains spurious power leaked from small scales, the inferred correlation will be systematically biased. Since the mode-mixing is geometry-dependent, different PTA configurations analyzing the same sky will infer systematically different (and systematically incorrect) large-scale power spectra, creating apparent inconsistencies between datasets that arise not from cosmic variance but from the different leakage patterns of each array.

Perfect reconstruction of LSS without prior knowledge of small-scale content would require either complete absence of small-scale power ($\mathbf{a}_{\mathrm{SS}} = \mathbf{0}$), achieving extremely high resolution such that $\mathcal{R}_{\mathrm{SS}} \ll \mathcal{R}_{\mathrm{LS}}$ for all pairs, or constructing estimators with intrinsic orthogonality between large- and small-scale contributions (for instance, $\Gamma_{\mathrm{LS}}^+ \Gamma_{\mathrm{SS}} = \mathbf{0}$). {The first two} conditions {cannot} be met for realistic PTAs. Addressing small-scale leakage will require explicit modeling of the unresolved power in statistical pipelines through joint inference of $\mathbf{a}_{\mathrm{LS}}$ and parametrized models for the small-scale spectrum, hierarchical Bayesian frameworks that marginalize over small-scale realizations, or orthogonal estimators designed to suppress mode-mixing. As PTAs improve their sensitivity and expand to larger arrays, understanding and mitigating this systematic effect will be essential for reliable anisotropy detection and interpretation.

\subsection{Recent Results and Future Prospects}
\label{sec:anisotropy_results}

To date, dedicated searches for GWB anisotropy have been carried out by EPTA~\cite{taylor15}, NANOGrav~\cite{Agazie2023_Anisotropy}, MPTA~\cite{Grunthal2025}, and PPTA~\cite{Chen2026}.

The NANOGrav 15-yr anisotropy search~\cite{Agazie2023_Anisotropy} placed a 95\% upper limit on the dipole power ($C_1/C_0 \lesssim 27\%$). However, the search was restricted to low frequencies ($f < 10$\,nHz), where millions of sources contribute per frequency bin and the background is expected to be nearly isotropic. Furthermore, constraints on higher multipoles ($\ell > 1$) effectively returned the Bayesian prior, indicating that current low-frequency data is insensitive to small-scale structure.

In contrast, recent results from the MPTA offer a glimpse into the high-frequency regime where anisotropy is physically expected. Using the radiometer (\eq{eq:radiometer}) on only 4.5 years of data but with high cadence, MeerKAT probed frequencies up to $\sim 21$\,nHz. At these frequencies, the source population thins out, and ``popcorn'' shot noise should begin to dominate. 
{No statistically significant anisotropy was found}~\cite{Grunthal2025}.

The PPTA DR3 analysis by Chen et al.~\cite{Chen2026} employed both radiometer and square-root spherical harmonic bases on {30 millisecond pulsars} spanning up to 18 years, performing a per-frequency search in the lowest five frequency bins. No statistically significant anisotropy was found: the angular power spectra in the square-root spherical harmonic basis remain below the $3\sigma$ detection threshold at all multipoles and frequencies. However, a potential hotspot at 5.26\,nHz was identified in both bases ($p = 0.016$ in the radiometer, $p = 0.021$ in the spherical harmonic analysis), though it lies in a region of low array sensitivity and has no corresponding CW signal in PPTA DR3 searches.

These results align with recent theoretical modeling by Gardiner et al.~\cite{Gardiner2024} and Lemke et al.~\cite{Lemke2025}. They demonstrate that while the low-frequency GWB is smooth, the popcorn noise from individual sources becomes the dominant feature at $f \gtrsim 10$\,nHz, consistent with the analytical derivation from \cite{m13}, \eq{eq:mingarelli_29_updated}. Their models predict that anisotropy {can} be {substantial} in this regime, which may help turn high-frequency anisotropy searches into a tool for identifying sub-threshold CW candidates.

Schult et al.~\cite{Schult2025} have introduced a ``frequency-resolved anisotropy'' search designed to exploit this regime. Unlike standard GWB searches that integrate over a broad frequency band assuming a power-law spectrum, this ``spike-pixel'' approach searches for excess power localized in both sky position and frequency bin, allowing pixel-based methods to identify candidate CW sources as hotspots of monochromatic power.

Future analyses may therefore pivot to this transition zone. The intrinsic anisotropy {generally increases} with frequency ({\sect{sec:scaling_laws}}), while extended baselines improve frequency resolution ($\Delta f = 1/T_{\mathrm{obs}}$) and expanding the array increases $\ell_{\mathrm{max}} \sim N$. Per-frequency estimators such as {the anisotropic per-frequency optimal statistic}~\cite{Gersbach2025} are specifically designed for this regime, localizing anisotropy to individual frequency bins while properly accounting for inter-pair covariance and cosmic variance~\cite{Konstandin:2024fyo}. A systematic frequentist forecast by Konstandin et al.~\cite{Konstandin2026} quantifies the prospects and limitations of such searches, finding that{, for narrowband anisotropic signals,} broadband searches lose sensitivity compared to per-frequency approaches. By combining the long baselines of the IPTA with the high cadence and low white noise of the SKAO and the future DSA-2000, we can move beyond upper limits and begin to map the discrete sources that comprise the GW sky.

\section{Continuous Gravitational Waves}
\label{sec:CW}

The search for individual, inspiraling SMBHBs complements the search for GWB anisotropy and the evidence for the monopole (or the isotropic GWB). While the GWB arises from the incoherent superposition of millions of unresolvable sources, CWs are deterministic, monochromatic signals from the loudest individual binaries in the cosmic population. {Detecting and identifying these binaries would allow us to study their properties and evolution and, where {EM} counterparts are available, connect them to their host galaxies.} 

Theoretical work has shown that the isotropy of the GWB is an approximation; individual realizations of the Universe produce a jagged strain spectrum {in which individual frequency bins can be} dominated by the few nearest or most massive sources \citep{Sesana2008PTAoccupancy,  m13, RosadoSesana2015, M17, CaseyClyde2022, Mingarelli2026}. The resulting anisotropy (\sect{sec:anisotropy}) and polarization of the background are intimately connected to the CW source population: a single bright binary contributes both a deterministic CW signal and a localized ``hotspot'' in the GWB intensity and polarization maps (\sect{sec:single_source_anisotropy}). {Searches for these individual sources} typically {follow} two distinct strategies: (1) All-Sky Searches, which scan the entire sky for unknown sources, and (2) Targeted Searches, which leverage {EM} catalogs to test specific binary hypotheses.

\subsection{{EM} Counterparts}
\label{sec:EM_counterparts}

{EM} counterparts to SMBHBs are essential for realizing the multi-messenger potential of PTAs (for recent reviews, see \cite{Bogdanovic2022,DOrazioCharisi2023,Mingarelli2025_KITP,BurkeSpolaor2025}). With PTA CW detections expected to have poor sky localization~\cite{Petrov2024}, an {EM} association is often the only practical route to identifying a unique host galaxy and measuring a redshift. Beyond localization, {EM}  data provide independent constraints on the astrophysical environment (gas-rich versus gas-poor), the accretion state, and the dynamical mechanisms that govern binary hardening (\sect{sec:modeling_gwb}; \eq{eq:binarylife}).

In gas-rich systems, the leading expectation is that a SMBHB excavates a low-density cavity while accreting through streams that feed mini-disks. Time-dependent accretion can imprint periodic variability across {optical, ultraviolet, and X-ray bands} \citep{MacFadyenMilosavljevic2008,DOrazioHaiman2017}. Observationally, time-domain surveys have produced large samples of periodically varying quasars (e.g., \citep{Graham2015,Charisi2016}), though these are subject to high false-positive rates from red noise and quasi-periodic disk variability. Spectroscopy offers a complementary handle via offset broad lines or drifting velocity centroids \citep{Eracleous2012,Shen2010}, while direct imaging with e.g.\ VLBI can resolve dual nuclei at wider ({parsec to kiloparsec}) separations (\fig{fig:binary_lifecycle}). However at those separations, the black holes are not yet emitting significant gravitational radiation~\citep{BurkeSpolaor2011}.

\subsection{Signal Model}
\label{sec:signal_model}

The timing residual induced {by a CW source} is the {time integral of the GW-induced fractional frequency shift}. For a source located at sky position $\hat{\Omega}$, the residual $s(t)$ in a pulsar $p$ is given by:

\begin{equation}
s(t) = \sum_{A=+,\times} F^{A}(\hat{\Omega}) {\int_0^t} \left[ h_{A}({t'}) - h_{A}({t'_p}) \right]{\,dt'},
\end{equation}

where {$t'_p = t' - L(1+\hat{\Omega}\cdot\hat{p})$} is the retarded time at the pulsar (see \sect{subsec:pta_response} for the general Earth--Pulsar term decomposition), and $F^{A}$ are the antenna pattern functions. The two terms in brackets sample the binary at two epochs separated by the light-travel time {$L \sim 10^3$}--$10^4$\,yr; we exploit this in \sect{subsec:time_machine} to measure binary evolution over millennial timescales.

A frequent point of confusion in the literature is the treatment of the polarization angle, $\psi$. In some conventions, $\psi$ appears explicitly in the antenna patterns $F^{A}(\hat{\Omega}, \psi)$; in others, it is absorbed into the time-dependent waveform components $h_{A}(t, \psi)$. These two approaches are mathematically equivalent, representing a choice of basis vectors for the polarization tensor.

\subsubsection{The Polarization Angle as a Basis Rotation}
\label{sec:polarization_angle}

The GW polarization is defined relative to a set of principal axes perpendicular to the propagation direction $\hat{\Omega}$. Let us define a fixed celestial basis $(\hat{u}, \hat{v})$ in the plane of the sky (e.g., aligned with Right Ascension and Declination). The principal axes of the source, denoted $(\hat{m}, \hat{n})$, are rotated relative to this fixed basis by the polarization angle $\psi$:

\begin{equation}
\begin{aligned}
\hat{m} &= \hat{u} \cos\psi + \hat{v} \sin\psi \\
\hat{n} &= -\hat{u} \sin\psi + \hat{v} \cos\psi
\end{aligned}
\end{equation}

The metric perturbation can be written in the source frame $(\hat{m}, \hat{n})$ using the intrinsic amplitudes $\mathcal{A}_+$ and $\mathcal{A}_\times$:
\begin{equation}
h_{ab}(t) = \mathcal{A}_+(t) (\hat{m}_a \hat{m}_b - \hat{n}_a \hat{n}_b) + \mathcal{A}_\times(t) (\hat{m}_a \hat{n}_b + \hat{n}_a \hat{m}_b).
\end{equation}
Substituting the rotation relations for $\hat{m}$ and $\hat{n}$ into this equation projects the waveform onto the fixed sky basis $(\hat{u}, \hat{v})$. This operation mixes the plus and cross components via the double-angle terms $\cos 2\psi$ and $\sin 2\psi$. This leads to two distinct conventions for writing the signal model:

\vspace{1mm}

\noindent\textit{Convention I: $\psi$ in the Waveform}

\noindent In this convention, we fix the antenna patterns to the celestial basis $(\hat{u}, \hat{v})$. The antenna patterns $F^+(\hat{\Omega})$ and $F^\times(\hat{\Omega})$ depend only on the source and pulsar positions. The polarization angle $\psi$ is carried by the waveform terms $h_{+,\times}(t)$.
\begin{equation}
\begin{aligned}
h_+(t) &= {\mathcal{A}_+(t)} \cos 2\psi - {\mathcal{A}_\times(t)} \sin 2\psi \\
h_\times(t) &= {\mathcal{A}_+(t)} \sin 2\psi + {\mathcal{A}_\times(t)} \cos 2\psi
\end{aligned}
\end{equation}

\vspace{1mm}

\noindent\textit{Convention II: $\psi$ in the Antenna Pattern}

\noindent Alternatively, one can define the antenna patterns aligned with the source axes $(\hat{m}, \hat{n})$. In this case, the waveform $h_{+,\times}(t)$ describes the intrinsic strain in the binary frame (independent of $\psi$), and the antenna patterns become functions of $\psi$, related to the Convention~I patterns $F^{+,\times}(\hat{\Omega})$ by:
\begin{equation}
\begin{aligned}
F^+(\hat{\Omega}, \psi) &= \cos 2\psi \, F^+(\hat{\Omega}) + \sin 2\psi \, F^\times(\hat{\Omega}) \\
F^\times(\hat{\Omega}, \psi) &= -\sin 2\psi \, F^+(\hat{\Omega}) + \cos 2\psi \, F^\times(\hat{\Omega})
\end{aligned}
\end{equation}
{Both conventions give the same timing residual.} For clarity, we recommend Convention I: treating $F^A$ as fixed geometric coefficients and including the $\psi$ rotation matrices directly in the signal templates.

\subsection{All-Sky Searches}

\begin{figure}
    \centering
\includegraphics[width=0.7\linewidth]{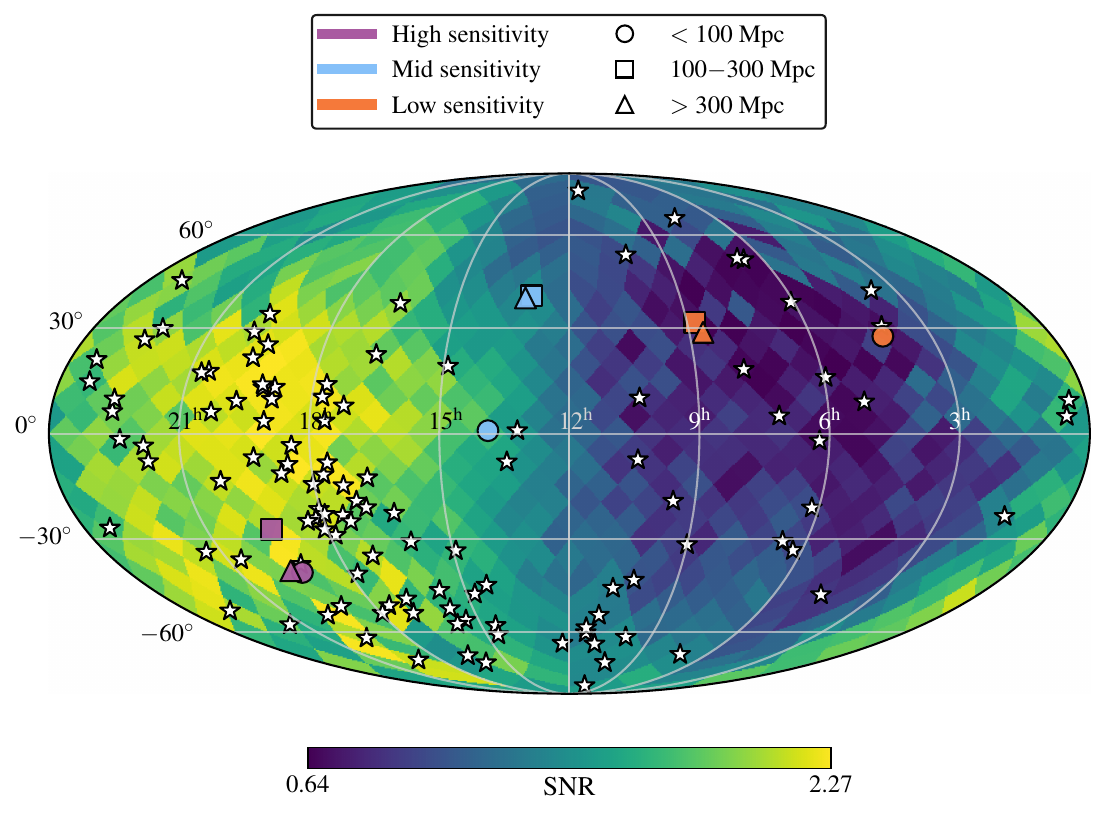}
    \caption{Skymap of the simulated PTA sensitivity, where each pixel is colored by the expected S/N of a CW source at fixed $d_L = 150$ Mpc, averaged over GW frequencies (2, 6, 20 nHz), chirp masses ($10^8$ and $10^9,M_\odot$), and 100 random draws of $\cos\iota$, $\psi$, and $\Phi_0$. Pulsar positions are shown as white stars. We inject into nine galaxies, grouped by sensitivity into high (purple), mid (light blue), and low (orange) regions, demarcated by ellipses of the same color. Within each group, injections span three distance bins: near ($< 100$ Mpc, circles), mid (100-300 Mpc, squares), and far ($> 300$ Mpc, triangles). Figure by P. Petrov}
    \label{fig:all-sky}
\end{figure}

All-sky searches attempt to detect CWs without prior spatial or frequency information (\fig{fig:all-sky}). {These searches} must {account for unknown} inclination, polarization, and phase.

The sensitivity of these searches has improved steadily, but they have yet to yield a detection. The most recent constraints are summarised in tab.~\ref{tab:CW_limits}. The NANOGrav 15-yr~\cite{NG15_CW}, IPTA DR2~\cite{Falxa2023}, and PPTA DR3~\cite{PPTACW2025} searches report sky-averaged 95\% upper limits on the strain amplitude $h_0$ at specific frequencies where each array is most sensitive. The PPTA DR3 is sensitive to binaries with $\mathcal{M} \geq 10^9\,M_\odot$ out to luminosity distances of 50--200~Mpc, depending on sky direction. The EPTA DR2 analysis~\cite{EPTACW2023} {also reports} per-frequency upper limits {using a CURN model. Marginalizing over frequency in the restricted 1--14~nHz follow-up band,} it {obtains} $\log_{10} h_{95} = -13.75$ for a model that includes a {HD}-correlated GWB, {per-pulsar custom noise, and a single CW source}. {In the CURN analysis, the} EPTA array is most sensitive near 20~nHz, where binaries with $\mathcal{M} > 10^9\,M_\odot$ are detectable out to the Virgo and Fornax clusters (16.5 and 19.3~Mpc, respectively).

\begin{table}[t]
\centering
\caption{Current sky-averaged 95\% upper limits on the CW strain amplitude from all-sky searches. For NANOGrav, IPTA, and PPTA the frequency column lists the single frequency of peak sensitivity; for EPTA DR2 it lists the {restricted frequency} band over which the {quoted} limit is {marginalized, using a log-uniform strain prior} (see text).}
\label{tab:CW_limits}
\begin{tabular}{lcccc}
\hline\hline
PTA & Data set & $h_0^{95\%}$ & $f$ (nHz) & Ref. \\
\hline
NANOGrav & 15-yr & $8 \times 10^{-15}$ & 6 & \cite{NG15_CW} \\
IPTA & DR2 & $9.1 \times 10^{-15}$ & 10 & \cite{Falxa2023} \\
PPTA & DR3 & $7 \times 10^{-15}$ & 10 & \cite{PPTACW2025} \\
EPTA & DR2 & $17.8 \times 10^{-15}$ & 1--14 & \cite{EPTACW2023} \\
\hline\hline
\end{tabular}
\end{table}

A critical systematic hurdle for all-sky searches is ``GWB confusion.'' As array sensitivity improves, the threshold for detecting a single source drops into the regime of the stochastic background itself. The EPTA DR2 analysis recently identified a candidate at $\sim 4.6$ nHz with a Bayes factor favoring a CW against uncorrelated noise. However, when a {HD}-correlated GWB component was added to the model, the Bayes factor for the CW dropped to $\sim 1$, indicating that the {data did not favor adding a CW to the GWB model}. A promising path forward is to exploit the deterministic cross-correlation fingerprint of a CW source, which differs qualitatively from the {HD} curve of the isotropic GWB (\sect{subsec:deterministic_fingerprints}).

\subsection{Targeted Searches}
\label{subsec:targeted_searches}

Targeted searches fix {or place informative priors on} the source parameters---sky location ($\hat{\Omega}$), luminosity distance ($D_L$), and often the GW frequency ($f_{\mathrm{GW}}$)---based on {EM} evidence. This creates a highly conditional, falsifiable test: \textit{if} a specific galaxy hosts a binary as described by {EM} models, is it consistent with PTA data?

A recent comprehensive study using the NANOGrav 15-yr dataset applied this method to 114 SMBHB candidates selected {using} {the Catalina Real-time Transient Survey (CRTS)} and radio {observations} \cite{Agarwal2026}. By fixing source parameters, the study achieved a median sensitivity improvement of ${2.2}\times$ (and up to $15\times$) relative to all-sky limits at the same frequency. These constraints are now beginning to cut into the astrophysically plausible parameter space for objects like 3C 66B, ruling out {some} high-mass {circular} binary models that were previously allowed {by} Iguchi et al.~\citep{Iguchi2010}. The analysis highlighted two candidates, J1536+0441 and J0729+4008, whose Bayes factors {were the largest in the sample}. {Follow-up using the} systematic detection protocol described below {found both candidates consistent with noise}.

\subsection{The first steps of a protocol for CW detection}

A {large} Bayes factor {has not been established as a standalone criterion} for claiming a CW detection. {For targeted searches, all available GW and {EM} information should be used to test a self-consistent model of the binary system.} Based on the analysis of candidates in the NANOGrav 15-yr targeted search, Agarwal et al.~\cite{Agarwal2026} define a suite of ten tests divided into three categories---{EM consistency checks}, GWB consistency checks, and statistical validation checks---which we reproduce {and extend} here as a concrete example of the multi-messenger vetting required for CW claims (tab.~2 of \cite{Agarwal2026}). This checklist applies specifically to the case of targeted searches where priors are drawn from periodic light curves in an {EM} catalog, though the framework is likely useful more broadly.

\medskip
\noindent\textit{{EM} consistency checks.}---These tests verify that the {EM} properties of the candidate are consistent with the binary hypothesis:
\begin{enumerate}
    \item \textbf{Extended Periodicity:} The optical or infrared periodicity that originally flagged the candidate {should be checked for persistence} in extended monitoring. A periodicity that fades or changes character in new data {can weaken} the binary hypothesis.

    \item \textbf{Spectral Features:} Broad emission line profiles (e.g., H$\alpha$, H$\beta$) should {be checked for} velocity shifts or profile changes consistent with orbital motion over the baseline of spectroscopic monitoring.
\end{enumerate}

\noindent\textit{GWB consistency checks.}---These tests assess whether the candidate is consistent with the measured GWB and astrophysical population models:
\begin{enumerate}
\setcounter{enumi}{2}
    \item \textbf{CRTS--GWB Consistency:} The number of periodic AGN candidates in surveys such as CRTS that are consistent with being SMBHBs should be compatible with the number predicted by the measured GWB amplitude and SMBHB population models.

    \item \textbf{Population Synthesis:} The GW-inferred total mass and frequency of the candidate should fall within the credible region of simulated SMBHB populations drawn from galaxy merger rates and scaling relations.

    \item \textbf{GWB Anisotropy:} A sufficiently loud CW source should produce a detectable anisotropy in the GWB at its sky location and frequency (\sect{sec:single_source_anisotropy}), and should leave a flat imprint in the GWB {angular power} spectrum {$C_\ell$} across all $\ell\mathrel{{>}}{0}$  at the source frequency.
\end{enumerate}

\noindent\textit{Statistical validation checks.}---These tests probe whether the GW signal itself is robust:
\begin{enumerate}
\setcounter{enumi}{5}
    \item \textbf{Distinct from GWB:} The candidate must remain significant when a {HD}-correlated GWB component is included in the noise model. A signal that is absorbed by the GWB ($\mathcal{B}_{\mathrm{HD}}^{\mathrm{CW+HD}} \approx 1$) may be a fluctuation of the stochastic background rather than a discrete source.

    \item \textbf{Signal Coherence:} A true CW induces a coherent phase evolution across the array. This is tested by scrambling the phases or sky positions of the pulsars \citep{Becsy2025}; a detection is robust only if the unscrambled Bayes factor is a significant outlier relative to the scrambled null distribution.

    \item \textbf{Dropout Analysis:} The support for the signal should be distributed across the array, rather than driven by a single noisy pulsar. A dropout analysis \citep{Aggarwal_2019} {measures} each {pulsar's support for the candidate} signal {and can help identify} noise {artifacts} (e.g., unmodeled red noise or ISM events).

    \item \textbf{Random Targeting:} The candidate should not be consistent with noise after accounting for the number of targets searched. If {randomized targets frequently} produce comparable Bayes factors, the {candidate} is not significant {in this test}.

    \item \textbf{Software Cross-check:} Results must be reproduced with independent analysis pipelines (e.g., \texttt{enterprise} and \texttt{QuickCW}) to guard against software-specific systematics.

    \item {\textbf{Spatial Correlations:} We add evidence for the spatial correlations predicted for an individual binary as a further detection criterion~\cite{Mingarelli2026_fingerprints}. This test was developed after, and motivated by, the targeted-search study~\cite{Agarwal2026}; we discuss it in Section~\ref{subsec:deterministic_fingerprints}.}
\end{enumerate}

The IPTA has established a CW Detection Task Force to develop and formalize such protocols for the global array, and future efforts in this vein will be essential as sensitivity improves and the first credible candidates emerge.

\subsection{Deterministic Fingerprints of CWs}
\label{subsec:deterministic_fingerprints}

{Standard GWB searches use spatial cross-correlations, while coherent CW searches fit a common binary waveform across the array. An individual binary also produces a characteristic spatial correlation pattern that can help distinguish it from an isotropic GWB.}

{This pattern is described by the single-source ORF, $\Upsilon_{ab}$, derived in closed analytic form by Mingarelli et al.~\cite{Mingarelli2026_fingerprints}.} For a single monochromatic source at sky location $\hat{\Omega}_0$, the {cycle-averaged} cross-correlation of the timing residuals between pulsars $a$ and $b$ is, in the Earth-term limit, {proportional to} the function $\Upsilon_{ab}$. It is purely geometric: it depends on the source direction $\hat{\Omega}_0$, the binary inclination $\iota$, polarization angle $\psi$, and the positions of pulsars $a$ and $b$, but not on the source amplitude or frequency. It is the kernel of the {HD} curve: the isotropic stochastic ORF $\Gamma_{ab}$ introduced in \sect{sec:HD} is the sky average of $\Upsilon_{ab}$ over source directions (\eq{eq:hd_as_average}). Note that the integrand $F^+_a F^+_b + F^\times_a F^\times_b$ used to derive the {HD} curve in \sect{sec:HD} is the same expression {up to an overall normalization} after averaging over $\iota$ and $\psi$; the full $\Upsilon_{ab}$ retains this orientation dependence. Unlike the {HD} curve, $\Upsilon_{ab}$ explicitly breaks statistical isotropy.

Working in a computational frame where pulsar $a$ lies on the $+\hat{z}$ axis and pulsar $b$ lies in the $x$--$z$ plane at angular separation $\zeta$, Mingarelli et al.~\cite{Mingarelli2026_fingerprints} derive a closed-form expression for the single-source ORF of a GW propagating {in} direction $(\theta,\phi)$:
\begin{align}
\Upsilon_{ab}(\theta,\phi,\zeta,\iota,\psi)
  &= \frac{1}{8}\bigl(1+6\cos^2\!\iota+\cos^4\!\iota
     +\sin^4\!\iota\,\cos4\psi\bigr)\,
     \mathcal{K}\;\mathcal{N}_+ -\frac{1}{4}\sin^4\!\iota\;\sin4\psi\;
     \mathcal{K}\;\mathcal{N}_\times\,,
\label{eq:upsilon-final-psi}
\end{align}
with
\begin{align}
    \mathcal{K}(\theta,\phi,\zeta) &\equiv \frac{1-\cos\theta}{1+\cos\theta\cos\zeta+\sin\theta\sin\zeta\cos\phi}\,,\label{eq:Kdef}\\
    \mathcal{N}_+(\theta,\phi,\zeta) &\equiv (\cos\phi\cos\theta\sin\zeta - \sin\theta\cos\zeta)^2 - \sin^2\!\phi\;\sin^2\!\zeta\,,\label{eq:Nplusdef}\\
    \mathcal{N}_\times(\theta,\phi,\zeta) &\equiv \sin\zeta\sin\phi\,(\cos\theta\sin\zeta\cos\phi - \sin\theta\cos\zeta)\,.\label{eq:Ncrossdef}
\end{align}
Here $\mathcal{K}$ is a geometric kernel encoding the Earth-term timing response, and $\mathcal{N}_+$, $\mathcal{N}_\times$ capture the $+$- and $\times$-polarization responses respectively~\cite{Mingarelli2026_fingerprints}. Unlike the {HD} curve, which depends only on pulsar separation, \eq{eq:upsilon-final-psi} depends on the specific sky location of the GW source relative to each pulsar pair as well as the binary inclination and polarization angle. Each SMBHB therefore imprints a distinct geometric fingerprint on the array, encoding its sky position and polarization content.

Explicitly,
\begin{equation}
\Gamma_{ab} \propto \int \frac{d\hat{\Omega}}{4\pi}\,\Upsilon_{ab}(\hat{\Omega},\iota,\psi),
\label{eq:hd_as_average}
\end{equation}
so deviations from {HD} in realistic SMBHB populations can be interpreted as incomplete averaging over a finite number of bright fingerprints rather than as a failure of the stochastic framework. {Beyond population studies, the same single-source cross-correlation formalism underpins precision {tests of GR with} individual binaries: Zheng et al.~\cite{Zheng2026} develop {a CW framework} to probe additional polarization states, modified dispersion relations, and parity-violating birefringence from resolved SMBHBs.}

\begin{figure}
    \centering
    \includegraphics[width=0.8\columnwidth]{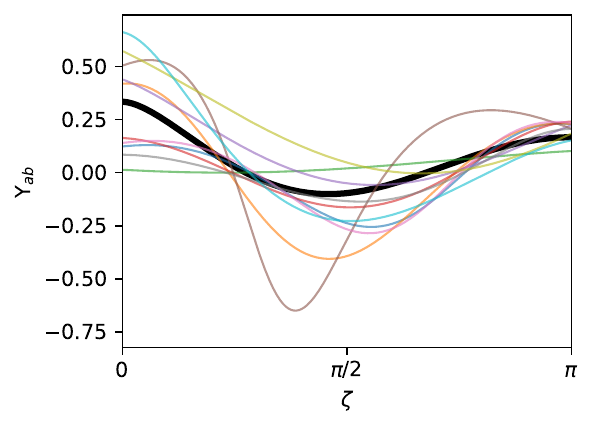}
    \caption{Single-source fingerprints $\Upsilon_{ab}(\zeta)$ {(averaged over $\iota$ and $\psi$)} for 10 randomly drawn GW source sky locations, each shown as a colored, un-normalized curve. The {HD} curve (black) is the sky-average of these fingerprints (\eq{eq:hd_as_average}). Individual fingerprints can differ markedly from {HD}, illustrating that a single bright SMBHB imprints a geometry-dependent correlation pattern on the array that is generically non-{HD}. Figure adapted from Ref.~\cite{Mingarelli2026_fingerprints}.}
    \label{fig:fingerprints}
\end{figure}

It is important to distinguish this pair-level fingerprint from the separation-averaged correlation studied by Cornish and Sesana~\cite{cornishsesana2013}, who demonstrated that even a single bright binary reproduces the {HD} curve once the cross-correlations are binned by angular separation and averaged over many pulsar pairs. For a sufficiently large, isotropically distributed array, this pair-averaging is equivalent to integrating over source directions, so the mean correlation as a function of $\zeta$ converges to Hellings and Downs regardless of whether the signal is a Gaussian background or a single quadrupolar source. The fingerprint $\Upsilon_{ab}$ addresses a complementary question: what is the exact cross-correlation between a specific pair of pulsars for a source at a particular sky location? This pair-level pattern generically departs from the {HD} curve and depends on the full set of angles $(\theta,\phi,\zeta,\iota,\psi)$. The Hellings and Downs limit is recovered only upon averaging $\Upsilon_{ab}$ over source directions or, equivalently, over pulsar positions at fixed separation~\cite{cornishsesana2013,Allen2023}. The scatter and variance around the Hellings and Downs curved observed in population-level simulations~\cite{Becsy2022, cornishromano2015, Schult2025} can therefore be traced to incomplete averaging over a finite number of single-source fingerprints.

The practical consequence is that $\Upsilon_{ab}$ provides a direct test between a single CW and an isotropic background even when the spectral models are identical. A CW can masquerade as a GWB in analyses that only fit for a common red process with a fixed Hellings and Downs ORF. 

Injection-and-recovery studies in simulated PTA data compare four cross-correlated spectral models for a single CW source~\cite{Mingarelli2026_fingerprints}: the full $\Upsilon_{ab}(\theta,\phi,\zeta,\iota,\psi)$ model; {the spike-pixel model}, \sect{sec:pixel_methods}; a {HD}-correlated model, \eq{eq:HD_split}; and a purely auto-correlated (diagonal) model. At template S/N~$=30$, the Bayes factors {for the $\Upsilon_{ab}$ model} over {the {HD}-correlated and diagonal models} are $\mathcal{B}\approx 1611$ {and} $\mathcal{B}\approx 159${, respectively}. The $\Upsilon_{ab}$ model is favored {over the spike-pixel model} at all S/N ($\mathcal{B}^{\Upsilon}_{\mathrm{SP}} = 2.14\pm0.02$ at S/N~$=30$), reflecting the additional constraining power of the inclination and polarization angles (\sect{sec:polarization_angle}). The {HD}-correlated search recovers excess power at the correct GW frequency but absorbs it into an isotropic GWB component---the degeneracy discussed in \sect{sec:CW} and noted by Cornish and Sesana~\cite{cornishsesana2013}. At S/N~$=15$, the $\Upsilon_{ab}$-based models yield $\mathcal{B}^{\mathrm{CW}}>1$ while the {HD} and auto-correlated models do not. Importantly, {at S/N~$=30$,} sky localization improves by a factor of $\sim 11 \times$ when cross-correlations are included.

This is especially relevant for the weak-signal regime that PTAs will inhabit for first evidence of individual binaries, where the total signal power can be dominated by auto-correlations. In that limit, time-domain coherence tests that require precise pulsar distances are difficult: exploiting pulsar terms coherently demands {either} distance knowledge at the level of a GW wavelength {or fitting the pulsar distances jointly with the source parameters~\cite{corbincornish2010}}. In contrast, Earth-term fingerprints require only spatial cross correlations, in analogy with the Hellings and Downs curve. Cross-correlated spectral searches based on $\Upsilon_{ab}$ therefore provide a robust intermediate step between stochastic detection and fully coherent CW matched filtering.

Operationally, one can view $\Upsilon_{ab}$ as the natural statistic for an early detection protocol: (i) detect excess power; (ii) demonstrate spatial coherence against a conservative null (e.g., an auto-correlated CW-like model); (iii) test the CW fingerprint against the isotropic {HD} hypothesis or together with it; and (iv) only then attempt full coherent waveform inference with pulsar terms (\sect{subsec:time_machine}) where warranted by data quality and distance precision (\sect{subsec:distances}).

\subsection{Strong Lensing}
\begin{figure}[b]
    \centering
    \includegraphics[width=\textwidth]{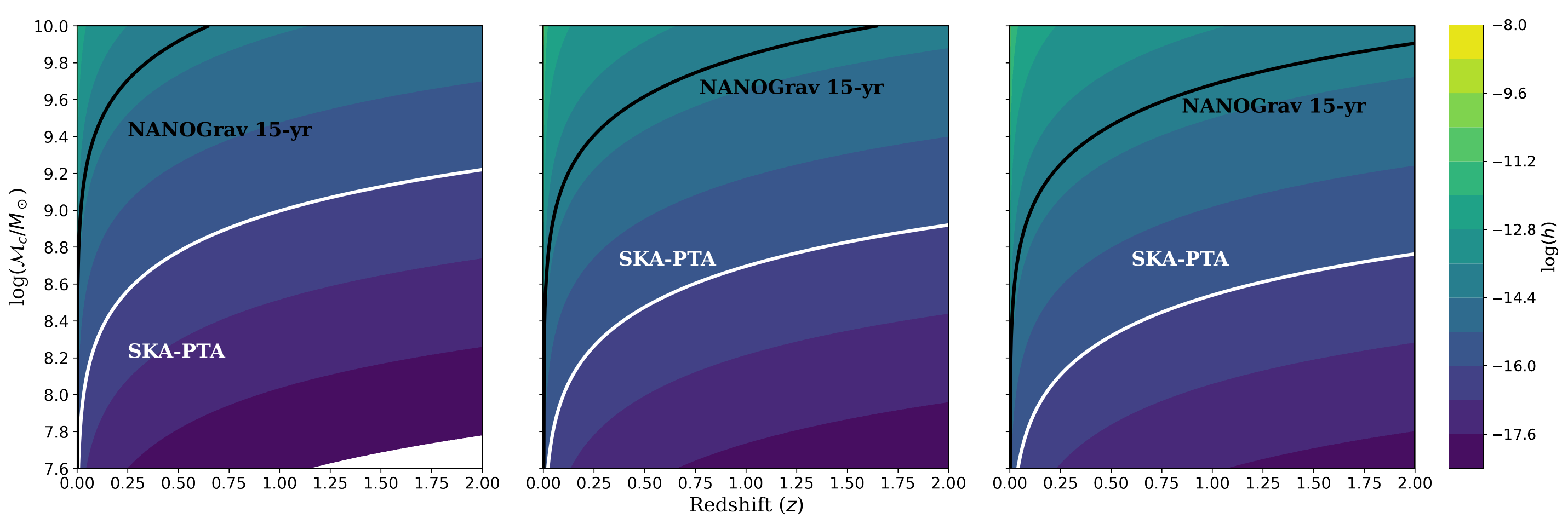}
    \caption{Strong lensing of GWs improves the PTA detection volume by sufficiently amplifying the strains of signals from distant and/or not massive enough binary systems. These GWs can be detected by current and future PTAs, namely NANOGrav (15-yr dataset strain upper limit \cite{NG15_CW}, black) {and a future PTA using the SKAO (SKA-PTA), see} Xin et al.~\cite{Xin_2021} (white). For a fixed $f_{\mathrm{GW}} = 6$ nHz (presently the most sensitive GW frequency \cite{NG15_CW}), we demonstrate the broadening of detectable $z$---$\mathcal{M}$ parameter space under three representative lensing scenarios: (left) a simplified spherical foreground lens model with a modest $\mu=3$, (center) an elliptical lens model with $\mu=30$, and (right) an elliptical lens model with optimistic $\mu=100$.}
    \label{fig:SL_h_param_map}
\end{figure}

Just like {EM} signals, GWs from distant sources can be lensed by foreground mass distributions such as stars, galaxies, and galaxy clusters. {In the geometric optics limit,} {EM} {fluxes increase} by a factor of $\mu$---the magnification factor associated with the lens surface mass distribution---{while} the strains of GW signals from individual sources experience amplification by a factor of $\sqrt{\mu}$ such that $h_{\mathrm{lensed}} = \mu^{1/2} h$, see e.g.  Ezquiaga~\cite{rootmu}. Strong lensing in particular also gives rise to multiple images of the background source. Khusid et al.~\cite{LensedGWs} investigates strong gravitational lensing as a tool for broadening the PTA sensitive volume and extracting richer multimessenger information from SMBHB systems.

Lensing improves PTA detection prospects two-fold, by amplifying otherwise subthreshold CW signals produced by sources that are too far away  or not massive enough, see  \eq{eq:SMBHB_h}. For example, {in the Euclidean approximation,} a constant $\mu=3$ due to strong lensing not only increases the PTA detection volume by a factor of $\sim 5$ for a fixed set of intrinsic SMBHB properties, but it also permits the detection of GWs from systems that are $3^{3/10} \approx 1.4$ times less massive in chirp mass for fixed $D_L(z)$ and $h$. The analysis of the NANOGrav 15-yr dataset \cite{NG15_CW} places a {sky-averaged} $95\%$ upper limit of $8 \times 10^{-15}$ on CW strain amplitude {at $6\,\mathrm{nHz}$}, while {future SKA-PTA datasets} are expected to be sensitive to $h \gtrsim 10^{-16}$ \cite{Xin_2021}. \fig{fig:SL_h_param_map} gives three different cases of magnification---$\mu=3, \, 30, \, \mathrm{and} \, 100$, for lensing configurations ranging from modest to exceptional, respectively---to demonstrate how the amplification of CW strain from strong lensing effectively improves PTA sensitivity. 

\begin{figure}[b]
    \centering
    \includegraphics[width=0.8\textwidth]{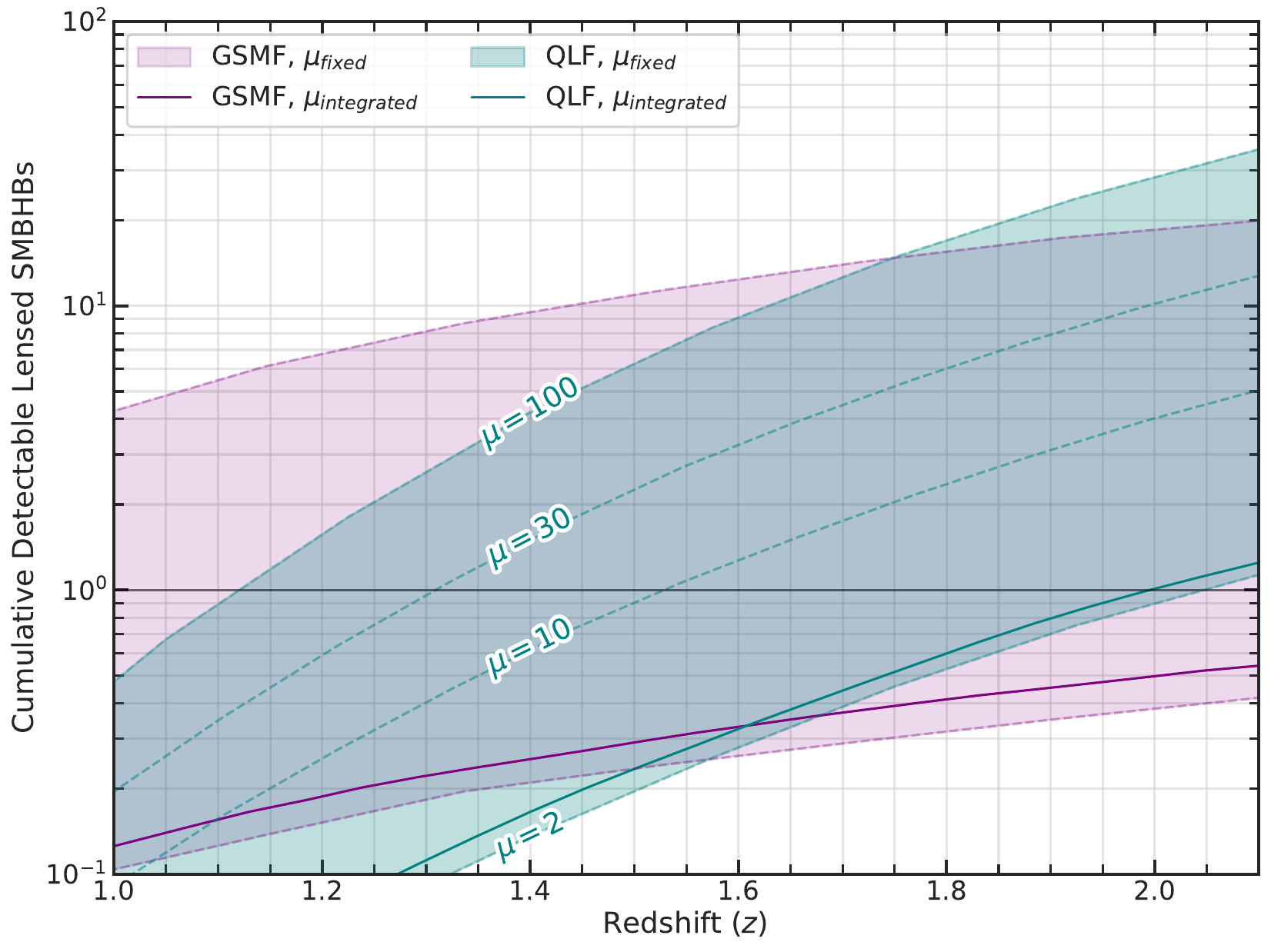}
    \caption{The predicted cumulative distribution of detectable strongly lensed SMBHBs in the SKA-PTA era, computed using both the GSMF (purple) and the QLF (teal). The wide bands show the spread of the population for different magnification $2 \leq \mu \leq 100$, fixed-$\mu$ examples given by the dashed curves. The solid curves represent a representative population obtained by {sampling} $\mu$ in each $z$---$\mathcal{M}$ bin according to the probability density function given in Dai et al.~\cite{Dai2017} (see Sec.~4.4.2 of \cite{LensedGWs}). The most realistic distribution---the solid teal curve---yields approximately a few detectable strongly lensed systems within $z \approx 2$.}
    \label{fig:Nlensed}
\end{figure}

Ref.~\cite{LensedGWs} predicts the number of expected SKA-PTA-detectable strongly lensed SMBHBs as a function of redshift and the additional astrophysical information that such sources offer by virtue of lensing effects. The population calculation entails binning the $z$---$\mathcal{M}$ parameter space, populating those bins with SMBHBs emitting detectable lensed GWs ($h_{\mathrm{lensed}} \geq 10^{-16}$) {using a black hole mass function} derived from either a GSMF \cite{Muzzin2013} or a QLF \cite{hopkins_observational_2007}, and applying cuts according to binary fraction and strong lensing probabilities. The authors take the result calculated using the QLF to be representative, since there exists sufficient evidence for quasars as tracers of galaxy mergers, hence SMBHB mergers~\cite{CaseyClyde2025}, and find an expected $\sim$ few detectable strongly lensed SMBHBs within $z \approx 2$ as shown in \fig{fig:Nlensed}.

Those detectable strongly lensed sources serve as exciting candidates for extracting multi-messenger astrophysics. In particular, the binary separation of a sufficiently low-$f_{\mathrm{GW}}$ lensed SMBHB might be amplified enough for {EM} counterparts of the individual SMBHs {to be resolved within each image} \cite{LensedGWs} by multi-wavelength instruments {such as the Next Generation Very Large Array}~\cite{Reid2018}, VLBI~\cite{VLBI_Brisken2009}, and the SKAO itself~\cite{Dewdney2009}. If such a binary is in an opportune caustic configuration with respect to a foreground elliptical galaxy lens, see e.g. Barnacka~\cite{Barnacka2018}, sufficient time delays between images would permit the tracking of its frequency evolution across different PTA frequency bins \cite{LensedGWs}. These ``golden" binaries will be some of the most rich multi-messenger systems in the Universe upon their detection

\subsection{Joint Likelihood Multi-Messenger Analysis}
\label{sec:joint_likelihood}

The standard approach to multi-messenger characterization has been to use {EM} surveys to identify candidates, which are then used to construct priors for a targeted PTA search. While effective, this priors-only method {can discard} information {if the EM priors neglect parameter correlations}. A more rigorous framework, recently introduced by Charisi et al.~\cite{Charisi2025}, employs a joint likelihood analysis that simultaneously fits the {EM} light curves and the pulsar timing residuals as a single data stream.

In this framework, the total likelihood is the product of the individual instrument likelihoods, $\mathcal{L}_{\mathrm{tot}} = \mathcal{L}_{\mathrm{em}} \times \mathcal{L}_{\mathrm{PTA}}$, linked by common physical parameters (e.g., chirp mass $\mathcal{M}$, period $P_{\mathrm{orb}}$, and inclination $\iota$). Ref.~\cite{Charisi2025} demonstrated that this approach significantly tightens parameter constraints compared to separate analyses. By leveraging the sensitivity of light curves {and PTA signals to different combinations of binary parameters}, the joint fit {reduces} degeneracies between e.g. {total} mass and {inclination} that are inherent in single-messenger studies.

{An additional} promise of the joint likelihood approach is its {potential} to extract measurements from signals that would be sub-threshold in either detector alone. However, the method is highly model-dependent. It relies on the assumption that the {EM} variability (e.g., Doppler boosting or accretion modulation) is correctly modeled; a mismatch between the assumed {EM} mechanism and the physical reality could bias the recovered GW parameters. It also {can benefit from} high-cadence {EM} monitoring contemporaneous with PTA observations to track the orbital phase coherently~\citep{Charisi2025}.

\subsection{Cosmology with Bright Sirens}
\label{sec:bright_sirens}

Beyond characterizing binary astrophysics, CW detections offer a novel probe of cosmology. A SMBHB detected by a PTA {can act} as a standard siren: the GW signal provides a direct measurement of the luminosity distance $D_L$, independent of the cosmic distance ladder \citep{Schutz1986, HolzHughes2005}. If the binary has an identified {EM} counterpart (a ``bright siren''), the host galaxy provides the redshift $z$.

The combination of a GW-derived $D_L$ and {an EM-derived $z$} allows for a direct measurement of the Hubble constant $H_0$~\citep{Wang2025}. Constraining the dark energy equation of state $w$ requires sources at redshifts where the $D_L(z)$ relation is sensitive to the expansion history beyond $H_0$, which is likely beyond the reach of current PTAs but may become accessible in the SKAO era. Unlike standard candles (e.g., Cepheids, Supernovae Ia), standard sirens do not require calibration against local anchors, making them an invaluable cross-check on the current ``Hubble tension'' between early- and late-Universe measurements~\cite{Riess2021,Planck2018Cosmo}.

{Sardana et al.~\cite{Sardana2026} develop this approach through targeted PTA searches (\sect{subsec:targeted_searches}) of {EM} SMBHB candidates with known host galaxies and redshifts. They jointly fit the Earth and pulsar terms, including pulsar distances as fitted parameters. The pulsar terms sample the binary at earlier epochs, so their phase and frequency differences from the Earth term constrain the chirp mass and break its degeneracy with $D_L$ in the GW amplitude (\sect{subsec:time_machine}). Their simulated CPTA observations yield an $H_0$ uncertainty of about $2\,\mathrm{km\,s^{-1}\,Mpc^{-1}}$ for one loud source at $z=0.07$, using 40 pulsars observed over 30 years and {assuming a flat cosmology with cold dark matter and a cosmological constant, with fixed $\Omega_{\mathrm{m}}$}. They also examine host misidentification, finding low confusion with other binaries for their best-measured target in the population model studied.}

{Even with the chirp mass constrained, the} precision of a bright siren measurement depends critically on breaking the degeneracy between the binary inclination $\iota$ and $D_L$ that is inherent in the GW amplitude. An independent {EM} constraint on $\iota$---for example from spectroscopic radial-velocity modelling of a binary AGN---would significantly tighten $D_L$, but such measurements remain challenging and model-dependent~\citep{DOrazioCharisi2023}.

\subsection{Binary Evolution: The PTA Time Machine}
\label{subsec:time_machine}
While the Pulsar term acts as incoherent self-noise for stochastic background searches (\sect{subsec:pta_response}), it is a vital signal for single-source science. For a CW source, the binary evolves (chirps) due to GW emission. The light-travel delay {$\tau_a \sim L_a \sim 10^3$}--$10^4$\,yr means that the Pulsar term provides a snapshot of the binary's state thousands of years in the past. If the binary is inspiraling ($df/dt > 0$), the frequency imprinted on the Pulsar term, $f_P$, will be lower than the frequency observed at the Earth term, $f_E$. For a circular{, GW-driven} inspiral, the two are related by \citep{Mingarelli12}:
\begin{equation}\label{eq:pulsar_term_freq}
{\omega(t_p) = \omega_0 \left[ 1 + \frac{256}{5} \mathcal{M}^{5/3} \omega_0^{8/3} (1+\hat{\Omega}\cdot\hat{p})L \right]^{-3/8},}
\end{equation}
where $\omega_0$ is the Earth-term {orbital} angular frequency.

For nearby, massive binaries, the frequency difference $\Delta f = f_E - f_P$ may be measurable. This allows PTAs to function as a dynamical time machine: by comparing the signal at the Earth and the pulsar, we can directly measure the binary's evolution over millennial timescales \citep{Mingarelli12, corbincornish2010}. {Assuming GW-driven evolution, this} measurement breaks degeneracies between the chirp mass $\mathcal{M}$ and the luminosity distance $D_L$ that are otherwise coupled in the Earth-term-only signal.

Beyond measuring the inspiral rate, Mingarelli et al.~\cite{Mingarelli12} showed that the pulsar term encodes higher-order {pN corrections} to the orbital phase, enabling measurements of SMBHB spins. In particular, the spin--orbit (1.5pN) and spin--spin (2pN) coupling terms modify the phase evolution between the Earth and pulsar terms, so that the frequency difference $\Delta f$ depends not only on $\mathcal{M}$ but also on the individual spin magnitudes and orientations. With sufficient pulsar distance precision {and S/N}, PTAs {could} therefore {constrain the binary masses and a combination of spin parameters} through a {coherent} comparison of the waveform at each pulsar and the Earth. However, extracting this physics requires knowing the pulsar distance $L_a$ to high precision.

{Whether the Earth and pulsar terms can be connected in phase} depends on the distance precision relative to the GW wavelength {$\lambda_{\mathrm{gw}} = 1/f$}. If the distance uncertainty $\delta L_a \gtrsim \lambda_{\mathrm{gw}}$, the pulsar-term phase is effectively unconstrained and must be marginalized over. In this regime---which describes most current PTA pulsars at nanoHertz frequencies---the Earth term alone provides robust constraints on the GW frequency and total mass {in targeted searches with the source sky position and distance fixed}, while being cheaper computationally~\cite{Charisi2024}. Conversely, when $\delta L_a \ll \lambda_{\mathrm{gw}}$, the pulsar term can be modeled coherently: Corbin \& Cornish~\cite{corbincornish2010} showed that including it roughly doubles the effective signal power, while Lee et al.~\cite{Lee2011} demonstrated via Fisher matrix analysis that it is essential for accurate sky localization at high S/N, and can even be used in reverse to improve pulsar distance measurements from a strong CW detection. In a future PTA with precise distances for a subset of the array (see \sect{subsec:distances}), a hybrid strategy may be useful: use the Earth term to drive detection and define the spatial correlation pattern (\sect{subsec:deterministic_fingerprints}), while selectively exploiting pulsar terms where distance precision is better than the detected GW wavelength (\sect{sec:resolution}).

{B\'{e}csy \& Siwek~\cite{BecsySiwek2026_pulsarterms} develop a Fisher framework that uses independently measured pulsar-term frequencies to reconstruct binary evolution without connecting the Earth and pulsar terms in phase. Pulsar-distance uncertainties then become uncertainties in the look-back times. Their forecasts for circular binaries in a 116-pulsar PTA spanning 30 years show how this can constrain environmental effects. The modeled gas torques are generally difficult to identify, although sufficiently strong torques can bias the inferred chirp mass if neglected.}

{Pulsar terms could reveal past SMBHB mergers~\cite{QuelquejayLeclere2026_pulsarterms} and connect PTAs to higher-frequency GW observations~\cite{Criswell2026_orphaned}. Zheng et al.~\cite{ZhengBecsyMingarelli2026_echoes} find {a connection between PTAs and the Laser Interferometer Space Antenna unlikely because typical sources in its band produce pulsar terms too weak to detect}; a future $\mu$Ares mission~\cite{Sesana2021_muAres} offers better prospects for nearby, massive binaries. Their forecasts also show that pulsar-term detections and nondetections across the array can help localize the source without requiring phase connection (\sect{sec:resolution}).}

\subsection{Pulsar Distances}
\label{subsec:distances}
To {constrain the pulsar-term phase from its distance}, we must know the pulsar distance $L_a$ well enough that {its} uncertainty {is small compared to the} gravitational {wavelength}: {$2\pi f \,\delta L_a < 1$}, i.e.\ {$\delta L_a < 1/(2\pi f)$}. This criterion is therefore frequency-dependent: at $f = {1}$\,nHz it requires $\delta L_a \lesssim 1$\,pc, but at picohertz frequencies the GW wavelength is so long that the constraint relaxes by orders of magnitude, and many pulsars already satisfy {$2\pi f L_a < 1$} (see \sect{subsec:picohertz}). Although $\delta L_a \lesssim 1$\,pc is often quoted as a blanket requirement, this applies specifically to the nanoHertz band.

Achieving parsec-level precision at nanoHertz frequencies is a major observational challenge. Traditional pulsar-timing parallax, which measures the {semiannual} modulation in arrival times due to the Earth's orbit, is {most sensitive} to nearby sources because the parallax signature scales as $1/L$ and {becomes harder to measure} for more distant pulsars~\cite{Toscano}. VLBI astrometry offers {high} precision \cite{Brisken, VLBA} but is observationally expensive and limited to bright radio sources. Indirect methods, such as Dispersion Measure (DM) distances derived from Galactic electron density models (e.g., NE2001 \cite{NE2001} or YMW16 \cite{YMW16}), are available for all pulsars but suffer from systematic uncertainties of $\sim 20-{40}\%$~\cite{Moran2023}. The recently released NE2025 model~\cite{NE2025} updates the Galactic free electron density using an expanded set of independent distance measurements and improved modeling of spiral structure, which should reduce DM-based distance uncertainties for some PTA pulsars.

In the \emph{Gaia} era, new approaches are opening possibilities. Jennings et al.~\cite{Jennings2018} used \emph{Gaia} DR2 parallaxes of binary pulsar companions to obtain independent distance measurements.  \emph{Gaia} parallaxes can also be combined with radio timing to further refine the distance estimate, see Moran et al.~\cite{Moran2023}. Improving these distance measurements is a priority for the field, as it directly enhances the sensitivity of PTAs to CW sources and enables tests of GR that rely on {correcting for} the distance-dependent Shklovskii effect \cite{Shk70, GenRelTest}.

\section{The Angular Resolution of PTAs}
\label{sec:resolution}

{Angular resolution} in {PTAs has long been a source of confusion, partly because the term} is {used for several different quantities. The ability to distinguish nearby sources,} the angular {width} of a {reconstructed point-source response, and the uncertainty in an individual source's position are related but distinct. These must also be distinguished from detection sensitivity and} the number of {sources that can be fitted simultaneously. Here we clarify these concepts and explain how different resolution estimates depend on} the number {and distribution} of {pulsars,} the {S/N,} and {whether pulsar-term phase information} is {available.}

\subsection{The Incoherent Regime}

In a standard PTA, pulsar distances are not known to within a GW wavelength ({$\lambda_{\mathrm{GW}} = 1/f \sim 1$}\,pc at 10\,nHz). The pulsar term therefore cannot be phased coherently and averages to noise, so the array relies solely on the Earth term for sky localization.

In this regime, adding pulsars improves two things at different rates. The number of CW sources that can be simultaneously resolved scales linearly with $N$: Boyle \& Pen~\cite{BoylePen2012} showed that {a PTA provides} $2N$ independent constraints {per frequency bin} ({amplitude and phase} per pulsar). {For monochromatic Earth-term signals at a known frequency, six parameters per binary give a counting limit of} $M_{\mathrm{sources}} \approx {N/3}$. This was confirmed numerically by Babak \& Sesana~\cite{BabakSesana2012} {in noiseless data}.

The angular resolution---the {effective beam width} of a reconstructed sky map---does not scale with $N$ in the same way. Jow et al.~\cite{Jow2026} {estimated that Earth-term-only matched-filter maps for uniformly distributed, equally sensitive pulsars have} ${\approx}\,32$ {effective resolution elements per frequency} ($16$ {sky positions} $\times$ $2$ polarizations), corresponding to {an effective beam width} of ${\approx}\,58^\circ$. This {width is approached} once $N \gtrsim 20$; adding further pulsars improves the S/N per {resolution element} but does not {appreciably narrow the beam. The number of distinguishable sources also depends on S/N.}

Sesana \& Vecchio~\cite{SesanaVecchio2010} used a Fisher matrix analysis to show that the {uncertainties on most parameters of} a single CW source {change little} beyond ${\approx}\,20$ pulsars at fixed total S/N, {while} the sky localization error box $\Delta\Omega$ {continues to improve} (see their Table~I). {The two studies} address different questions. Sesana \& Vecchio quantify how well a single source can be localized, finding $\Delta\Omega \approx 40\,\mathrm{deg}^2$ for 100 {uniformly distributed} pulsars at S/N~$= 10$. Jow et al.\ {study the effective beam resolution of} a sky map, and further extend the analysis to polarization.

\subsection{The Coherent Regime}

{When} pulsar distances are measured {precisely enough to constrain the pulsar-term phases} (see \sect{subsec:distances}), {the Earth and pulsar terms can be modeled coherently. Their relative phase depends on the source direction, allowing the PTA to act as} an interferometer with Galactic-scale {baselines} $D_{\mathrm{array}} \sim \mathrm{kpc}$.

{In} this coherent regime, {the diffraction angular scale is $\lambda_{\mathrm{GW}}/D_{\mathrm{array}}$, while the localization uncertainty of a bright CW source can be smaller by a factor of S/N} {\citep{BoylePen2012,Tsai2025}}:
\begin{equation}
    \delta \theta_{{\mathrm{loc}}} \approx \frac{1}{\mathrm{S/N}} \frac{\lambda_{\mathrm{GW}}}{D_{\mathrm{array}}}\,.
\end{equation}
{In the idealized maps of Tsai et al.~\cite{Tsai2025}, localization improves rapidly with} the number of pulsars with {precisely known distances}, $N_{\mathrm{dist}}$, {until the localization scale above is reached.}
With $N_{\mathrm{dist}} \approx {10}$ pulsars at {300\,pc and their distances treated as exact, Tsai et al.\ estimate ${\sim}\,1$\,arcmin} localization {for $\lambda_{\mathrm{GW}}\approx1$\,pc and} $\mathrm{S/N} = 10$ {in the matched-filter map}, which {could substantially narrow the search for EM} counterparts of individual CW sources~\cite{Tsai2025}. This result underscores the importance of ongoing pulsar astrometry programs (\sect{subsec:distances}) for PTA science.

\subsection{Implications for Searches}
\label{sec:practical}

These resolution limits have direct consequences for search pipeline design. For GWB anisotropy searches (\sect{sec:anisotropy}), the sky power is decomposed into spherical harmonics $Y_{\ell m}$, and a central question is the maximum recoverable multipole $\ell_{\mathrm{max}}$.  A PTA with $N$ pulsars provides $N(N{-}1)/2$ {distinct} cross-correlations, which constrain the {spherical harmonic} coefficients ${c}_{\ell m}$ {of the sky power}~\cite{m13, ani2020}. The rank of the response matrix {gives a necessary counting condition for jointly recovering all coefficients through $\ell_{\mathrm{max}}$}~\cite{Domcke:2025esw,ani2020}:
\begin{equation}
    (\ell_{\mathrm{max}} + 1)^2 \lesssim \frac{N(N-1)}{2}\,.
\end{equation}
{This bound permits $\ell_{\mathrm{max}}$ to grow linearly with $N$, but the multipoles measurable in practice also depend on S/N and array geometry.}

For CW searches, it is essential to distinguish resolution (the minimum angular separation at which two sources can be told apart) from localization (how precisely a single source can be pinpointed). {The broad Earth-term beam discussed above} (${\sim}\,60^\circ$) {describes the width of a point source's response in the map. For a bright CW source, the uncertainty in its position} can be {much smaller than this width.}  {In the simulations of} Goldstein et al.~\cite{Goldstein2019}, the {90\%} localization area {decreases approximately as}
\begin{equation}
    \Delta \Omega \propto \frac{1}{{(\mathrm{S/N})}^2}\,
\end{equation}
{for total $\mathrm{S/N}\gtrsim10$. In coherent analyses, uncertainties in pulsar distances introduce phase errors that can shift the inferred source position}~\cite{Tsai2025}.

{Even at fixed S/N,} localization {is generally poorer} in regions {with} sparse pulsar coverage. {In the simulations of} Petrov et al.~\cite{Petrov2024}, {the 90\% localization areas at $\mathrm{S/N}=8$ ranged from a few hundred square degrees to most of the sky.}

\section{Search for New Physics}
\label{sec:new_physics}

While the astrophysical population of SMBHBs provides a natural explanation for the nanoHertz GWB, PTAs also serve as a unique probe of the very early Universe. In the era between inflation and {{Big Bang nucleosynthesis}}, the Universe may have undergone violent processes that generated substantial gravitational radiation. {For example, {FOPTs} at MeV--GeV temperatures can produce signals in the PTA band, with the observed frequencies depending on the transition dynamics and cosmological redshift. PTAs therefore provide a window onto early-Universe physics complementary to terrestrial experiments.}

Distinguishing these cosmological signals from the astrophysical foreground relies fundamentally on their spectral shapes and angular power distribution. While circular{, GW-driven} SMBHBs produce a power-law spectrum with a characteristic strain $h_c(f) \propto f^{-2/3}$ (spectral index $\gamma = 13/3$; \eq{eq:strain_spectrum_circular_bhbs}), cosmological sources often predict distinct spectral slopes, broken power laws, or peaked spectra~\cite{Afzal2023}.

\subsection*{Cosmic Strings}
Cosmic strings are one of the most compelling candidates for a cosmological GWB. These one-dimensional topological defects can form during symmetry-breaking phase transitions in the early Universe (e.g., at the Grand Unification scale). As the string network evolves, it forms closed loops that oscillate and decay via gravitational radiation~{\cite{Ghoshal2023,Chang2020}}.

The resulting GWB spectrum depends on the {dimensionless string tension, $\mu$}, and the loop production efficiency. {In a power-law approximation over the} PTA band, cosmic string spectra {can have} $\gamma$ {above or below} $13/3$, {depending on the string model and tension}~{\cite{Afzal2023}}. {For} stable Nambu-Goto strings, {the NANOGrav 15-yr analysis gives 95\% upper limits on $\mu$ of order $10^{-10}$ when an SMBHB contribution is included; the precise bound depends on the loop emission model}~{\cite{Afzal2023}}. If the {NANOGrav 15-yr GWB} is interpreted entirely as a {stable-string} signal, it favors {$\mu\sim10^{-10}$, although Bayesian model comparison favors SMBHBs. For metastable strings and cosmic superstrings, the preferred tension also depends on the network lifetime and reconnection probability, respectively}~\cite{Afzal2023}.

\subsection*{ Cosmological Phase Transitions}
The early Universe may have experienced {FOPTs} as it cooled. In a FOPT, bubbles of the new vacuum state nucleate, expand, and collide. This process generates GWs through three dominant mechanisms: (1) the collision of bubble walls, (2) sound waves generated in the primordial plasma, and (3) subsequent {magnetohydrodynamic turbulence} \cite{Weir2018, Neronov2021}.

The {observed} peak frequency {depends on both the transition temperature} $T_*$ {and the source's characteristic length scale, such as the bubble size. PTAs can therefore probe a range of FOPT scenarios, including transitions near the} {quantum chromodynamics scale} ($T_* \sim 100$ MeV) {and in hidden sectors}~{\cite{Afzal2023, Brandenburg2021, Moore2021}}. Recent proposals, such as a ``Dark Big Bang,'' suggest that dark matter could have been generated in a separate phase transition, producing a detectable GWB signal {while remaining consistent with {Big Bang nucleosynthesis constraints}}~\cite{Freese2023}.

\subsection*{Primordial Gravitational Waves and Inflation}
Standard slow-roll inflation predicts a background of {primordial GWs} with a red-tilted spectrum (tensor spectral index $n_T < 0$)~{\cite{Liddle2000,Kamionkowski2016}}, which would be unobservably faint at nanoHertz frequencies~{\cite{Boyle2005,Planck2018Inflation}}. However, a detection {of an inflationary GWB} by PTAs would be transformative, implying non-standard inflationary physics.

{Blue primordial tensor spectra} ($n_T > 0$) {can arise in non-standard inflation or in alternatives such as ekpyrotic bouncing cosmologies}~\cite{Lasky2016}. {Second-order} GWs induced by large scalar perturbations {provide another possible source of a GWB in the PTA band}~{\cite{Afzal2023}}. The latter mechanism is particularly intriguing as it connects the GWB to the formation of Primordial Black Holes. {The same scalar perturbations can also form primordial black holes that contribute to} the dark matter{, but their masses and abundance depend on the perturbation spectrum, its statistics, and the collapse model}~{\cite{Afzal2023}}.

\subsection*{Dark Matter}
{The astrophysical GWB carries information about both SMBHBs and their environments. As discussed in \sect{sec:modeling_gwb}, stars and gas can accelerate binary evolution and alter the strain spectrum. Dark matter surrounding the binary can also extract orbital energy through dynamical friction. Alonso-\'{A}lvarez et al.~\cite{AlonsoAlvarez2024_SIDM} showed that dense concentrations, or ``spikes,'' of self-interacting dark matter can sustain this process, overcoming the final parsec problem and producing a low-frequency turnover in the characteristic strain spectrum. In their models with velocity-dependent self-interactions, scattering replenishes the spike while the halo core absorbs the binary's orbital energy.}

PTAs also function as direct detectors for local dark matter. {ULDM}, composed of scalar or vector bosons with masses $m \sim 10^{-23}$ eV, behaves as a coherent wave on {roughly kiloparsec} scales. This oscillating field interacts with standard matter via gravity, inducing periodic fluctuations in the gravitational potential of the Milky Way and the spacetime metric itself. The theoretical framework for detecting this signal with pulsar timing was established by Khmelnitsky \& Rubakov~\cite{KhmelnitksyRubakov2014}, who showed that an oscillating scalar field produces a monochromatic timing residual at a frequency set by the particle mass ($f \approx 2m/h$). Porayko \& Postnov~\cite{PoraykoPostnov2014} derived the first constraints using existing pulsar data.

{The scalar-field signal described above has a monopolar Earth term and does not follow the {HD} correlations of} the stochastic GWB (\sect{sec:stochastic_background}){. Dedicated searches therefore model its oscillatory timing signature and spatial correlations}~{\cite{Afzal2023}}. Porayko et al.~\cite{pzl+18} placed upper limits {on the ULDM density} using PPTA {data, as did} NANOGrav {with its} 15-yr dataset~\cite{Afzal2023} and the EPTA {with its} {DR2~\cite{Smarra2025}}. None have found evidence for ULDM, but together they place competitive constraints on the local dark matter density across the mass range $m \sim 10^{-24}$--$10^{-22}$\,eV. The PPTA DR3 has pursued a complementary approach, using pulsar polarization data to search for axion-like dark matter via the cosmic birefringence effect~\cite{PPTADR3_birefringence}. {Porayko et al.~\cite{Porayko2025Polarimetry} likewise searched EPTA pulsar polarization for oscillations induced by axion--photon coupling.}

{Timing searches can also probe ULDM with oscillation periods longer than the observing span.} Unal et al.~\cite{Unal2023_ULDM} {forecast sensitivity to scalar and vector ULDM down to} $m \sim 10^{-26}$\,eV {for representative current PTA parameters: over this range, the increasing signal amplitude can compensate for reduced sensitivity} below $1/T_{\mathrm{obs}}$. Their analysis also {includes tensor ULDM and predicts improved sensitivity {with longer observations and the Square Kilometre Array}}. 

\subsection*{Tests of General Relativity and Alternative Polarizations}
\label{sec:non_gr_polarizations}

Beyond searching for exotic sources, PTAs offer a precision laboratory for testing the fundamental nature of gravity itself. {GR} predicts that GWs possess only two tensor polarization modes: plus ($+$) and cross ($\times$), which propagate at the speed of light. These modes generate the specific quadrupolar spatial correlation pattern described by the {HD} curve ($\Gamma_{ab}^{\mathrm{HD}}$; \sect{subsec:pta_response}, \eq{eq:HD_split}, \fig{fig:HDcurve}).

However, many extensions to GR{,} such as scalar-tensor theories, $f(R)$ gravity, or massive gravity{,} predict the existence of up to four additional polarization modes: two scalar modes and two vector modes~\cite{Lee2008, Chamberlin2012}. {For an isotropic GWB, the} scalar breathing mode {produces a mixture of monopolar and dipolar correlations, with} $\Gamma_{ab}^{\mathrm{ST}} = \tfrac{1}{{8}}(3 + \cos\zeta)$ {for distinct pulsars with incoherent pulsar terms}~{\cite{NANOGrav15yr_transverse}}. {When the pulsar terms average away, the transverse tensor and breathing modes have ORFs that depend only on angular separation. For the non-transverse scalar longitudinal and vector modes, these terms can remain important, introducing frequency and distance dependence. The response can be enhanced by orders of magnitude for nearly co-aligned pairs, especially for the scalar longitudinal mode}~\cite{Chamberlin2012}.

By decomposing the observed spatial correlations (\sect{subsec:pta_response}) into these tensor, scalar, and vector basis functions, PTAs can place stringent upper limits on the amplitude of non-Einsteinian modes \cite{Lee2008, Chamberlin2012}. {The NANOGrav 15-yr search found no compelling evidence for a scalar breathing component in addition to the tensor GWB~\cite{NANOGrav15yr_transverse}. A reanalysis of the same dataset with customized chromatic noise models strengthened the preference for {HD} correlations over a purely scalar breathing background and tightened the upper limit on the scalar amplitude~\cite{NG15_CNM_GWB}.}

{The interpretation of polarization searches depends on the assumed angular distribution of the GWB. An anisotropic tensor background can change the correlations between individual pulsar pairs without introducing additional GW polarizations (\sect{sec:anisotropy}). Departures from the {HD} expectation should therefore also be assessed against anisotropic GR models before being attributed to non-GR polarizations.}

\section{Pulsar Noise Models}\label{subsec:noise}

The inherent timing stability of millisecond pulsars and the ability to cross-correlate signals between pulsars (\sect{sec:stochastic_background}) is tantamount for PTAs to detect GW signals~\citep{Verbiest2009}.
However, the success of PTA experiments is also inherently tied to the challenges posed by various noise processes.
Sources of pulsar noise can originate from intrinsic processes at the pulsar itself, propagation effects through the {ISM}, or clock errors and solar system ephemeris errors \citep{Cordes2013, Tiburzi2016, NG15_detchar}.

The impacts of noise on a GW analysis can be understood in terms of sensitivity curves \citep{Hazboun2019, NG15_detchar}.
GW searches using PTAs are fundamentally limited at certain frequencies due to degeneracies with known physical effects, which are accounted for by a deterministic timing model for each pulsar.
For example, fits for the spindown of a pulsar resulting from by magnetic dipole radiation act as a quadratic high-pass filter, while fits for each pulsar's precise sky location remove power at frequencies of $1/\mathrm{yr}$ due to the Earth's orbit.
Unresolved stochastic ``red'' and ``white'' noise in each pulsar further limits sensitivity at low and high frequencies, respectively.
Intrinsic pulsar red noise limits sensitivity to low-frequency signals, such as from a GWB (\sect{sec:stochastic_background}), while white noise is the dominant limitation at higher frequencies, where PTAs are more likely to resolve individual CW sources (\sect{sec:CW}).

In practice, accurate noise modeling is a deep and complicated endeavor which introduces many challenges for GW measurements.
The properties of any noise process may vary considerably from one pulsar to another \citep{Lentati2016, Goncharov2021, Chalumeau2022, Hazboun+2025}, and seemingly subtle noise modeling decisions have in fact played crucial roles in GW analyses.
For example, failing to account for unmodeled noise in certain pulsars was found to bias GWB parameter estimation in a time-slicing analysis of the NANOGrav 11 yr data set~\citep{Hazboun2020_NG11}.
It is also challenging to characterize pulsar noise without biasing the GW analyses.
For instance, the past use of uniform priors on the amplitude of the intrinsic red noise in individual pulsars has been shown to result in overly low GWB amplitudes and upper limits \citep{Hazboun2020}.
Meanwhile, the failure to account for ensemble properties of intrinsic red noise across millisecond pulsars has been shown to result in spurious detections of GWB-like signals \citep{Goncharov2021_CRN, Zic2022, Goncharov2022, GoncharovSardana2024, vanHaasteren2024}.
These challenges ultimately introduce an opportunity, as improvements to pulsar noise analysis and mitigation is a promising means to improve PTA sensitivity to GWs \citep{NG15_detchar}.
Furthermore, better characterization of pulsar noise can provide valuable scientific insights on the underlying astrophysics, such as ISM turbulence~\citep{Jones2017} and variations in the solar wind density \citep{Hazboun2022, Susarla2024}.
This dual-purpose analysis broadens the scope of pulsar noise analysis efforts beyond GW detection.

Of the types of noise sources affecting pulsar timing, perhaps the richest and most diverse are ``chromatic,'' meaning the magnitude of the timing delay depends on the radio frequency of the pulsar's emission \citep{CordesShannon2010, Cordes2013, Hazboun+2025}.
Chromatic effects typically arise due to ISM propagation effects, and they must be mitigated by observing TOAs at several radio frequencies \citep{Alam2021_wideband, NG15_dataset}.
However, multiple factors make the complete removal of chromatic noise difficult.
These include the finite range and resolution of radio frequency coverage \citep{Lam2015, Sosa2023} and the potential of mismodeling either the radio-frequency dependence or time-dependence of the chromatic delay \citep{Alam2021_wideband, Goncharov2021, Larsen2024}.

\begin{figure}
    \centering
    \includegraphics[width=\textwidth]{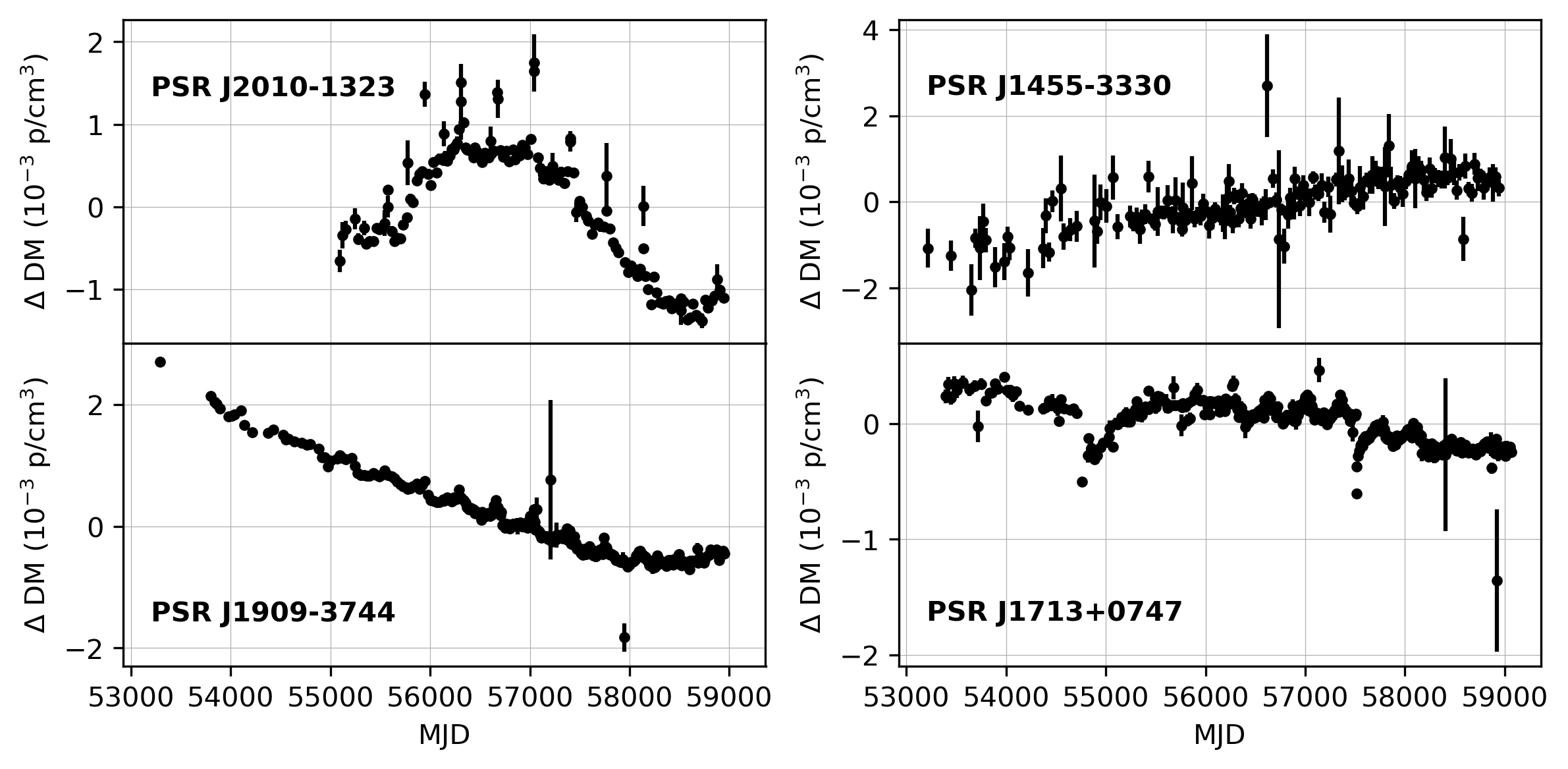}
    \caption{DM variations for four pulsars from the NANOGrav 15 yr data set \citep{NG15_dataset}. DM variations can be linear, periodic, or stochastic, and may manifest very differently in different pulsars. DM cusps with annual periodicity originate from the solar wind. Mismodeled DM variations will propagate errors into the infinite-frequency TOA, reducing sensitivity to GWs.}
    \label{fig:DM}
\end{figure}

Here we highlight DM and scattering variations as two prominent sources of chromatic noise.
DM variations are changes in the integrated electron density content of the ISM and the solar wind along the line of sight.
These are observed in nearly all pulsars and the time delay $\Delta t$ from DM depends on radio frequency $\nu$ as $\Delta t \propto \nu^{-2}$ \citep{RankinRoberts1971, Keith2013, Jones2017}.
DM variations can be fit using a piecewise constant model~\citep{Demorest2013}, modeled as a Gaussian Process \citep{Lentati2014, vanHaasterenVallisneri2014}, or measured directly alongside the TOA using wideband timing methods \citep{Pennucci2014, Curylo2023}, each method coming with advantages and disadvantages~\citep{Iraci2024}.
For example, the overall impact of the piecewise constant fit for DM variations is a broadband reduction in GW sensitivity across low and high GW frequencies \citep{Hazboun2019}.
Gaussian process models for DM variations impose a smaller penalty on sensitivity as they produce more precise DM time series and occupy a much smaller prior volume in a Bayesian sense \citep{Iraci2024, Gitika+2025}, but they come with higher computational overhead and require accurate modeling for the time-dependent covariance of the variations.
\fig{fig:DM} shows examples of DM time series from four pulsars from the NANOGrav 15-yr data set \citep{NG15_dataset}.
These showcase the variety and complexity of DM trends that may appear in different pulsars, which will propagate errors into the infinite-frequency TOA if imperfectly estimated \citep{Lam2015}.
In particular, the solar wind introduces quasi-periodic structures which are not well-modeled as Gaussian perturbations to the DM time series itself, instead requiring additional detailed models for the local solar wind electron density \citep{Hazboun2022, PPTA_noise, Susarla2024, Iraci2025, LarsenCNM2025}. Importantly, the solar wind affects all pulsars viewed through the inner heliosphere, so solar wind--induced DM variations are correlated between pulsars at similar solar elongations, introducing a source of spatially correlated chromatic noise that must be distinguished from the GWB~\citep{Tiburzi2016, Hazboun2022}.

We also note two sudden dips in the DM variations for PSR J1713+0747 in \fig{fig:DM}.
These highlight one of the bigger challenges for DM modeling and pulsar noise modeling in general, as these events are currently believed to be related either to intrinsic magnetospheric disturbances or complicated ISM phenomena such as plasma lensing, rather than a dispersive process \citep{Shannon2016, Lam2018, Goncharov2021}.
However, since these events are intrinsically chromatic, they can be easily mismodeled as DM variations, at the cost of excess time-dependent noise that leaks into the achromatic noise channel \citep{Hazboun2020_NG11, Larsen2024}.

Time-variable scattering is another chromatic effect which results from frequency-dependent refraction and diffraction as the pulse propagates through inhomogeneities in the ISM \citep{CordesRickett1998, Hemberger2008, CordesShannon2010}.
While scattering is not expected to affect TOAs as strongly as DM variations, it is significantly more difficult to model, described mathematically as the convolution of the pulse profile with a pulse broadening function.
The first order effect of scattering is to delay the TOA proportionally to the pulse broadening function, which scales as $\Delta t \propto \nu^{-4.4}$ for a Kolmogorov medium.
However, a much broader range of chromatic dependencies may arise depending on the geometry of the ISM, the strength of the scattering, and the interplay between the pulse broadening function and the intrinsic pulse shape \citep{Shannon2017, NG15_detchar, Geiger2025}.
If scattering is not accounted for, it will be absorbed by the DM variations model, propagating time-and-DM-correlated errors into the achromatic delay and inflating the red and white noise levels \citep{Lentati2016, Larsen2024}.
Gaussian processes have been used to partially correct for scattering using an appropriate covariance prior (which thereby reduces the degeneracy with DM variations) and a fixed prescription for the TOA scaling such as $\Delta t \propto \nu^{-4}$ or $\Delta t \propto \nu^{-\chi}$ with $\chi$ as a free parameter \citep{Lentati+2017, Lam2018, Goncharov2021, Larsen2024, Miles2025_noise}, but this may still be insufficient for strongly scattered pulsars \citep{Geiger2025, Kulkarni2025, LarsenCNM2025, cordes2026fundamental_noise}.
A more promising technique for precise measurement and mitigation of the scattering delay is deconvolution of the pulse broadening function from the intrinsic pulse shape, see Young et al.~\citep{Young2023}, in particular using precise scintillation measurements from cyclic spectroscopy \citep{Demorest2011, Dolch2021, Turner2023}.
In practice, both methods will complement each other; cyclic spectroscopy applies best for the most highly scattered pulsars \citep{Dolch2021}, whereas Gaussian processes should work better for weakly-scattered pulsars where the TOA shifts are more closely proportional to the scaling of the pulse broadening function \citep{NG15_detchar, Geiger2025}.
Beyond scattering, an inhomogeneous ISM may produce a variety of higher-order effects which are inextricably linked with scattering, including angle-of-arrival variations, refractive scintillation, diffractive scintillation, and frequency-dependent DM due to multipath propagation \citep{CordesShannon2010, Stinebring2013, Cordes16, NG15_detchar, cordes2026fundamental_noise}.

\begin{figure}
\centering
\includegraphics[width=\textwidth]{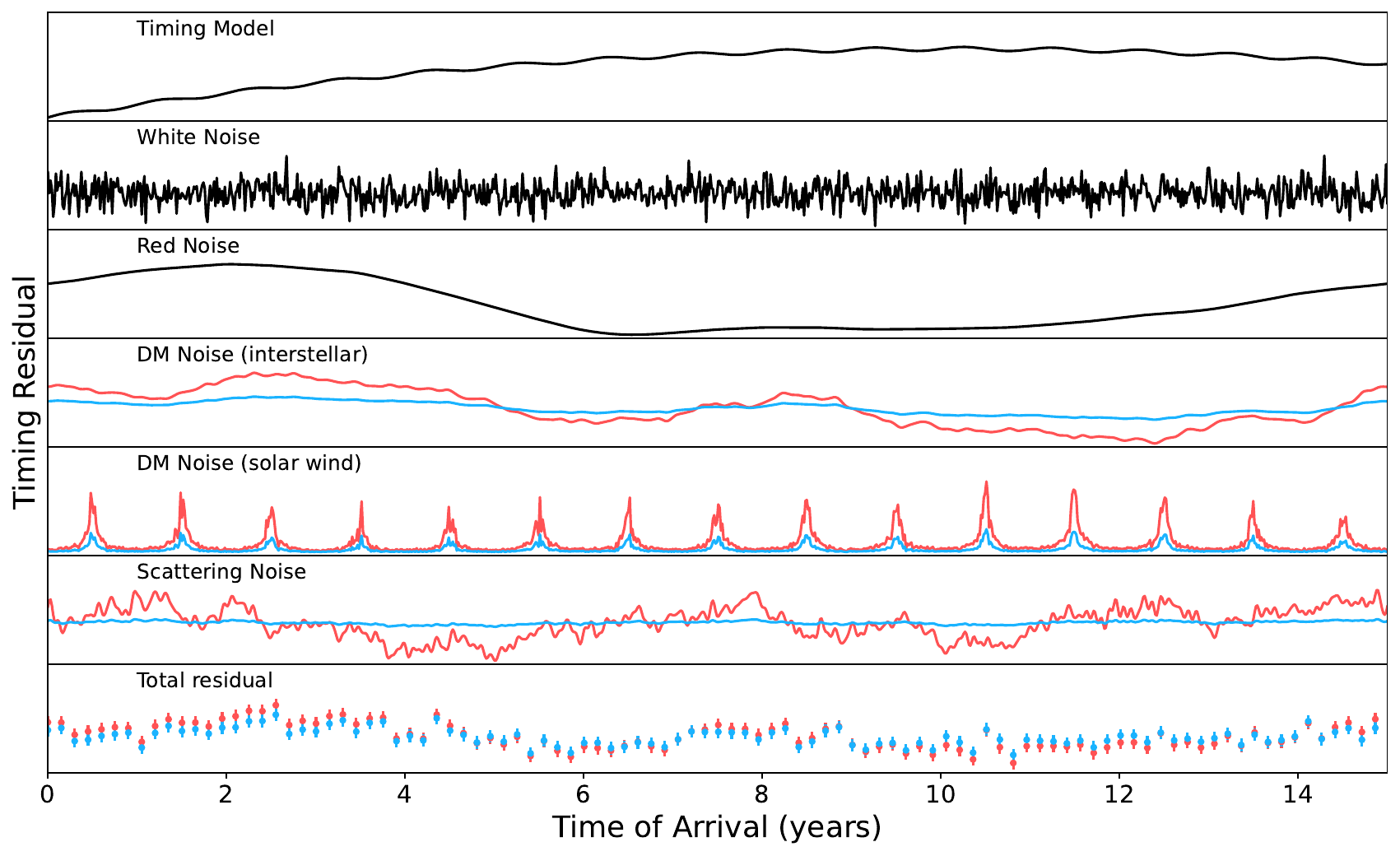}
\caption{Simulated timing residuals as produced by different types of sources of noise. The top three panels show example contributions to a pulsar's achromatic noise budget. First, a quadratic plus annual sinusoidal trend arising from perturbations to a simple timing model, white (time-uncorrelated) noise, and red (time-correlated) noise. The next three panels show examples of chromatic noise, in which the delay is larger for TOAs measured at lower radio frequency (shown in red) than a higher radio frequency (blue). In order we have stochastic DM variations from the {ISM}, annually-modulated DM variations from the solar wind, and scattering variations which have a different radio-frequency dependence than DM. The total timing residuals are shown in the bottom panel, including the sum of all noise sources and a finite cadence of observations. The simulated residuals in this example assume each type of noise is described by a Gaussian process.}
\label{fig:GPs_example}
\end{figure}

In contrast to chromatic noise, ``achromatic'' noise is not dependent on radio frequency.
Achromatic noise can be caused by stochastic fluctuations in pulsar spindowns due to e.g., torques from the pulsar magnetosphere \citep{Cheng1987, Lyne2010}, variable coupling between the neutron star crust and superfluid interior \citep{Jones1990, MelatosLink2014}, or the presence of an asteroid belt \citep{Shannon2013, Jennings2020}.
Mismodeled or unmodeled chromatic noise can arise as an achromatic noise process which can be potentially mitigated \citep{Lam2015, Hazboun2020_NG11, Sosa2023, LarsenCNM2025}.
However, intrinsic sources of achromatic noise are less straightforward to mitigate, directly impacting GW sensitivity or potentially masquerading as a GW signal if the achromatic noise is present in enough pulsars \citep{Zic2022, vanHaasteren2024}. 
Thankfully, the value of a PTA lies in the ability to distinguish GW signals from achromatic noise, since GW signals are correlated between different pulsars, while any achromatic noise which originates from processes intrinsic to the pulsar will be uncorrelated between different pulsars.
\fig{fig:GPs_example} illustrates an example set of simulated pulsar timing residuals, decomposed into individual sources of chromatic and achromatic noise.

{Noise originating outside individual pulsars can also be correlated across a PTA. Clock and solar-system ephemeris errors are achromatic, producing monopolar and approximately dipolar correlations, respectively}~\citep{Tiburzi2016}.
{Solar-wind variations produce chromatic correlations that depend on the lines of sight through the heliosphere and can contaminate inferred achromatic signals if inadequately modeled}~\citep{Tiburzi2016,NG15_detchar}.
{These different spatial patterns help distinguish such effects from the HD ORF expected for an isotropic, unpolarized tensor GWB} (\sect{subsec:pta_response}; \eq{eq:HD_split}; \cite{HD83}).
{In the NANOGrav 15-yr data, HD correlations are favored over purely monopolar or dipolar models}~\citep{NG15_gwb}.

\begin{figure}
\centering
\includegraphics[width=0.38\textwidth]{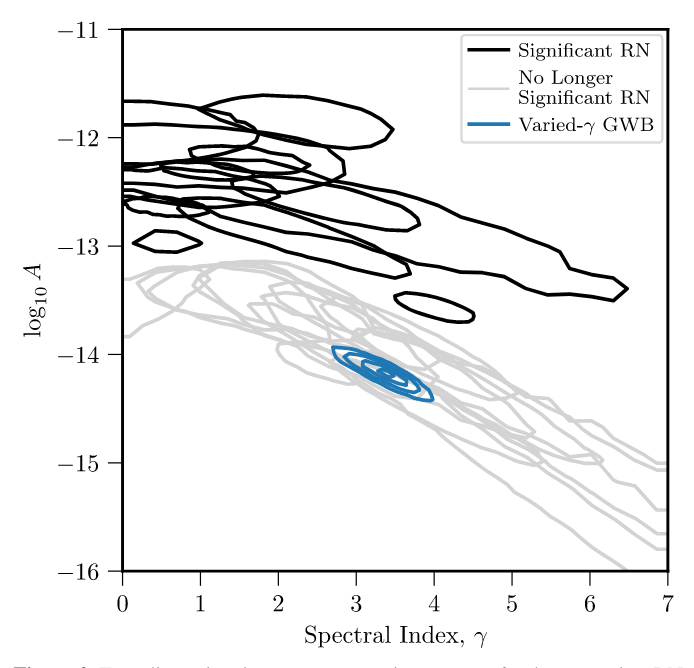}
\includegraphics[width=0.6\textwidth]{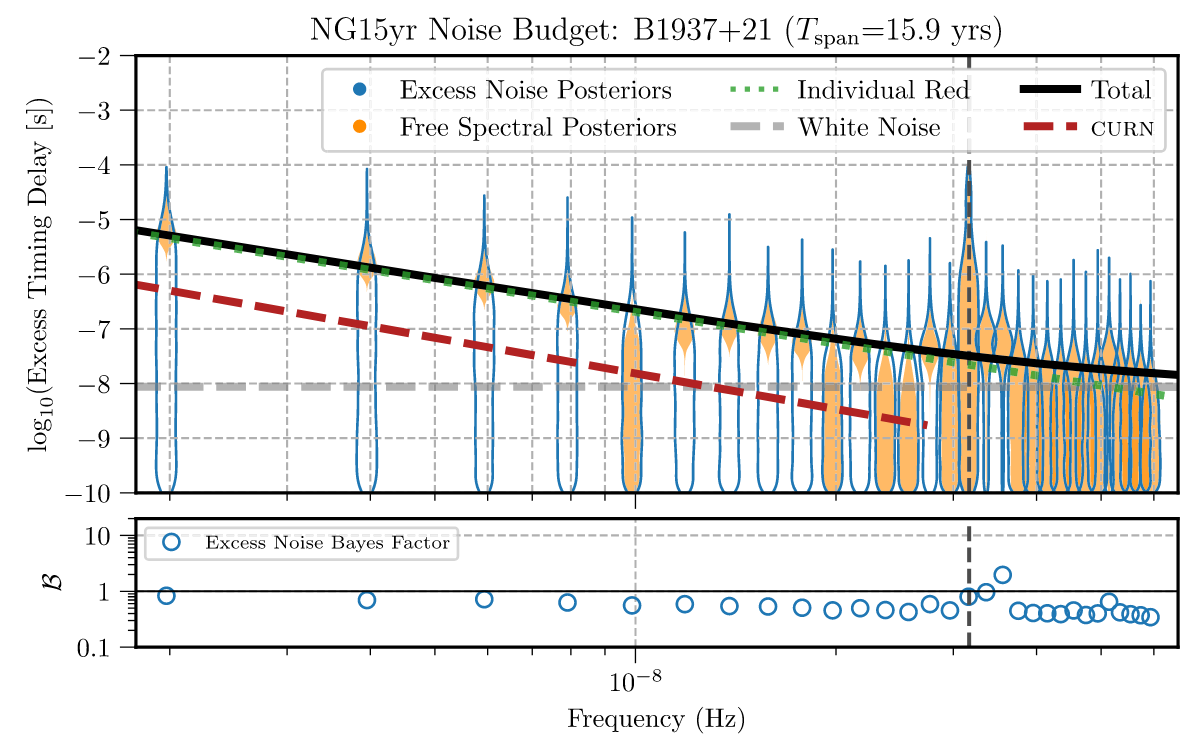}
\caption{\emph{Left:} {{Achromatic red noise (ARN) amplitude} and spectral-index posteriors from single-pulsar analyses of the} NANOGrav 15-yr data set \citep{NG15_dataset}.
{Grey contours identify pulsars whose individual ARN is no longer significant after including CURN in the full PTA analysis; black contours identify those retaining significant additional ARN. Blue contours show the common-process posterior.}
\emph{Right:} {PSR B1937+21 is an example of a pulsar with strong additional ARN. Its individual ARN (green) dominates at low frequencies over CURN (red), while white noise (grey) dominates at high frequencies; black shows their combined model. Orange violins show a fit with independent amplitudes at each frequency. Blue violins and the lower-panel Bayes factors test for excess noise beyond the fixed combined model and show no significant excess.}
Figures from \citep{NG15_detchar}.}
\label{fig:red_noise}
\end{figure}

Overall, a rich phenomenology of achromatic and chromatic noise processes have been observed across PTA datasets from different collaborations, with ever-increasing levels of comprehensiveness and model fidelity \citep{Lentati2016, lam_nanograv_2017, Goncharov2021, Chalumeau2022, Srivastava2023, NG15_detchar, EPTA_noise, PPTA_noise, Larsen2024, Miles2025_noise, Iraci2025, Nobleson2025, Hazboun+2025, LarsenCNM2025}. Focusing the present discussion on the NANOGrav 15-yr dataset, ref.~\cite{NG15_detchar} first performed a detailed analysis of the {achromatic red noise (ARN)} in the pulsars from the NANOGrav 15-yr dataset \citep{NG15_dataset}.
They found the pulsars could be grouped into three main categories: 1) pulsars with loud, shallow-spectral index ARN, 2) pulsars whose ARN disappears when modeling the GWB, and 3) pulsars where ARN was not detected (\fig{fig:red_noise}, left side).
The one outlier is PSR B1937+21, which exhibits a loud red noise process with a steep spectral index (\fig{fig:red_noise}, right side).
Ref.~\citep{NG15_detchar} suggest that for the pulsars in group 1, the majority of ARN likely results from mismodeled chromatic noise.
Meanwhile, the dominant source of ARN in group 2 appears to be the GWB.
Additionally, \citep{NG15_detchar} found no significant evidence of excess noise on top of a power law red noise spectrum for all pulsars in the NANOGrav 15-yr data set.

A more recent custom noise model analysis of the NANOGrav 15-yr dataset \citep{LarsenCNM2025}, in which detailed chromatic models are specified for each pulsar as driven by each pulsar's dataset, provides striking confirmation and refinement of this picture.
By replacing the standard DMX (piecewise-constant DM) model with customized chromatic noise models, the analysis found significant changes to the {ARN} in 19 of 67 pulsars.
The results illuminate each of the three ARN categories identified by \cite{NG15_detchar}.
Among the 12 pulsars {previously classified as} dominated by {intrinsic red noise}, five {no longer showed significant ARN} (PSRs B1953+29, J1747$-$4036, J1802$-$2124, J1946+3417, and J2145$-$0750), {supporting the interpretation that mismodeled chromatic noise contributed to their apparent ARN.}
Among the 13 pulsars dominated by {common red noise}, four lost their ARN detection (PSRs J0437$-$4715, J1600$-$3053, J1738+0333, and J1744$-${1134}), while PSR J1713+0747---one of the most precisely timed pulsars in the array---showed the largest shift, with $\log_{10} A_{\mathrm{RN}} = -14.7 \pm 0.4$ and $\gamma_{\mathrm{RN}} = 3.7 \pm 1.2$ under the custom noise model, compared to a higher amplitude and lower spectral index under DMX.
This shift is driven by the explicit modeling of the two chromatic events (at {modified Julian dates} 54758 and 57509) that had previously leaked power into the {ARN spectrum} and the DM variations (\fig{fig:DM}).
Among {the 42 pulsars previously dominated by white noise, ARN became significant in five} (PSRs J0023+0923, J0636+5128, J1455$-$3330, J1944+0907, and J2043+1711){. These new detections may reflect improved sensitivity to ARN or residual chromatic noise that remains imperfectly modeled.}
Furthermore, 24 of the 37 pulsars without ARN detections showed reduced 95\% upper limits on the red noise amplitude, reflecting the improved noise separation.
The custom noise model analysis \citep{LarsenCNM2025} also found notable changes to the white noise parameters under customized chromatic models.

\begin{figure}
\centering
\includegraphics[width=0.27\textwidth]{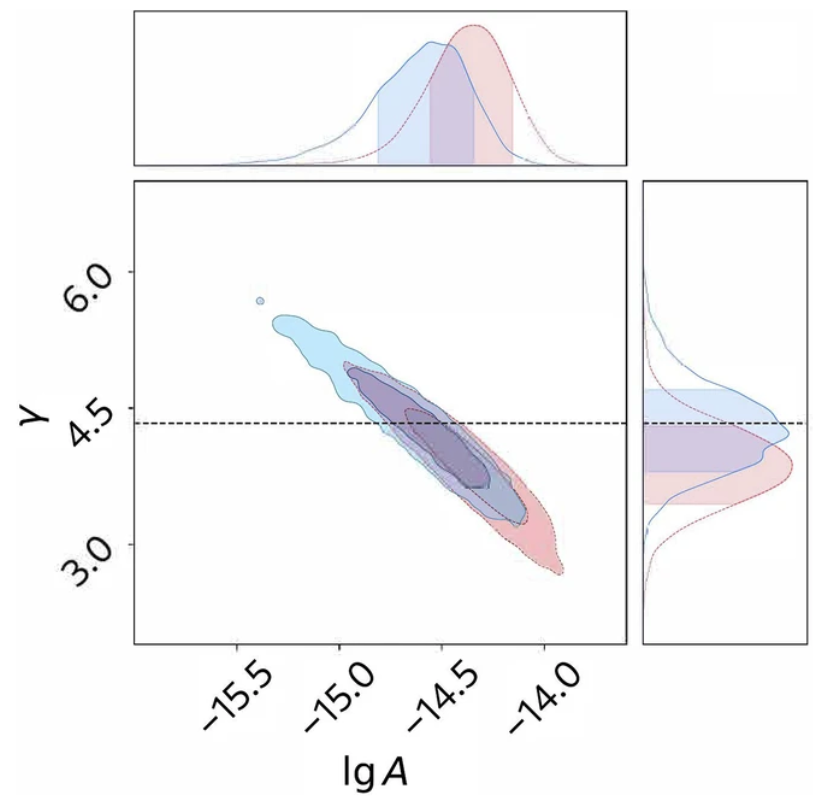}
\includegraphics[width=0.4\textwidth]{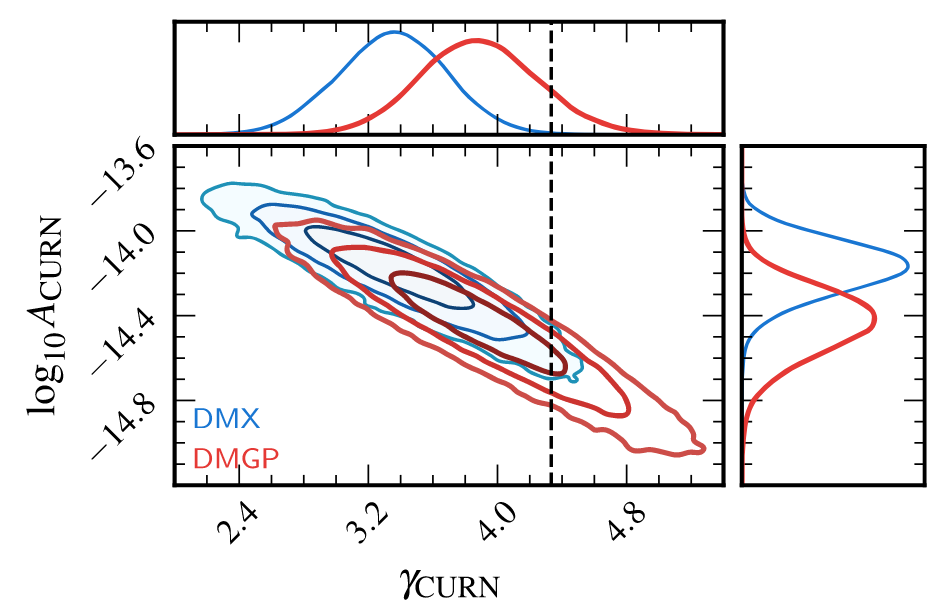}
\includegraphics[width=0.28\textwidth]{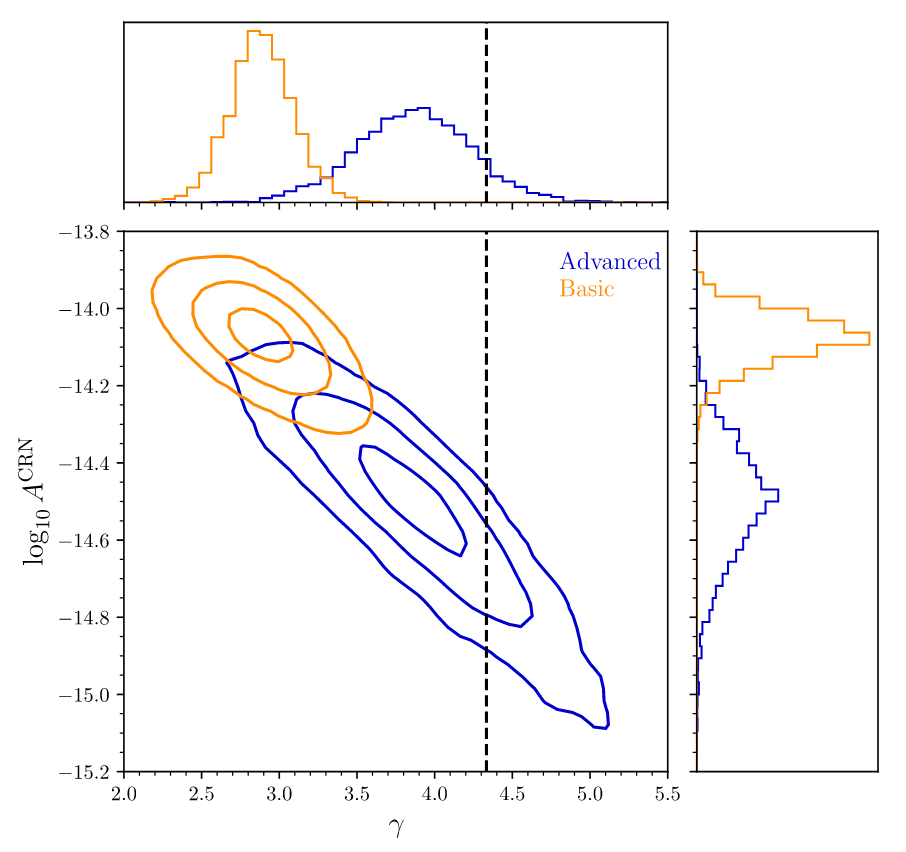}
\caption{How GWB spectral parameter inferences are affected under different noise modeling assumptions, for three different datasets.
\emph{Left:} For EPTA DR2 \citep{EPTA_DR2}, the inclusion of hierarchical priors and additional white noise models (blue) produce lower amplitude and higher spectral index GWB measurements than the original use of only achromatic and chromatic analysis (red) \citep{Goncharov+2025}.
\emph{Middle:} For the NANOGrav 15-yr dataset \citep{NG15_dataset}, a piecewise-constant DM variations model (blue) and a DM Gaussian process model with a power-law spectrum (red) also produce differences between recovered {CURN} parameters, with increased consistency with $\gamma_{{\mathrm{CURN}}} = 13/3$ using the Gaussian process model \citep{NG15_gwb}.
\emph{Right:} For PPTA DR3~\citep{PPTA_DR3}, a set of standard noise models (orange) and customized noise models (blue) result in stark differences between recovered {CURN} properties, again with the custom noise models favoring a steeper {CURN} spectrum much more consistent with $\gamma_{{\mathrm{CURN}}} = 13/3$ \citep{PPTA_noise}.
{The middle and right panels show CURN fits, which measure the common spectrum without including HD correlations.}
Figures from \citep{Goncharov+2025, NG15_gwb, PPTA_noise}.}
\label{fig:dataset_noise_differences}
\end{figure}

Choices in noise modeling continue to have strong bearing on current GW inferences.
\fig{fig:dataset_noise_differences} shows various examples of how different noise modeling choices impact inferred GWB properties across independent PTA datasets: EPTA DR2 \citep{EPTA_DR2, Goncharov+2025}, the NANOGrav 15-yr dataset \citep{NG15_dataset, NG15_gwb, NG15_CNM_GWB}, and PPTA DR3 \citep{PPTA_DR3, PPTA_noise}.
While the detailed prescriptions behind the noise models and the nature of the comparison differs for each of these analyses, the common link is that a deeper refinement of the pulsar noise models impacts the spectral characterization of the GWB.
One may notice the commonality that refined noise modeling tends to {drive} the characterization towards lower amplitudes, and towards the steeper spectral index $\gamma = 13/3$ predicted for SMBHBs; this {can partly arise} as many of the most difficult sources of noise to model are chromatic and produce errors with whiter spectra than low frequency GWs, leaking broadband power into GWB measurements.
{The lower amplitude quoted at $1\,\mathrm{yr}^{-1}$ also reflects covariance with the spectral index.}
However, there is nothing intrinsic in these analyses that {suggests} the GWB should be closer to $\gamma = 13/3$, since {the spectral index is allowed to vary in these fits}.

Overall, it is clear that pulsar noise characterization impacts the inferred power spectrum of the GWB (\eq{eq:strain_spectrum_circular_bhbs}), which ultimately impact understanding about the source of the GWB, as the GWB power spectrum is used to fit various SMBHB population parameters~\citep{NG15_astro} and used to test different hypotheses about the GWB's source \citep{Afzal2023}.
This observations motivate continued, rigorous development of more accurate pulsar noise models.
Sophisticated noise modeling approaches will only become more important for future prospects to identify individual binary sources of nanoHertz GWs (\sect{sec:CW}). As discussed in \sect{sec:noise_breakthrough}, the development of custom noise models has already shifted the inferred GWB spectral index toward the $\gamma = 13/3$ predicted for circular SMBHBs (\eq{eq:strain_spectrum_circular_bhbs}), underscoring the tight coupling between noise characterization and astrophysical interpretation.

{The most recent efforts in custom chromatic noise modeling using the NANOGrav 15 yr dataset {by Agarwal et al.}~\citep{NG15_CNM_GWB} {and Larsen et al.}~\citep{LarsenCNM2025} further highlight the scientific impacts of detailed pulsar noise modeling beyond the spectral characterization of the GWB.
{The custom noise models increased the HD/CURN Bayes factor by roughly a factor of seven relative to the original NANOGrav 15-yr DMX analysis. The nominal significance estimated using the frequentist optimal statistic increased only modestly, from $3.16\sigma$ to $3.32\sigma$. This illustrates that improved noise modeling need not produce a large increase in frequentist detection significance, and that additional high-precision timing data remain essential.}
{Noise modeling also affects tests of gravity, since mismodeled noise can mimic departures from the HD correlation pattern. Agarwal et al.}~\citep{NG15_CNM_GWB} {found that custom noise models reduced support for an additional scalar-transverse polarization and tightened the upper limit on its amplitude.}}

{Finally, refined pulsar noise models have already been found to strongly impact the search for {CWs from individual binaries} \citep{Baker2020_MDC, Falxa2023, NG15_CW}; {Agarwal et al.}~\citep{NG15_CNM_GWB} {revisited SDSS J0729+4008 and SDSS J1536+0441, the two candidates highlighted by the targeted search}~\citep{Agarwal2026}. {Their custom noise models reduced the evidence for CW emission from both systems and tightened the upper limits on their chirp masses.}
{In the all-sky search, support for a CW at 4--4.5\,nHz increased but remained statistically insignificant when modeled alongside an HD-correlated GWB.}}

\section{The International Pulsar Timing Array}
\label{sec:ipta}
\label{sec:data_combination}

The IPTA was established in 2010~\cite{Hobbs2010} to combine the datasets of the regional PTAs into a single array. Beyond GW searches, the IPTA has pursued a broad science program, including constructing a pulsar-based timescale~\cite{Hobbs2020_timescale} and constraining Solar system ephemerides and planetary masses, including limits on the mass of Ceres and a hypothetical Planet~9 \cite{Caballero2018_solar, Vallisneri2020}.

An important early effort was the IPTA Mock Data Challenge (MDC) program, which provided standardized simulated datasets to benchmark and cross-validate the analysis pipelines developed independently by each regional PTA. The first MDC~\cite{Ellis2012_MDC1,vanHaasteren2013_MDC1,Taylor2013_MDC1} tested the recovery of injected GWB and CW signals in realistic timing data, revealing significant differences in how groups handled noise estimation and signal extraction. The second MDC~\cite{Hazboun2019_MDC,Baker2020_MDC} introduced more complex scenarios, including non-Gaussian noise and multiple simultaneous signals. These exercises were instrumental in building confidence that the diverse software pipelines across the collaboration would yield consistent results when applied to real data---a prerequisite for the coordinated 2023 evidence announcements.

{The IPTA's DR1}~\cite{IPTA_DR1} combined data from all three founding PTAs and placed a GWB upper limit, with noise characterization by Lentati et al.~\cite{Lentati2016}. {DR2}~\cite{IPTA_DR2} expanded the array to 65 pulsars with baselines up to 30 years, and was followed by a GWB search~\cite{Antoniadis2022} and a CW search~\cite{Falxa2023}. In anticipation of a detection, the IPTA Detection Committee established a formal checklist of validation criteria~\cite{Allen2023_checklist} that informed the interpretation of the 2023 regional results, which were subsequently compared in a joint analysis~\cite{3P+paper}.

A persistent challenge for the IPTA is the long timescale of full data combination. Aligning time standards, harmonizing backend offsets, and fitting consistent noise models (\sect{subsec:noise}) across heterogeneous datasets is labor-intensive, causing IPTA releases to lag years behind the regional state of the art. However, GWB sensitivity scales with the number of pulsar pairs, and measurements of the full {HD} curve (\sect{sec:stochastic_background}), GWB anisotropy (\sect{sec:anisotropy}), and individually resolvable CW sources (\sect{sec:CW}) all benefit from a larger, globally distributed array. Several recent methods address the combination bottleneck, but first we summarize the quantitative scaling.

\subsection{The Need for Data Combination: Scaling Laws}
\label{sec:ipta_scaling}

The sensitivity of a PTA depends on the number of pulsars $N$, the observing timespan $T_{\mathrm{obs}}$, the timing precision $\sigma_{\mathrm{TOA}}$, and the observing cadence $c$ (or equivalently the mean interval $\Delta t = 1/c$). Tab.~\ref{tab:scaling_laws} collects the scaling relations for each signal class; we summarize the key points below.

\begin{table}[t]
\centering
\caption{S/N scaling laws for each PTA signal class. Here $N$ is the number of pulsars, $T_{\mathrm{obs}}$ is the observing timespan, $\sigma_{\mathrm{TOA}}$ the timing precision, and $c$ the observing cadence. {The GWB entries give asymptotic cross-correlation S/N scalings for a fixed background amplitude, $\gamma=13/3$, and comparable pulsars with white timing noise, regular sampling, and similar sky coverage. White noise dominates at all frequencies in the weak regime; in the intermediate regime, the GWB dominates at low frequencies and white noise at high frequencies}~\cite{siemens2013}. CW scalings are from~\cite{Ellis_2012, M17}{; scalings for a burst with memory (BWM) are from~}\cite{vanHaasteren2010, Madison2014}.}
\label{tab:scaling_laws}
\renewcommand{\arraystretch}{1.4}
\begin{tabular}{l l l}
\hline\hline
\textbf{Signal class} & $\textbf{S/N}~~\boldsymbol{(\rho)}$ & \textbf{References} \\
\hline
GWB (weak signal) & $N \, T_{\mathrm{obs}}^{{13/3}} \left({c}/{\sigma_{\mathrm{TOA}}^2}\right)$ & \cite{siemens2013} \\
GWB (intermediate) & $N \, T_{\mathrm{obs}}^{{1/2}} \left({c}/{\sigma_{\mathrm{TOA}}^2}\right)^{{3/26}}$ & \cite{siemens2013} \\
CW (individual SMBHB) & $\left({N \, T_{\mathrm{obs}} \, c}/{\sigma_{\mathrm{TOA}}^2}\right)^{1/2}$ & \cite{Ellis_2012, M17} \\
{BWM} & $N^{1/2} \, T_{\mathrm{obs}}^{3/2} \left({c}/{\sigma_{\mathrm{TOA}}^2}\right)^{1/2}$ & \cite{vanHaasteren2010, Madison2014} \\
\hline\hline
\end{tabular}
\end{table}

\begin{figure*}[t]
\centering
    \includegraphics[width=\textwidth]{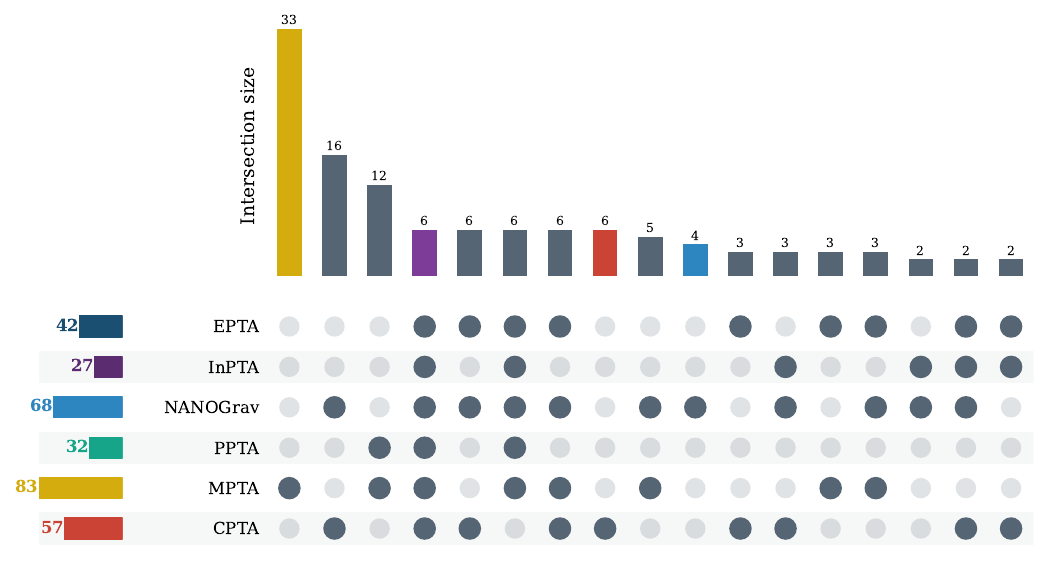}
    \caption{UpSet plot showing the overlap of pulsars currently being timed by regional PTAs contributing to the IPTA, based on each array's most extensive data release. For EPTA, this is Desvignes et al. (2016)~\cite{Desvignes2016}, not EPTA DR2. These pulsars have not necessarily been selected for IPTA DR3, which will be defined by its own selection criteria. Each column corresponds to pulsars timed by exactly the combination of PTAs indicated by the filled circles below; the bar height gives the count. Horizontal bars on the left show each PTA's total. The 13 intersections containing a single pulsar are listed in \tab{tab:ipta_overlap}.}
    \label{fig:ipta_upset}
\end{figure*}

\begin{table}[t]
\centering
\caption{Tabular companion to \fig{fig:ipta_upset}. A filled circle indicates membership in that PTA. The upper part lists intersections containing two or more pulsars; the lower part lists the 13 single-occupant intersections by name. The six PTAs collectively time 131 unique pulsars. This does not define IPTA DR3, which will be determined by its own selection criteria.}
\label{tab:ipta_overlap}
\renewcommand{\arraystretch}{1.1}
\setlength{\tabcolsep}{3pt}
\begin{tabular}{r l cccccc}
\hline\hline
\textbf{Count} & \textbf{Pulsar} & \textbf{EPTA} & \textbf{InPTA} & \textbf{NANOGrav} & \textbf{PPTA} & \textbf{MPTA} & \textbf{CPTA} \\
\hline
33 & & & & & & $\bullet$ & \\
16 & & & & $\bullet$ & & & $\bullet$ \\
12 & & & & & $\bullet$ & $\bullet$ & \\
 6 & & $\bullet$ & $\bullet$ & $\bullet$ & $\bullet$ & $\bullet$ & $\bullet$ \\
 6 & & & & & & & $\bullet$ \\
 6 & & $\bullet$ & $\bullet$ & $\bullet$ & $\bullet$ & $\bullet$ & \\
 6 & & $\bullet$ & & $\bullet$ & & $\bullet$ & $\bullet$ \\
 6 & & $\bullet$ & & $\bullet$ & & & $\bullet$ \\
 5 & & & & $\bullet$ & & $\bullet$ & \\
 4 & & & & $\bullet$ & & & \\
 3 & & $\bullet$ & & $\bullet$ & & $\bullet$ & \\
 3 & & & $\bullet$ & $\bullet$ & & & $\bullet$ \\
 3 & & $\bullet$ & & & & $\bullet$ & \\
 3 & & $\bullet$ & & & & & $\bullet$ \\
 2 & & $\bullet$ & $\bullet$ & $\bullet$ & & & $\bullet$ \\
 2 & & & $\bullet$ & $\bullet$ & & & \\
 2 & & $\bullet$ & $\bullet$ & & & & $\bullet$ \\
\hline
 1 & J0437$-$4715  & & $\bullet$ & $\bullet$ & $\bullet$ & $\bullet$ & \\
 1 & J0614$-$3329  & & & $\bullet$ & $\bullet$ & $\bullet$ & \\
 1 & J1455$-$3330  & $\bullet$ & $\bullet$ & $\bullet$ & & $\bullet$ & \\
 1 & J1614$-$2230  & & $\bullet$ & $\bullet$ & & $\bullet$ & \\
 1 & J1741+1351    & & & $\bullet$ & $\bullet$ & & $\bullet$ \\
 1 & J1744$-$1134  & $\bullet$ & $\bullet$ & & $\bullet$ & $\bullet$ & $\bullet$ \\
 1 & J1824$-$2452A & & & & $\bullet$ & & \\
 1 & J1832$-$0836  & & & $\bullet$ & $\bullet$ & $\bullet$ & $\bullet$ \\
 1 & J1857+0943    & $\bullet$ & $\bullet$ & $\bullet$ & $\bullet$ & & $\bullet$ \\
 1 & J1911$-$1114  & $\bullet$ & & & & $\bullet$ & $\bullet$ \\
 1 & J1939+2134    & $\bullet$ & $\bullet$ & $\bullet$ & $\bullet$ & & \\
 1 & J2150$-$0326  & & & & & $\bullet$ & $\bullet$ \\
 1 & J2234+0944    & & & $\bullet$ & & $\bullet$ & $\bullet$ \\
\hline
131 & & \multicolumn{6}{l}{\textbf{Total unique pulsars}} \\
\hline\hline
\end{tabular}
\end{table}

{Longer observations initially improve GWB sensitivity rapidly because they reach lower frequencies, where binaries are more numerous per frequency bin (\eq{eq:delta_N}) and the background is stronger (\eq{eq:strain_spectrum_circular_bhbs}). Once the GWB dominates the lowest-frequency bins, S/N growth slows, eventually approaching $T_{\mathrm{obs}}^{1/2}$, with only a weak dependence on cadence and timing precision. In both regimes, the cross-correlation S/N scales as $\rho\propto N$ for comparably well-timed pulsars}~\cite{siemens2013} (tab.~\ref{tab:scaling_laws}).{\footnote{{We follow Siemens et al.'s terminology~\cite{siemens2013}. The intermediate regime is included in the ``strong-signal'' discussion of Sesana \& Figueroa~\cite{Sesana2025}. In the formal limit where the GWB dominates all sampled frequencies up to $c/2$, the scaling becomes $\rho\propto N(cT_{\mathrm{obs}})^{1/2}$, independent of timing precision.}}} {Compared with the intermediate GWB regime}, CW and BWM sensitivities scale as $(c/\sigma_{\mathrm{TOA}}^2)^{1/2}$, so investments in wideband timing~\cite{Pennucci2014, Lentati2017_wideband, Alam2021_wideband, Nobleson2022, Sharma2022, Curylo2023, Agazie2025_wideband, Susobhanan2025_vela, Susobhanan2025_revisiting}, which lowers $\sigma_{\mathrm{TOA}}$ by averaging over broader frequency coverage and better constraining DM variations, or increased cadence have a stronger effect on the S/N. The BWM scaling carries an additional $T_{\mathrm{obs}}^{3/2}$ factor, making long uninterrupted baselines critical for memory searches.

Combining datasets across PTAs {can improve} all of these quantities: it increases $N$, extends $T_{\mathrm{obs}}$, and for pulsars observed by multiple PTAs, this increases the effective cadence $c$ and broadens the radio frequency coverage, which improves noise characterization, and thereby lowers the effective $\sigma_{\mathrm{TOA}}$. {For the GWB in the intermediate regime, the priority is to time many pulsars, with relatively modest gains from further improving their timing precision. CW searches place greater emphasis on exquisitely well-timed pulsars, with a weaker dependence on pulsar number} {(tab.~\ref{tab:scaling_laws})~\cite{siemens2013}. For a targeted CW search, additional observing time can be concentrated on pulsars with low noise at the source frequency and favorable antenna responses to its sky position} (\sect{subsec:pta_response}){~\cite{M17}. Well-timed pulsars close to the source on the sky can also improve its localization~\cite{Petrov2024}.}

Beyond raw sensitivity, a global PTA provides qualitatively better sky coverage. Regional PTAs {observe overlapping pulsar samples due to their telescope locations}. The IPTA combines these into a nearly all-sky array. {For GWB anisotropy, estimates of the monopole, dipole, quadrupole and higher multipoles of the sky power are generally statistically correlated: their pulsar-pair responses are not orthogonal under the noise weighting of a finite, unevenly distributed array~\cite{ani2020,ani2021}. Broader coverage can help separate these components, but their reconstruction must still use the actual array response and account for leakage from unresolved angular structure (\sect{sec:fisher_matrix} and \sect{sec:anisotropy_challenges}).}

The degree of overlap between arrays is illustrated in the UpSet plot of \fig{fig:ipta_upset}, with the full intersection data listed in \tab{tab:ipta_overlap}. {The} MPTA times 83 pulsars in total, but 33 are unique to that array. Of InPTA's 27 pulsars, all are also observed by at least one other PTA---InPTA has no unique pulsars, and neither does the EPTA. Six pulsars are observed by all six PTAs (the column with all six circles filled). Only one pulsar is unique to the PPTA: J1824$-$2452A (PSR~B1821$-$24A). This extensive overlap motivates the \textsc{Lite} framework's {figure of merit (FoM) approach} {(\sect{sec:lite})} as a first step to data combination: rather than merging all available TOAs, one selects the highest quality dataset per pulsar.

\subsection{The \textsc{Lite} method: three steps to combination}
\label{sec:lite}

Addressing this challenge, Larsen et al.~\cite{LarsenEtAl2025} proposed the \textsc{Lite} method, a principled framework for rapid PTA data combination. Rather than attempting to merge all datasets in one step, \textsc{Lite} decomposes the process into three deliberate stages, each of which yields scientifically valuable intermediate products.

\begin{enumerate}
    \item \textbf{Lite data sets.} At the first stage, for each pulsar observed by multiple PTAs, the single dataset with the highest {FoM} is selected. The FoM is a scaling proxy for the expected S/N of a given GW signal (Tab.~\ref{tab:scaling_laws}). This step immediately yields a ``best of'' compilation: a global array assembled from the highest-quality data per pulsar, without the need for detailed cross-calibration.

    \item \textbf{{Early combined data sets.}} The second stage prioritizes combination of the most sensitive pulsars, again ranked by their FoM. By combining the top-tier pulsars first, one can rapidly construct an intermediate dataset that already captures much of the total sensitivity. This staged approach mirrors the strategy of early data releases in other fields (e.g.\ Gaia), enabling preliminary science well before the full dataset is complete.

    \item \textbf{Fully combined data sets.} Finally, the traditional endpoint is reached, where all available data are harmonized and combined. This stage remains essential for definitive analyses, but by the time it arrives, much science has already been extracted from the earlier stages.
\end{enumerate}

There are several advantages to this step-by-step approach. It ensures that new PTA data can be rapidly incorporated into global analyses, reducing the time lag between data acquisition and astrophysical inference. It provides multiple checkpoints where results can be cross-validated across different stages, identifying biases or systematics early. And it allows flexibility: one can construct GWB-optimized Lite sets, CW-optimized Lite sets, or even burst-optimized sets, depending on the signal class of interest.

A key innovation of \textsc{Lite} is its reliance on FoMs to guide data selection. The per-pulsar FoMs {of Larsen et al.~\cite{LarsenEtAl2025} follow the timespan, cadence, and timing-precision dependences in} tab.~\ref{tab:scaling_laws}{, using the intermediate regime for the GWB}. Since the FoMs weight $T_{\mathrm{obs}}$, $\sigma_{\mathrm{TOA}}$, and cadence $c$ differently, the optimal Lite data set depends on the target signal. These metrics provide a quantitative basis for selecting among overlapping datasets.

Applying this framework to IPTA DR2, {Larsen et al.}~\cite{LarsenEtAl2025} showed that {DR2 Lite recovers} the {CURN} signal with amplitudes consistent with the fully combined dataset, albeit with somewhat broader posteriors. {With $\gamma=13/3$ fixed}, DR2 Lite yielded a {CURN} amplitude of $A \approx 4.8^{+1.8}_{-1.8} \times 10^{-15}$, compared to $A \approx 3.9^{+1.0}_{-1.0} \times 10^{-15}$ for the full DR2 {(95\% credible intervals)}. {Their tests removing individual pulsars suggest that intrinsic pulsar noise contributes to the higher Lite amplitude, although the difference is not statistically significant. With $\gamma$ free, constraints} on the spectral index are {weaker}, but still consistent. {Neither dataset provided strong evidence for HD correlations.}

\subsection{FrankenStat: likelihood-level data combination without merging TOAs}

A complementary approach avoids {fitting a single timing model to merged TOAs}. FrankenStat~\cite{WrightEtAl2025FrankenStat} combines {the timing residuals} for each pulsar observed by multiple PTAs, {modeling its intrinsic red noise and GW signals jointly while retaining separate timing-model perturbations for each dataset. The resulting} ``FrankenPulsar'' object {can} be dropped directly into {compatible} Bayesian pipelines without modifying the analysis code. {Constructing these objects takes} minutes on a laptop{; the subsequent noise and GW analyses must still be run}.

In simulation studies, the FrankenStat and traditionally combined datasets yield nearly identical S/N distributions for a {HD}-correlated background (\eq{eq:HD_split}), with sensitivity curves differing at the $\lesssim 1\%$ level when full noise models (\sect{subsec:noise}) are included~\cite{WrightEtAl2025FrankenStat}. FrankenStat does not replace fully harmonized IPTA releases, which remain essential for precision inference and careful treatment of cross-dataset systematics, but it provides a fast surrogate that enables routine cross-PTA checks and rapid iteration whenever a regional PTA updates a key pulsar or noise model.

\subsection{Fourier-space combination and global Gibbs sampling}

A third strategy {performs the global analysis} in Fourier space. Valtolina \& van Haasteren~\cite{ValtolinavanHaasteren2024} {approximate each pulsar's regularized posterior for its sine and cosine coefficients by a Gaussian, specified by its mean and full covariance. The regularization prior controls degeneracies between Fourier modes and the timing model. White noise and deterministic signals weakly covariant with the GWB are marginalized locally; intrinsic red noise and DM variations remain in the global inference. Writing the Gaussian summary as $q_i(\mathbf a_i)$ and its regularization prior as $p_{0,i}(\mathbf a_i)$, the global likelihood within this approximation is}
{
\begin{equation}
\mathcal L(\boldsymbol\theta)\propto
\int d\mathbf a\,p(\mathbf a\mid\boldsymbol\theta)
\prod_i\frac{q_i(\mathbf a_i)}{p_{0,i}(\mathbf a_i)}.
\end{equation}
}
{Here $\mathbf a$ collects the Fourier coefficients and $\boldsymbol\theta$ the GWB and noise hyperparameters. The product runs over pulsars. Dividing by $p_{0,i}$ removes the explicit regularization prior, while $p(\mathbf a\mid\boldsymbol\theta)$ introduces the joint covariance, including HD correlations. The Gaussian integral is analytic. Retaining the coefficient means and covariances preserves phase information and provides a common representation for radio and gamma-ray data.}

{\textsc{Lite} initially selects one dataset per pulsar, leaving other datasets for later combination. FrankenStat combines multiple datasets for the same pulsar, retaining separate timing-model perturbations. The Fourier method changes the information passed to the global analysis: coefficient means and covariances replace residuals. These approaches therefore address different stages and can, in principle, be used together.}

Complementing this, Laal et al.~\cite{Laal2024_Gibbs} developed a global Gibbs sampling scheme that {estimates the red-noise power in each pulsar and the cross-power between every pulsar pair at each Fourier frequency. Assuming stationary Gaussian red noise, it alternates draws of the Fourier coefficients and their covariance without imposing a power-law spectrum or HD correlations. This allows the HD pattern to be checked frequency by frequency, showing where it is constrained by the data and whether one correlation curve adequately describes the full band. Their demonstration on simulated data held white-noise parameters fixed.}

\subsection{Outlook for data combination}

The trajectory from cross-PTA validation~\cite{3P+paper} to rapid combination methods captures a broader shift: the immediate priority has moved from establishing the existence of a common signal to precision characterization of the {HD} curve (\sect{sec:stochastic_background}), GWB anisotropy (\sect{sec:anisotropy}), and individually resolvable CW sources (\sect{sec:CW}). {Rapid combination methods can bring new observations into global analyses sooner and demonstrate the scientific gains that motivate the greater investment of time and resources required for full data combination.} {For example, by combining NANOGrav's nine-year data with EPTA and PPTA observations, IPTA DR2 achieved constraints on the CURN spectrum comparable to those from NANOGrav's 12.5-year release~\cite{Antoniadis2022}.}

{Full combination also deepens our understanding of the pulsars and the signals in their timing data. For PSR~J1713+0747, adding NANOGrav and PPTA observations to the EPTA data resolved additional structure in an approximately 100-day DM event and better constrained the underlying DM variations~\cite{Lentati2016}. For PSR~J1730$-$2304, adding PPTA observations broke the degeneracy between intrinsic spin noise and Nan\c{c}ay system noise: the preferred combined model required system noise, with no detected spin noise~\cite{Lentati2016}. Improved timing models can also sharpen pulsar parallax measurements, tightening the distance constraints needed to exploit pulsar terms in CW searches (\sect{subsec:distances})~\cite{IPTA_DR2}.} Integrating these approaches will allow the community to preserve the robustness that established the first evidence for the GWB while unlocking the full discovery space of a truly global PTA.

\section{Complementary Science}
\label{sec:complementary}

\subsection{Picohertz GW Searches}
\label{subsec:picohertz}
{PTA observations can also probe GWs with periods much longer than the observing span, $fT_{\mathrm{obs}}\ll1$. Their slowly varying signals are largely absorbed into fitted timing parameters. Searches therefore use GW-induced contributions to the binary orbital-period derivative $\dot{P}_b$ and the second derivative of the spin period $\ddot{P}$ to constrain CWs and, through correlations across pulsars, the GWB~\cite{DeRocco2023, DeRocco2024, Zheng2025}. These searches could probe wider SMBHBs whose evolution is driven by stars or gas (\sect{sec:modeling_gwb}), and low-frequency cosmological signals (\sect{sec:new_physics}).}

{The GW-induced redshift $v_{\mathrm{GW}}$ is given by \eq{eq:pta_redshift}, with the antenna pattern defined in \eq{eq:beam_pattern}.  When $2\pi fL_a\lesssim1$, both terms must be retained: GWB correlations then depend on frequency and pulsar distances, and cancellation between the Earth and pulsar terms suppresses the response toward lower frequencies~\cite{DeRocco2023}.}

\begin{figure}
    \centering
    \includegraphics[width=\linewidth]{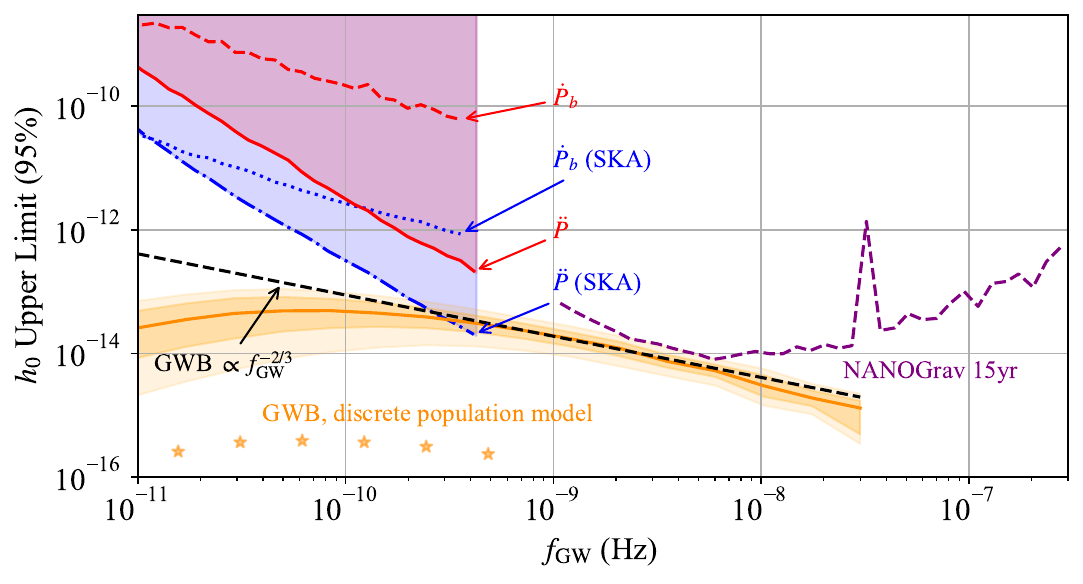}
    \caption{{95\% upper limits on CW strain $h_0$ for circular SMBHBs from $\dot{P}_b$ (red dashed) and $\ddot{P}$ (red solid), with SKAO forecasts using more pulsars at current measurement precision (blue) and NANOGrav 15-yr limits (purple)~\cite{Zheng2025}. Stars show the loudest individual CWs per frequency bin in the quasar-based population model (\sect{sec:modeling_gwb}). Its GWB characteristic strain $h_c$ (orange, with uncertainty bands) and the $f^{-2/3}$ extrapolation (black dashed) are normalized to the NANOGrav GWB at $1\,\mathrm{yr}^{-1}$.}}
    \label{fig:CGW_upbd}
\end{figure}

{A pHz GW contribution to $\dot{P}_b$ must be separated from intrinsic orbital decay, the Shklovskii effect and Galactic acceleration (\eq{eq:pbdot}; \sect{subsec:galacc}). For $\ddot{P}$, red noise must also be included in the uncertainty budget. Zheng et al.~\cite{Zheng2025} searched the two observables separately, obtaining the CW upper limits in \fig{fig:CGW_upbd}. Reaching the modeled sources with the forecast SKAO sample requires substantially better timing-parameter precision.}

\subsection{Mapping Galactic Acceleration}\label{subsec:galacc}

For pulsars in binary orbits, precise PTA data allows us to measure not only the orbital periods of the systems (denoted ${P}_b$), but the change in these periods over time (${\dot{P}}_b$). The change we measure, $\dot{P}_\mathrm{b}^\mathrm{Obs}$, is the combined effect of several factors~\cite{Bell}:

\begin{equation}\label{eq:pbdot}
\dot{P}_\mathrm{b}^\mathrm{Obs} = \dot{P}_\mathrm{b}^\mathrm{Kin} + \dot{P}_\mathrm{b}^\mathrm{GW} + \dot{P}_\mathrm{b}^\mathrm{Gal} + \dot{P}_\mathrm{b}^\mathrm{n} .
\end{equation}

The kinematic or Shklovskii term $\dot{P}_\mathrm{b}^\mathrm{Kin}$ {arises from the changing Doppler shift as the pulsar system moves across the sky}~\cite{Shk70}. It depends on the proper motion and distance of the pulsar system. $\dot{P}_\mathrm{b}^\mathrm{GW}$ is the change in orbital period  due to GW radiation, and depends on the pulsar and companion masses and orbit eccentricity, see~\cite{GW_test_Shapiro_64}. $\dot{P}_\mathrm{b}^\mathrm{Gal}$ is the change due to galactic acceleration, and $\dot{P}_\mathrm{b}^\mathrm{n}$ is a nuisance term which characterizes all other contributions to the time derivative.

The galactic term is of particular interest because it can be used as in~\cite{Chakrabarti_pb_gal, Moran_pbdot} to determine the line-of-sight relative acceleration between the pulsar and the Earth via:

\begin{equation}\label{eq:agal}
{\mathbf{a}_{\mathrm{Gal}}\cdot\mathbf{\hat{r}}=\frac{\dot{P}_\mathrm{b}^\mathrm{Gal}}{P_\mathrm{b}},}
\end{equation}

\begin{figure}
    \centering
    \includegraphics[width=0.8\linewidth]{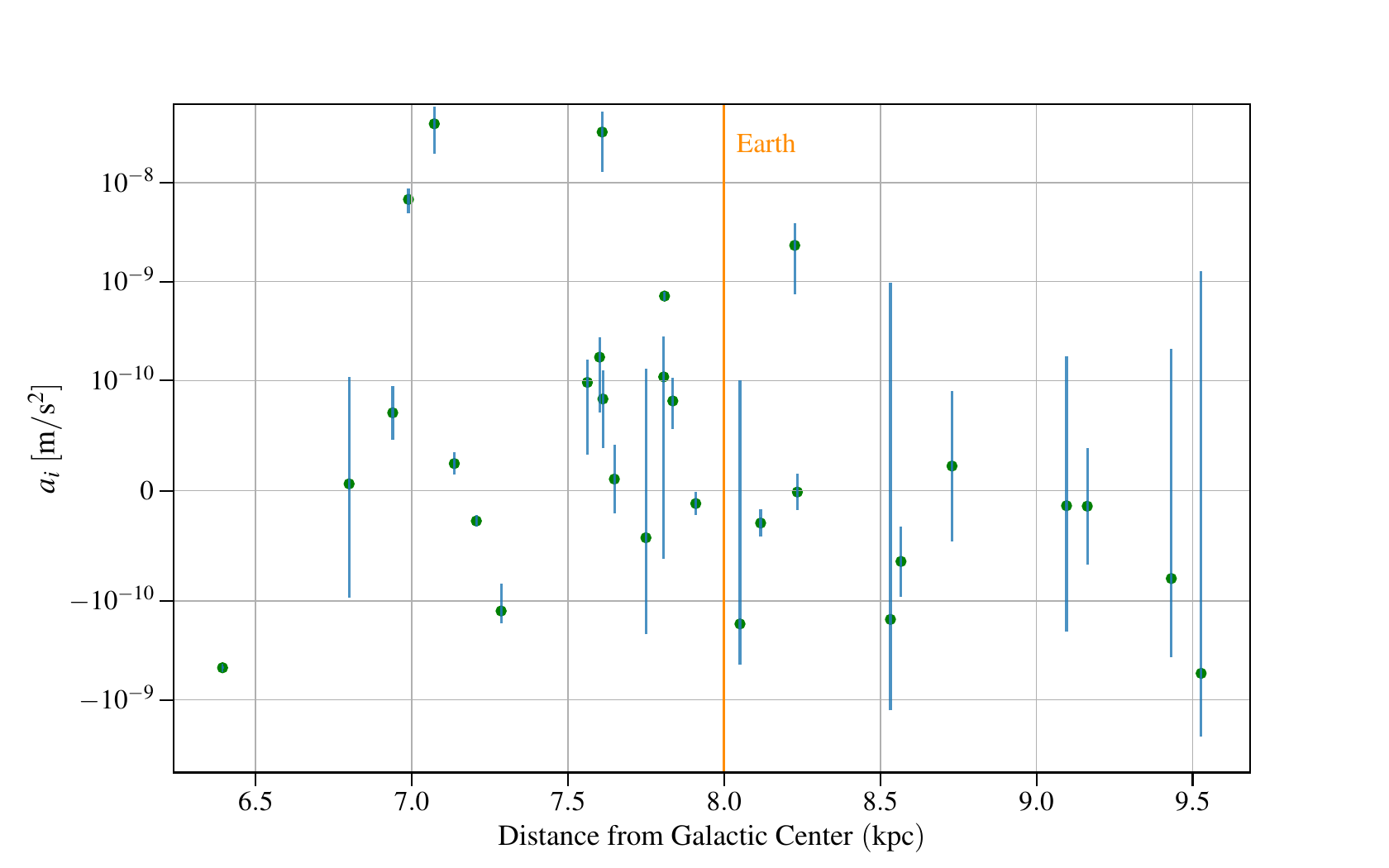}
    \caption{The line-of-sight relative acceleration between the Earth and each binary system from Ref.~\cite{Moran_pbdot} as a function of distance from the center of the Milky Way. The Earth is shown for reference. Error bars do not account for acceleration noise encapsulated in $\dot{P}_\mathrm{b}^\mathrm{n}$.}
    \label{fig:gal_acc_plot}
\end{figure}

{where $\mathbf{a}_{\mathrm{Gal}}$ is the pulsar's Galactic acceleration relative to the Earth and $\mathbf{\hat{r}}$ is the unit vector pointing from the Earth to the pulsar.}  
{Subtracting the Shklovskii and intrinsic GW contributions from the observed $\dot{P}_b$ leaves $\dot{P}_\mathrm{b}^\mathrm{Gal}+\dot{P}_\mathrm{b}^\mathrm{n}$. \fig{fig:gal_acc_plot} shows the corresponding line-of-sight acceleration estimates for 29 binary pulsars~\cite{Moran_pbdot}. Even after excluding known contaminants such as accreting binaries, remaining noise contributions must be accounted for statistically when fitting a Galactic acceleration model.}

{To interpret these data, Ref.~\cite{Moran_pbdot} fits the line-of-sight projection (\eq{eq:agal}) of a local model for the relative Galactic acceleration:}
\begin{equation}\label{eq:gal_pot}
{\mathbf{a}_{\mathrm{Gal}} = a^{\prime}_x x\,\hat{\mathbf{x}} + a^{\prime}_z z\,\hat{\mathbf{z}}.}
\end{equation}

{Here $x$ and $z$ are the pulsar's displacement components relative to the Earth along the fixed unit directions $\hat{\mathbf{x}}$, away from the Galactic Center, and $\hat{\mathbf{z}}$, toward the Galactic North Pole. The coefficients $a^{\prime}_x$ and $a^{\prime}_z$ are the corresponding acceleration gradients.} 

\begin{figure}
    \centering
    \includegraphics[width=0.85\linewidth]{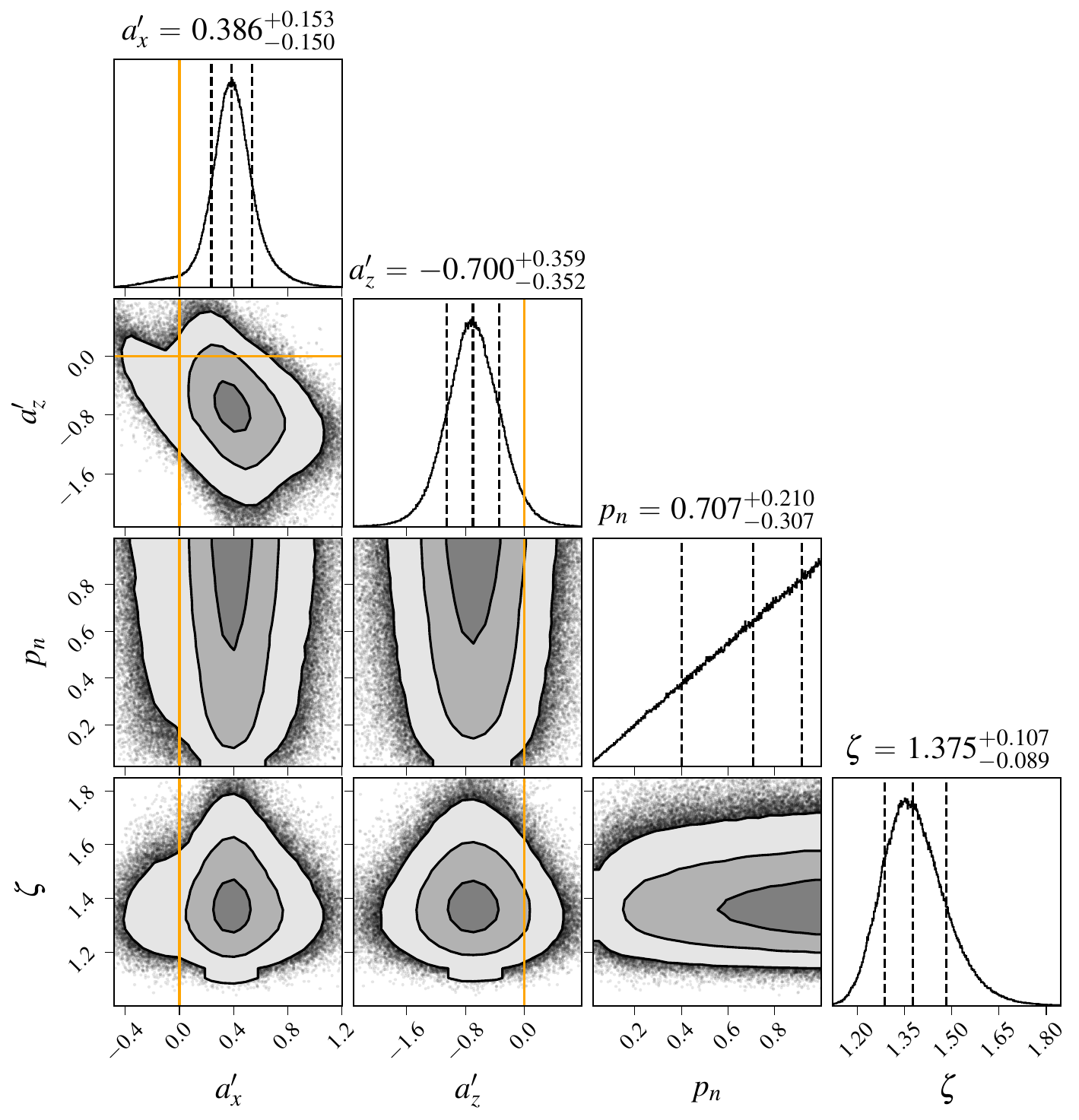}
    \caption{Posterior distributions for the galactic acceleration gradients $a^{\prime}_x$ and $a^{\prime}_z$ and the noise parameters $p_n$ and $\zeta$. The 16th, 50th, and 84th percentiles are shown with black lines. The acceleration gradient values shown have units of $[10^{-10}\ \mathrm{m}\ \mathrm{s}^{-2} \ \mathrm{kpc}^{-1}]$. The orange lines show the null hypothesis where $a^{\prime}_x=a^{\prime}_z=0$. }
    \label{fig:gal_acc_MCMC}
\end{figure}

{Ref.~\cite{Moran_pbdot} models the residual noise using $p_n$, the probability of an additional nuisance acceleration for a pulsar, and $\zeta$, the power-law index of its magnitude distribution. {A Markov chain Monte Carlo analysis} of the 29 pulsars favors this additional noise while retaining evidence for both acceleration gradients (\fig{fig:gal_acc_MCMC}).} 

We use Poisson  equation to write the vertical acceleration gradient as:
\begin{equation}\label{eq:rhod}
    {a^{\prime}_z \simeq -4\pi \rho_d}
\end{equation}
{where $\rho_d$ is the density of the Milky Way disk.} 
The value of $\rho_d$ corresponding to the best-fit $a^{\prime}_z$ from Ref.~\cite{Moran_pbdot} is shown alongside other measurements in tab.~\ref{tab:rhod_vals}. Chakrabarti et al.~\cite{Chakrabarti_pb_gal} independently applied this technique to measure galactic acceleration using binary pulsars. Both methods use \eq{eq:agal} to extract the line-of-sight acceleration from the galactic contribution to $\dot{P}_b$. The key difference is in the treatment of systematics: Chakrabarti et al.\ fit the acceleration gradients directly to their pulsar sample, while Moran et al.\ introduce the nuisance parameters $p_n$ and $\zeta$ to statistically account for pulsars whose $\dot{P}_b$ may be contaminated by unmodeled effects. 

{Precision radial-velocity measurements of stars offer a complementary way to measure Galactic accelerations~\cite{Ravi2019}. The inferred $\rho_d$ is the total local mass density; isolating dark matter requires an independent estimate of the baryonic contribution. The pulsar estimates in tab.~\ref{tab:rhod_vals} are less precise than the quoted Jeans estimates, but do not require an assumption of dynamical equilibrium. Moran et al.~\cite{Moran_pbdot} report differences of up to $3.1\sigma$ from these Jeans estimates, and their measurement does not yet meaningfully constrain dark matter. Better timing and independent pulsar distances, together with improved baryonic density estimates, can strengthen these tests of Galactic mass models and their sensitivity to dark matter.}

\begin{table}[h]
    \centering
    \renewcommand{\arraystretch}{1.3} 
    \begin{tabular}{l l}
        \hline\hline
        Reference & $\rho_d$ ($10^{-2} M_{\odot} / \mathrm{pc}^3$) \\
        \hline
        Ref.~\cite{Moran_pbdot} & {$4.0^{+2.0}_{-2.0}$} \\
        Ref.~\cite{Kipper_model} & $5.6^{+3.1}_{-3.1}$ \\
        Ref.~\cite{holmberg2000local} & $10.2^{+1.0}_{-1.0}$ \\
        Ref.~\cite{Creze} & $7.6^{+1.5}_{-1.5}$ \\
        Ref.~\cite{Chakrabarti_pb_gal} & $8^{+5}_{-2}$ \\
        \hline\hline
    \end{tabular}
    \caption{Comparison of $\rho_d$ values from the literature. Refs.~\cite{Moran_pbdot, Chakrabarti_pb_gal} use the binary pulsar method described here. Ref.~\cite{Kipper_model} uses the orbital arc method, and Refs.~\cite{holmberg2000local, Creze} use Jeans analyses.}
    \label{tab:rhod_vals}
\end{table}

\section{Conclusion and Future Outlook}
\label{sec:conclusion}

The independent evidence of a common red-noise process with {HD} correlations by the NANOGrav, EPTA, InPTA, PPTA, CPTA, and MPTA collaborations marks the opening of the nanoHertz GW window. After decades of precision monitoring, the local Galaxy has been successfully transformed into a kiloparsec-scale GW detector. As we transition from evidence to characterization, {eight} themes that mirror the structure of this review define the roadmap for the next decade:

\smallskip
\noindent\textbf{The GWB.}
The recovered GWB amplitude lies at the upper end of early predictions, but there is still some tension with astrophysical GWB models. {If the GWB is produced by SMBHBs, it demonstrates that binaries can overcome the final parsec problem; how commonly they do so and which mechanisms drive them inward remain open questions (\sect{sec:gwb_history} and \sect{sec:modeling_gwb}).} {{Reconciling EM and PTA observations} requires consistent assumptions about black-hole demographics and growth. Sato-Polito et al.~\cite{SatoPolito2024_bigBH} find tension between PTA amplitudes and populations inferred from local scaling relations, with a limited enhancement from repeated mergers at fixed remnant demographics. {Using the Parametrisation of the Massive Black Hole Population for Comparison to Observations model, Toubiana} et al.~\cite{Toubiana2025} obtain joint fits to AGN luminosity functions and the EPTA background amplitude when allowing an enhanced high-redshift AGN population. Establishing consistency between these results requires comparing the predicted local black-hole mass functions and integrated mass densities under common observational constraints and radiative-efficiency assumptions. Revised little-red-dot luminosities provide an additional test of these population models~\cite{Greene2026_LRDluminosities}.} Indeed, new work is already using the GWB amplitude to constrain galactic feedback processes and massive black hole growth~\cite{Tillman2026} {and} infer the $M_{\rm BH}$--$M_{\rm bulge}$ relation out to $z\sim 1$~\cite{Matt2026}{.} {Energetic arguments place ceilings on {astrophysical GWBs}~\cite{Mingarelli2026}.} Beyond amplitude, the spectral shape of the GWB is itself becoming a measurable quantity: piecewise power-law reconstructions~\cite{Agazie2026_piecewise} and searches for a running spectral index~\cite{NANOGrav15yr_running} {test for departures from a single power law}, while systematic studies of individual pulsar contributions to the signal~\cite{NANOGrav15yr_removing} and harmonic analyses of the angular correlations~\cite{NANOGravHarmonic2024} are cross-checking the evidence for the GWB. A key diagnostic will be signs of spectral discreteness---departures from a smooth $f^{-2/3}$ power law at high frequencies due to the finite number of SMBHBs per frequency bin~\cite{NG15_discreteness} (\sect{sec:scaling_laws}). Such features would {help test the SMBHB origin of the signal and could reveal the contribution of individual binaries before either GWB anisotropy or individual CWs can be detected}.

\smallskip
\noindent\textbf{GWB anisotropy.}
An astrophysical GWB is guaranteed to be anisotropic, and detecting that anisotropy---or resolving the individual SMBHBs that produce it---is a central goal of the next decade (\sect{sec:anisotropy}). {The relative shot-noise anisotropy grows with frequency, with} $C_{\ell>0}/C_0 \propto f^{11/3}$ {in the low-frequency limit for GW-driven binaries (\sect{sec:scaling_laws}), but its detectability also depends on the GWB signal strength and pulsar noise}. {Cross-correlations with galaxy catalogs can reveal the LSS contribution and} constrain the source population and its redshift evolution{, provided the GWB maps have sufficient angular resolution and loud sources can be removed}~\cite{Semenzato2024, SahMukherjee2024} (\sect{sec:cross_correlation}). Complementary analytical estimates by Lin, Lidz \& Ma~\cite{LinLidzMa2026} confirm that shot-noise anisotropy dominates over the LSS contribution by two to three orders of magnitude, and that its frequency dependence can diagnose SMBHB residence times. Extracting these signals faces two systematic hurdles: small-scale leakage bias in current anisotropy estimators~\cite{semenzato2025} (\sect{sec:anisotropy_challenges}), which regularization alone does not fully resolve, and cosmic variance from a finite source population, which can produce realization-dependent deviations from {HD} correlations that mimic anisotropy~\cite{Konstandin:2024fyo}. Addressing both will require new estimator frameworks---per-frequency optimal statistics that account for inter-pair covariance and cosmic variance~\cite{Gersbach2025, Konstandin2026}, and joint inference over resolved and unresolved angular scales. Recent forecasts~\cite{Lemke2025, Konstandin2026} suggest that per-frequency searches offer substantially better prospects than broadband approaches {for narrowband anisotropic signals}.

{For GW-driven binaries, higher chirp masses shorten the residence time in each frequency bin; at fixed merger rate this reduces the source count (\eq{eq:delta_N}). This motivates searches for signs of discreteness in the GWB spectrum alongside searches for individual CWs (\sect{sec:discrete_population} and \sect{sec:CW}).}

\smallskip
\noindent\textbf{{Individual binaries.}}
The search for individual binaries has evolved from all-sky limits (\sect{sec:CW}) to rigorous hypothesis testing of {EM candidates}. Targeted searches for SMBHBs in known galaxy pairs are now underway~\cite{Agarwal2026}, and quasar variability studies suggest that quasars themselves may signpost close binary systems~\cite{CaseyClyde2025}. Coordinating nanoHertz GW follow-up with {multi-messenger EM campaigns}~\cite{BurkeSpolaor2025} will be essential for confirming the first CW detection. The path to detection requires a protocol to rule out red-noise false positives: candidates must survive phase scrambling, demonstrate robustness against correlated background models, and withstand dropout analyses across the array. The new single-source fingerprint $\Upsilon_{ab}$ (\sect{subsec:deterministic_fingerprints}) provides a geometry-dependent discriminant between a CW and the isotropic GWB, and may become a useful tool. {Beyond detection, identifying a CW host and exploiting pulsar terms could reveal the binary's long-term evolution and enable independent measurements of the cosmic distance scale.}

\smallskip
\noindent\textbf{{Angular resolution.}}
{Higher S/N can improve a bright CW's localization within a broad point-source response, while recovering GWB multipoles also depends on the array geometry. Sufficiently precise distance measurements can constrain pulsar-term phases and enable coherent analyses with much better localization. Together, these improvements can narrow EM counterpart searches and expand the angular structure accessible to PTAs.}

\smallskip
\noindent\textbf{New physics.}
The GWB spectrum may encode signals beyond the standard SMBHB picture. {Dark matter provides a further connection: self-interacting dark matter can alter SMBHB hardening and the GWB spectrum, while ULDM motivates dedicated timing and polarization searches (\sect{sec:new_physics}).} While the astrophysical background is expected to fracture into discrete sources at high frequencies (\sect{sec:scaling_laws}), a cosmological background from inflation or phase transitions would remain smooth and isotropic across the PTA band. Searches for signs of discreteness in the GWB~\cite{NG15_discreteness} and for GW memory events~\cite{NANOGrav15yr_memory} are already probing this boundary. Distinguishing astrophysical and cosmological scenarios requires measuring the spectral index for deviations from $\gamma = 13/3$, searching for persistent isotropy where astrophysical models predict resolvability, and testing for non-Einsteinian polarization modes (\sect{sec:non_gr_polarizations}). Searches for transverse polarization modes~\cite{NANOGrav15yr_transverse} have placed the first constraints on scalar and vector contributions---monopolar or dipolar correlations rather than the quadrupolar {HD} curve---which would signal alternative gravity theories or exotic early-Universe sources. A further guaranteed signal is the stochastic nonlinear GW memory background: Unal et al.~\cite{Unal2025_memory} computed this background for both astrophysical and cosmological {parent stochastic GWBs} and showed that the SKAO should be sensitive to the memory contribution from SMBHB mergers, with a characteristic $h_c \propto f$ scaling at the lowest frequencies where the parent spectrum turns over.

\smallskip
\noindent\textbf{Custom noise models.}
Accurate per-pulsar noise characterization underpins every result discussed above (\sect{subsec:noise}). As datasets grow in length and sensitivity, noise misspecification becomes the dominant systematic: it can bias the inferred GWB amplitude, inflate or suppress CW candidates, and corrupt anisotropy maps. Advances in chromatic noise modeling with time-domain kernels~\cite{Hazboun+2025} and the treatment of spatially correlated noise sources such as the solar wind are essential to keep pace with improving data quality. Custom pulsar noise models will remain foundational for PTA science.

\smallskip
\noindent\textbf{{IPTA and data combination.}}
{Rapid combination frameworks---\textsc{Lite}~\cite{LarsenEtAl2025}, FrankenStat~\cite{WrightEtAl2025FrankenStat}, and Fourier-space combination~\cite{ValtolinavanHaasteren2024}---can deliver analyses of combined PTA data sooner and demonstrate the gains that motivate full data combination (\sect{sec:data_combination}). Full combination also improves our understanding of pulsar noise and can help improve pulsar distance measurements, supporting both GWB and CW searches.}

\smallskip
\noindent\textbf{{Complementary science.}}
{We can also extract more physics from the pulsar timing model, in addition to searching for the quadrupolar correlations of the GWB. Secular changes in pulse and orbital periods can constrain pHz GWs, but reaching the modeled wider SMBHB population requires more precise measurements. The same data can map Galactic acceleration and constrain the total local mass density without assuming dynamical equilibrium. Better timing and independent distances can strengthen both applications; interpreting the Galactic density in terms of dark matter also requires independent information about baryons.}

\smallskip
{Two next-generation facilities will transform the PTA landscape. The SKAO~\cite{Shannon2025_SKAPTA} will provide a factor of three to four improvement in {telescope} sensitivity over {existing southern-hemisphere facilities, allowing a much larger sample of millisecond pulsars to be timed precisely. The DSA, formerly} DSA-2000~\cite{Hallinan2019_DSA2000,DSA2000_ScienceBook}, {plans to time up to $\sim$200 millisecond pulsars, with higher-cadence observations of the best pulsars for CW searches.}} {Together, these facilities will improve PTA sensitivity across both hemispheres, strengthening GWB measurements, anisotropy searches, and the prospects for individual CW detections (\sect{sec:resolution} and \sect{sec:ipta_scaling}). Combining these timing data with sufficiently precise distances from timing parallaxes, \emph{Gaia} parallaxes of pulsar companions, and VLBI (\sect{subsec:distances}) could enable coherent analyses of the Earth and pulsar terms, making use of Galactic-scale baselines for much better CW localization (\sect{sec:resolution}). Longer observing spans will also improve measurements of secular changes in timing parameters and strengthen searches in the pHz band (\sect{subsec:picohertz})~\cite{Zheng2025}.}

\smallskip
Evidence for the GWB is just the beginning. The challenge for the next decade is to extract astrophysical and cosmological information from the signal, detect the first individual binary, and extend the GW spectrum to millihertz and picohertz frequencies. With next-generation facilities on the horizon and a growing body of new science already emerging from existing data, there is good reason to believe that PTAs will become a cornerstone of multi-band GW astronomy.

\section*{Acknowledgements}
{\looseness=-1
The authors thank M. Lam for his careful reading of this manuscript, providing a link to the DSA-2000 community document, and for useful discussions about GWB anisotropy and scaling laws. They thank P. Petrov for providing fig. 11. They also thank IPTA colleagues for sending comments on this review, and especially Ma{\l}gorzata Cury{\l}o and Boris Goncharov for sharing additional GWB models for fig.~5 and detailed comments on custom noise models, and G. Shaifullah and J. Verbiest for their comments on fig. 19.  C.\ M.\ F.\ M.\ thanks Ue-Li Pen for useful discussions about the GWB polarization, Meg Urry for suggestion \fig{fig:quasarVmm}, and V. Ozolins for useful conversations. She thanks all her students, past and present, who are the co-authors of this work, for being a constant source of joy. C.\ M.\ F.\ M.\ {was supported in part by National Science Foundation grants PHY-1748958 and PHY-2309135 to the Kavli Institute for Theoretical Physics, where she carried out some of this work, and by National Aeronautics and Space Administration Laser Interferometer Space Antenna Preparatory Science grant 80NSSC24K0440.} C.M.F.M. also thanks the Center for Computational Astrophysics of the Flatiron Institute for support. The Flatiron Institute is supported by the Simons Foundation.
\par}
\newpage


\def\cqg{Class.~Quantum~Grav.}%
\def\natast{Nat.~Astron.}%
\def\grg{Gen.~Relativ.~Gravit.}%
\def\raa{Res.~Astron.~Astrophys.}%
\def\ptep{Prog.~Theor.~Exp.~Phys.}%
\def\expa{Exp.~Astron.}%
\def\aj{AJ}%
\def\actaa{Acta Astron.}%
\def\araa{ARA\&A}%
\def\apj{ApJ}%
\def\apjl{ApJL}%
\def\apjs{ApJS}%
\def\ao{Appl.~Opt.}%
\def\apss{Ap\&SS}%
\def\aap{A\&A}%
\def\aapr{A\&A~Rev.}%
\def\aaps{A\&AS}%
\def\azh{AZh}%
\def\baas{BAAS}%
\def\bac{Bull. astr. Inst. Czechosl.}%
\def\caa{Chinese Astron. Astrophys.}%
\def\cjaa{Chinese J. Astron. Astrophys.}%
\def\icarus{Icarus}%
\def\jcap{J. Cosmology Astropart. Phys.}%
\def\jrasc{JRASC}%
\def\mnras{MNRAS}%
\def\memras{MmRAS}%
\def\na{New A}%
\def\nar{New A Rev.}%
\def\pasa{PASA}%
\def\pra{Phys.~Rev.~A}%
\def\prb{Phys.~Rev.~B}%
\def\prc{Phys.~Rev.~C}%
\def\prd{Phys.~Rev.~D}%
\def\pre{Phys.~Rev.~E}%
\def\prl{Phys.~Rev.~Lett.}%
\def\pasp{PASP}%
\def\pasj{PASJ}%
\def\qjras{QJRAS}%
\def\rmxaa{Rev. Mexicana Astron. Astrofis.}%
\def\skytel{S\&T}%
\def\solphys{Sol.~Phys.}%
\def\sovast{Soviet~Ast.}%
\def\ssr{Space~Sci.~Rev.}%
\def\zap{ZAp}%
\def\nat{Nature}%
\def\iaucirc{IAU~Circ.}%
\def\aplett{Astrophys.~Lett.}%
\def\apspr{Astrophys.~Space~Phys.~Res.}%
\def\bain{Bull.~Astron.~Inst.~Netherlands}%
\def\fcp{Fund.~Cosmic~Phys.}%
\def\gca{Geochim.~Cosmochim.~Acta}%
\def\grl{Geophys.~Res.~Lett.}%
\def\jcp{J.~Chem.~Phys.}%
\def\jgr{J.~Geophys.~Res.}%
\def\jqsrt{J.~Quant.~Spec.~Radiat.~Transf.}%
\def\memsai{Mem.~Soc.~Astron.~Italiana}%
\def\nphysa{Nucl.~Phys.~A}%
\def\physrep{Phys.~Rep.}%
\def\physscr{Phys.~Scr}%
\def\planss{Planet.~Space~Sci.}%
\def\procspie{Proc.~SPIE}%
\let\astap=\aap
\let\apjlett=\apjl
\let\apjsupp=\apjs
\let\applopt=\ao

\end{document}